\documentclass[a4paper,fleqn,usenatbib]{mnras}

\usepackage[T1]{fontenc}
\usepackage{ae,aecompl}

\usepackage{graphicx}	
\usepackage{amsmath}	
\usepackage{amssymb}	

\newcommand{\blue}{\textcolor{blue}}

\title[Radiative turbulence]{Kinetic turbulence in shining pair plasma: intermittent beaming and thermalization by radiative cooling}

\author[V. Zhdankin et al.]{
Vladimir Zhdankin,$^{1}$\thanks{Einstein fellow} \thanks{E-mail: zhdankin@princeton.edu}
Dmitri A. Uzdensky,$^{2}$
Gregory R. Werner,$^{2}$
\newauthor
Mitchell C. Begelman$^{3,4}$ \\
$^{1}$Department of Astrophysical Sciences, Princeton University, Peyton Hall, Princeton, NJ 08544, USA\\
$^{2}$Center for Integrated Plasma Studies, Department of Physics, 390 UCB, University of Colorado, Boulder, CO 80309, USA\\
$^{3}$JILA, University of Colorado and NIST, 440 UCB, Boulder, CO 80309, USA\\
$^{4}$Department of Astrophysical and Planetary Sciences, 391 UCB, Boulder, CO 80309, USA\\}

\date{Accepted XXX. Received YYY; in original form ZZZ}

\pubyear{2020}

\begin{document}
\label{firstpage}
\pagerange{\pageref{firstpage}--\pageref{lastpage}}
\maketitle

\begin{abstract}
High-energy astrophysical systems frequently contain collisionless relativistic plasmas that are heated by turbulent cascades and cooled by emission of radiation. Understanding the nature of this radiative turbulence is a frontier of extreme plasma astrophysics. In this paper, we use particle-in-cell simulations to study the effects of external inverse Compton radiation on turbulence driven in an optically thin, relativistic pair plasma. We focus on the statistical steady state (where injected energy is balanced by radiated energy) and perform a parameter scan spanning from low magnetization to high magnetization ($0.04 \lesssim \sigma \lesssim 11$). We demonstrate that the global particle energy distributions are quasi-thermal in all simulations, with only a modest population of nonthermal energetic particles (extending the tail by a factor of $\sim 2$). This indicates that nonthermal particle acceleration (observed in similar non-radiative simulations) is quenched by strong radiative cooling. The quasi-thermal energy distributions are well fit by analytic models in which stochastic particle acceleration (due to, e.g., second-order Fermi mechanism or gyroresonant interactions) is balanced by the radiation reaction force. Despite the efficient thermalization of the plasma, nonthermal energetic particles do make a conspicuous appearance in the anisotropy of the global momentum distribution as highly variable, intermittent beams (for high magnetization cases). The beamed high-energy particles are spatially coincident with intermittent current sheets, suggesting that localized magnetic reconnection may be a mechanism for kinetic beaming. This beaming phenomenon may explain rapid flares observed in various astrophysical systems (such as blazar jets, the Crab nebula, and Sagittarius A*). \\
\end{abstract}




\section{Introduction} \label{sec1}

High-energy astrophysical systems often contain dilute, hot plasmas in a turbulent state. Examples include pulsar wind nebulae, supernovae remnants, radiatively inefficient accretion flows onto either stellar mass black holes in X-ray binaries (XRBs) or supermassive black holes in active galactic nuclei (AGN), jets emanating from XRBs or AGN, giant radio lobes, the intracluster medium, and gamma-ray bursts (GRBs). More exotic possibilities include the mass flows resulting from tidal disruption events and compact object (e.g., neutron star) mergers. These plasmas often contain relativistic particles that are efficient emitters of radiation, providing an energy sink to the system and yielding radiative signatures that are ultimately observed from Earth. To develop accurate physical models of these systems, it is important to understand the role of radiative cooling on collisionless plasma turbulence. Likewise, to interpret observations and constrain the physical conditions in these systems, it is crucial to understand the radiative signatures of turbulence.

Radiative emission in these environments is commonly due to synchrotron, external inverse Compton (IC), and synchrotron self-Compton (SSC) mechanisms (in contrast to collisional mechanisms such as Bremsstrahlung). For an optically thin plasma, the resulting photons channel energy out of the system. A detailed understanding of the radiative emission requires statistical knowledge of charged particles at the microscopic level, where the plasma is generally dynamic (due to kinetic turbulence and other physical processes) and out of thermal equilibrium (since the collisional timescale is much longer than the relevant dynamical timescales). Thus, a proper study of radiative collisionless plasma demands use of a first-principles kinetic model.

The consequences of radiative cooling have recently been investigated in kinetic particle-in-cell (PIC) plasma simulations for a number of high-energy astrophysical settings and processes. This includes relativistic magnetic reconnection \citep{jaroschek_hoshino_2009, cerutti_etal_2013, cerutti_etal_2014b, cerutti_etal_2014, kagan_etal_2016a, kagan_etal_2016b, hakobyan_etal_2019, werner_etal_2019, schoeffler_etal_2019}, decay of magnetostatic equilibria \citep{yuan_etal_2016, nalewajko_etal_2018}, pulsar wind \citep{cerutti_etal_2017}, and pulsar magnetospheres \citep{cerutti_etal_2016, philippov_spitkovsky_2018}. The (synchrotron and jitter) radiative signatures of collisionless shocks have also been explored \citep{medvedev_spitkovsky_2009, sironi_spitkovsky_2009, kirk_reville_2010, nishikawa_etal_2011}.

In this work, we incorporate radiative cooling into PIC simulations of driven turbulence in relativistic plasma for the first time. This is a numerically unexplored regime of turbulence, which we call {\it radiative turbulence\rm}. Recent analytic works studied radiative turbulence in various relativistic astrophysical contexts, including blazar jets (specifically, BL Lac objects) \citep{uzdensky_2018, sobacchi_lyubarsky_2019} and GRBs \citep{zrake_etal_2019}. There have also been a number of numerical works investigating radiation spectra from test particles in random magnetic fields, mimicking turbulence \citep[e.g.,][]{teraki_takahara_2011}. Our self-consistent numerical simulations of kinetic turbulence with the radiation backreaction force can address several questions that are beyond the scope of these previous studies.

One major question about radiative turbulence is: how does cooling affect the statistical properties of kinetic turbulence? The turbulent cascade progressively transfers energy from fluctuations at large scales (where energy is injected) to those at small scales (where collisionless dissipation mechanisms damp the fluctuations). The heating of the plasma is inhomogeneous in space, leading to localized hot spots where radiative cooling (from synchrotron, IC, and SSC mechanisms) may be important \citep[e.g.,][and references therein]{zhdankin_etal_2016b}. Radiative cooling can thus influence the small-scale dynamics, altering the properties of the kinetic cascade (at scales below the typical particle Larmor radius or inertial length) and coherent structures. Furthermore, the presence of strong cooling allows turbulence to be maintained in the {\it high-magnetization\rm} regime, where the magnetic energy density exceeds the total particle energy density (including rest mass). In this regime, the rapid conversion of magnetic energy to plasma kinetic energy can locally energize the plasma, driving compressions and relativistic flows. Shocks and magnetic reconnection may complicate the nature of turbulence and cause nonthermal particle acceleration. This regime of sustained relativistic turbulence has been studied in the framework of force-free electrodynamics \citep[e.g.,][]{thompson_blaes_1998, cho_2005, cho_lazarian_2013, zrake_east_2016}, which neglects the plasma inertia and does not self-consistently include dissipation channels. It has also been studied with simulations of relativistic magnetohydrodynamics (MHD) \citep{zrake_macfadyen_2011, zrake_2014, takamoto_lazarian_2016}. Our present work is the first to study fully developed turbulence in this relativistic regime with kinetic simulations.

A second major question about radiative turbulence is: what is the effect of radiative cooling on the kinetic aspects of collisionless plasma turbulence, i.e., on the momentum distribution of the charged particles? The particle distribution is linked to the spectrum and time-variability of outgoing radiation, and thus may provide useful constraints for future observational campaigns. Recent studies of kinetic turbulence in relativistic plasmas without radiative cooling have indicated that turbulence produces a significant population of high-energy, nonthermal particles with a power-law energy distribution extending to energies limited only by the outer scale of turbulence \citep{zhdankin_etal_2017, zhdankin_etal_2018b, comisso_sironi_2018, nattila_2019, comisso_sironi_2019}. Since radiative cooling acts mainly on particles in the high-energy tail of the distribution, it may suppress the nonthermal population above certain energies. For sufficiently strong cooling, the power-law tail may either steepen \citep[e.g.,][]{kardashev_1962} or be completely eliminated, leading to a narrow energy distribution. For stochastic particle acceleration described by a Fokker-Planck equation in momentum space with a radiative cooling term, analytic work has demonstrated that the steady-state particle distributions are generally quasi-thermal \citep{schlickeiser_1984, schlickeiser_1985, stawarz_petrosian_2008}. However, despite recent progress \citep[e.g.,][]{wong_etal_2019}, the validity and appropriate form of the Fokker-Planck equation for turbulence is not yet known \citep[e.g., ][]{isliker_etal_2017}, particularly in this physical regime, so first-principles kinetic simulations are necessary to test and further develop the models. One of the main objectives of our present work is to determine the functional form of the steady-state particle distribution, thus unveiling whether or not it can maintain a significant nonthermal component. 

In addition to the energy distribution of particles, another key kinetic quantity is the anisotropy of their momentum distribution, which is linked to the directions of outgoing radiation (assuming particles are relativistic). At any given time, high-energy particles may be coherently beamed in random directions due to localized bulk flows or asymmetric acceleration processes. This can potentially lead to distinct radiative signatures on a global scale, including rapid, intense flares when a beam crosses the line of sight of an observer. An important question is whether dissipative events are sufficiently intermittent and anisotropic to make the turbulent plasma appear inhomogeneous and temporally variable to a distant observer. This has implications for high-energy flares observed in a broad range of astrophysical systems. For example, the beaming of radiation by relativistic flows (minijets) powered by turbulence (or reconnection sites) has been invoked to explain blazar and GRB flares \citep{lyutikov_2006, giannios_etal_2009, narayan_kumar_2009, kumar_narayan_2009, giannios_etal_2010}. Kinetic beaming of high-energy particles has previously been observed in PIC simulations of relativistic magnetic reconnection, being proposed as an explanation for the Crab nebula flares \citep{cerutti_etal_2012, cerutti_etal_2013, cerutti_etal_2014} and blazar flares \citep{nalewajko_etal_2012}. Our present paper investigates the properties of intermittent high-energy particle beams in relativistic turbulence, which is a first step toward connecting to radiative signatures.

In this study, we focus on driven turbulence in relativistic pair plasma cooled by external IC radiation (deferring the cases of synchrotron and SSC cooling to future work). The plasma is assumed to be optically thin, so that radiated energy effectively evacuates the domain, allowing a rigorous statistical steady state to be achieved and maintained. In Section~\ref{sec2}, we analytically calculate the steady-state physical conditions (i.e., temperature) and describe the PIC simulation campaign. In Section~\ref{sec3}, we present an overview of the turbulence statistics in the simulations. In Section~\ref{sec4}, we describe the particle statistics, showing that radiative cooling efficiently thermalizes the bulk of the plasma, and that the steady-state distribution is well fit by a Fokker-Planck model for stochastic particle acceleration (associated with, e.g., the second-order Fermi mechanism or gyroresonant interactions). At the same time, for high magnetization cases, we identify a modest high-energy nonthermal population that is spatially correlated with current sheets and is beamed intermittently in direction. These results have important implications for high-energy astrophysical systems such as blazar jets, as we discuss in Section~\ref{sec5}. Finally, we summarize our results and conclude in Section~\ref{sec6}. This work builds on our previous studies of relativistic pair plasma turbulence without any cooling mechanism \citep{zhdankin_etal_2017, zhdankin_etal_2018a}.

\section{Methods} \label{sec2}

\subsection{Background on radiative turbulence} \label{sec:theory}

In this section, we provide a brief theoretical overview of the physical problem under consideration (turbulence in relativistic pair plasma with external IC radiative cooling) and formulate the physical conditions (in particular, temperature) necessary to maintain turbulence in statistical steady state \citep[see also][for the case of synchrotron and SSC radiation]{uzdensky_2018}. The statistical steady state is a novel aspect of radiative turbulence, in the sense that a rigorous steady state is absent in kinetic turbulence without an energy sink (since dimensionless parameters evolve in time due to the turbulent heating).

In this paper, given a field ${\mathcal F}(\boldsymbol{x},t)$, we will consider spatial and temporal averages, denoted as follows. We use an overbar, $\overline{{\mathcal F}}$, to denote a spatial average or average over all particles in the domain (if a kinetic quantity). We use angular brackets, $\langle {\mathcal F} \rangle$, to denote a temporal average, generally taken over the period in which the turbulence is in a statistical steady state (beginning after a few Alfv\'{e}n crossing times).

We consider an ultra-relativistic electron-positron (pair) plasma. Specifically, we assume that the mean particle Lorentz factor is large, $\overline{\gamma} \gg 1$, where $\gamma = (1-v^2/c^2)^{-1/2}$ for a particle of velocity $v$. Such a turbulent pair plasma can be characterized by two dimensionless physical parameters \citep[see discussion in, e.g.,][]{zhdankin_etal_2018a}, as follows.

The first parameter is the magnetization $\sigma$, which is defined to be the ratio of the magnetic enthalpy to relativistic plasma enthalpy: $\sigma = B_{\rm rms}^2/4\pi \overline{h}$, where $B_{\rm rms}$ is the characteristic (rms) magnetic field in the system and $\overline{h} = n_0 \overline{\gamma} m_e c^2 + \overline{P}$ is the characteristic relativistic enthalpy density, $n_0$ is the (electron plus positron) number density, and $\overline{P}$ is the average plasma pressure. In our case, approximating the particle distribution as isotropic, $\overline{P} = n_0 \overline{\gamma} m_e c^2/3$, so that $\sigma = 3 B_{\rm rms}^2/16 \pi n_0 \overline{\gamma} m_e c^2$. The magnetization sets the characteristic Alfv\'{e}n velocity, $v_A = c [\sigma/(\sigma+1)]^{1/2}$, and thus determines to what extent the turbulent motions (which are predominantly Alfv\'{e}nic) are relativistic. Note that the magnetization also determines the relationship between the two primary plasma kinetic scales: the characteristic Larmor radius,~$\rho_e = \overline{\gamma} m_e c^2/eB_{\rm rms}$, and the plasma skin depth,~$d_e = (\overline{\gamma} m_e c^2 / 4 \pi n_0 e^2)^{1/2}$, are related by~$d_e/\rho_e = [(4/3) \sigma]^{1/2}$. We also note that for an ultra-relativistic pair plasma, the plasma beta parameter is inversely proportional to magnetization, $\beta = 8 \pi \overline{P}/B_{\rm rms}^2 = 1/(2\sigma)$.

The second parameter is $L/2\pi\rho_e$, which is the ratio of the driving scale ($L/2\pi$) to the characteristic Larmor radius. In the case of $\sigma \lesssim 1$, $L/2\pi\rho_e$ essentially describes the extent of the large-scale (MHD) inertial-range cascade, which is generally terminated by collisionless plasma effects at scales comparable to and smaller than~$\rho_e$. In the case of~$\sigma \gg 1$, the transition may instead occur at scales comparable to $d_e$, making $L/2\pi d_e$ more representative of the inertial range extent \citep{chen_etal_2014b, boldyrev_etal_2015, franci_etal_2016}. For most collisionless plasmas in space and astrophysical systems, $L/2\pi\rho_e \gg 1$, a limit that is difficult to achieve in numerical simulations. It is thus important to do scaling studies to identify which quantitative properties of turbulence are sensitive to $L/2\pi\rho_e$ (in the limit of large values) and which are not.

External IC radiative cooling occurs due to the upscattering of low-energy seed photons (from an external radiation field) by relativistic particles \citep[see, e.g.,][]{blumenthal_gould_1970, rybicki_lightman_2008}. The energy density of the ambient photon field $U_{\rm ph}$ controls the strength of this cooling process. However, as we explain next, for turbulence in a statistical steady state, where injected energy is balanced by radiated energy, $U_{\rm ph}$ is only relevant for setting the steady-state mean particle energy, $\overline{\gamma}$. As long as $\overline{\gamma} \gg 1$ is maintained, the precise value of $\overline{\gamma}$ is dynamically irrelevant because it only sets an arbitrary energy scale (assuming $\sigma$ and $L/2\pi\rho_e$ are held fixed). Thus, we do not consider it to be an independent free parameter in our simulation campaign.

The IC emission process exerts a radiation backreaction force, $\boldsymbol{F}_{\rm IC}$, that is added to the Lorentz force describing the evolution of electrons and positrons. The motion of the $i$th particle is thus governed by 
\begin{align}
\frac{d \boldsymbol{p}_i}{dt} &= q_i \left[\boldsymbol{E}(\boldsymbol{x}_i,t) + \frac{\boldsymbol{v}_i}{c} \times \boldsymbol{B}(\boldsymbol{x}_i,t)\right] + \boldsymbol{F}_{\rm IC}(\boldsymbol{v}_i) \, , \label{eq:eom}
\end{align}
where $\boldsymbol{p}_i = \gamma_i m_e \boldsymbol{v}_i$ is the particle momentum, $q_i$ is the particle charge, $\boldsymbol{E}(\boldsymbol{x},t)$ and $\boldsymbol{B}(\boldsymbol{x},t)$ are the electric and magnetic fields, $\boldsymbol{x}_i$ is the particle position, and $\boldsymbol{v}_i = d\boldsymbol{x}_i/dt$ is the particle velocity. The IC radiation backreaction force is given by \citep{landau_lifshitz_1975}
\begin{align}
\boldsymbol{F}_{\rm IC}(\boldsymbol{v}) &= - \frac{4}{3} \sigma_T U_{\rm ph} \gamma^2 \frac{\boldsymbol{v}}{c} \, ,
\end{align}
where we assumed an isotropic external photon density (in the laboratory frame) and ultra-relativistic particles ($\gamma_i \gg 1$). Here, $\sigma_T = (8\pi/3) r_e^2$ is the Thomson cross section, with $r_e = e^2 / m_e c^2$ the classical electron radius.

In our case of IC radiative cooling acting on an optically thin, relativistic pair plasma ($v \approx c$), the radiative cooling rate (normalized to the total number of particles $N_{\rm part}$) is then given by
\begin{align}
\dot{{\mathcal E}}_{\rm rad} &= - \frac{1}{N_{\rm part}} \int d^3x d^3p f(\boldsymbol{x},\boldsymbol{p}) \boldsymbol{v} \cdot \boldsymbol{F}_{\rm IC} \nonumber \\
&= \frac{4}{3} \sigma_T c U_{\rm ph}  \frac{1}{N_{\rm part}} \int d^3x d^3p f(\boldsymbol{x},\boldsymbol{p}) \gamma^2 \nonumber \\
&= \frac{4}{3} \sigma_T c U_{\rm ph} \overline{\gamma^2} \, , \label{eq:rad}
\end{align}
where $f(\boldsymbol{x},\boldsymbol{p})$ is the combined (electron plus positron) distribution function.

On the other hand, the energy injection rate (from an external driving source) can be estimated by assuming that the turbulent magnetic energy $\delta B_{\rm rms}^2/8\pi$ dissipates within a cascade time that is comparable to the large-scale Alfv\'{e}n crossing time, $L/v_A$, where $L$ is the system size. We consider strong turbulence with a fluctuating magnetic field that is comparable to the mean field, so that the rms fluctuating magnetic field component is $\delta B_{\rm rms} \approx B_0$, implying a total rms field given by $B_{\rm rms}^2 = B_0^2 + \delta B_{\rm rms}^2 \approx 2 B_0^2$. The energy injection rate normalized to the number of particles is then
\begin{align}
\dot{{\mathcal E}}_{\rm inj} &= \eta_{\rm inj} \frac{B_0^2}{8 \pi n_0} \frac{v_A}{L} \, . \label{eq:inj}
\end{align}
We included a coefficient $\eta_{\rm inj} \sim 1$ that describes the efficiency of the external driving coupling to the plasma.

Because of the dependence of cooling on $\overline{\gamma^2}$, regardless of initial conditions, the plasma internal energy will adjust (by net heating or cooling) until the cooling compensates the energy injection, $\dot{{\mathcal E}}_{\rm rad} \sim \dot{{\mathcal E}}_{\rm inj}$. For a given $U_{\rm ph}$, the expected steady-state mean square particle Lorentz factor is obtained by balancing Eq.~\ref{eq:rad} with Eq.~\ref{eq:inj}, giving
\begin{align}
\overline{\gamma^2} &= \frac{3 \eta_{\rm inj}}{4} \frac{B_0^2}{8 \pi U_{\rm ph} \sigma_T n_0 L} \frac{v_A}{c} \nonumber \\
&= \frac{3 \eta_{\rm inj}}{4} \frac{B_0^2}{8 \pi U_{\rm ph}} \frac{1}{\tau_T} \frac{v_A}{c} \, , \label{eq:ss:gamma}
\end{align}
where the Thomson optical depth of the system is~$\tau_T \equiv n_0 \sigma_T L$. We consider the special case of a thermal (Maxwell-J\"{u}ttner) distribution with temperature~$T = \Theta m_e c^2$, in the ultra-relativistic limit ($\Theta \gg 1$),
\begin{align}
f_{0}(\boldsymbol{x},\boldsymbol{p}) &= \frac{n_0}{8 \pi \Theta^3 m_e^3 c^3} \exp{\left(-p/\Theta m_e c\right)} \, . \label{eq:mj}
\end{align}
In this case, the mean energy and mean squared energy are related to the temperature by~$\overline{\gamma} = 3 \Theta$ and $\overline{\gamma^2} = 12 \Theta^2$, respectively. The steady-state temperature for a thermal plasma can then be written as
\begin{align}
 \Theta_{ss} &= \eta_{\rm inj} \frac{m_e c^2}{16 \sigma_T U_{\rm ph} L} \sigma \frac{v_A}{c} \nonumber \\
 &= \frac{\eta_{\rm inj}}{16} \frac{1}{\ell} \frac{\sigma^{3/2}}{(1 + \sigma)^{1/2}} \, , \label{eq:ss}
\end{align}
where we defined the compactness parameter~$\ell \equiv \sigma_T U_{\rm ph} L / m_e c^2$ (describing the ratio of the system light crossing time to the particle cooling timescale, all divided by the particle's Lorentz factor). We thus initialize our simulations with a temperature close to $\Theta_{ss}$ in order to quickly arrive at a steady state. In general, the actual average energy during steady state may deviate from this estimate if the distribution is significantly nonthermal.

\subsection{Numerical simulations}

We perform the simulations with the explicit electromagnetic PIC code {\sc Zeltron} \citep{cerutti_etal_2013}, which incorporates the radiation backreaction force from IC emission. We use the same numerical set-up as in our previous studies of nonradiative relativistic plasma turbulence \citep[e.g.,][]{zhdankin_etal_2018a}. The domain is a periodic cube of side length $L$ with uniform background magnetic field $\boldsymbol{B}_0 = B_0 \hat{\boldsymbol{z}}$. We initialize electrons and positrons from a uniform Maxwell-J\"{u}ttner distribution (Eq.~\ref{eq:mj}). We drive turbulence by applying a fluctuating external current density $\boldsymbol{J}_{\rm ext}$ \citep{tenbarge_etal_2014}. This driving is characterized by four quantities: the wavevector $\boldsymbol{k}_0$, the frequency $\omega_0$, the decorrelation rate $\gamma_0$, and the amplitude $A_0$; we optimize these parameters based on simulations at $\sigma \sim 1$. We drive $J_{{\rm ext},z}$ at eight modes, ${\bf k}_0 L / 2 \pi \in \{ (1,0,\pm1)$, $(0,1,\pm1)$, $(-1,0,\pm1)$, $(0,-1,\pm1) \}$, and each of $J_{{\rm ext},x}$ and $J_{{\rm ext},y}$ at four modes to enforce $\nabla \cdot {\bf J}_{\rm ext} = 0$. We choose a driving frequency of $\omega_0 = 0.6 \cdot 2 \pi v_A / \sqrt{3} L$ and decorrelation rate $\gamma_0 = 0.5 \cdot 2 \pi v_A / \sqrt{3} L$. Finally, we fix $A_0$ such that the rms magnetic fluctuations are comparable to the background field, $\delta B_{\rm rms} \sim B_0$. The simulation timestep is fixed at $\Delta t \approx (\Delta x/c)/\sqrt{3}$ to satisfy the Courant-Friedrichs-Lewy condition. Most cases have $64$ total particles per cell and cell size $\Delta x \approx \min{(\rho_e/2, d_e/2)}$; however, for low-$\sigma$ cases, we choose a greater number of particles per cell to reduce the inherent PIC noise, which is high in this regime due to thermal energy dominating the energy in turbulent fluctuations. In particular, we choose $128$ particles per cell for cases with $\langle\sigma\rangle = 0.2$ and $256$ particles per cell for $\langle\sigma\rangle = 0.04$. We have optimized these numerical parameters by conducting a thorough convergence study with respect to resolution and number of particles per cell.

We initialize all simulations with a temperature of~$\Theta = 100$, corresponding to $\overline{\gamma} = 300$. We always choose $U_{\rm ph}$ such that the predicted steady-state temperature, calculated from Eq.~\ref{eq:ss} with initial parameters\footnote{In terms of physical (dimensional) parameters, required as input parameters to {\sc Zeltron}, we do the following procedure: the cell size $\Delta x$ is arbitrarily set, $B_0$ is chosen to get $\rho_e/\Delta x$ as determined by the resolution requirements, $n_0$ is chosen to obtain the prescribed $\sigma$, and then we set $U_{\rm ph} = \eta_{\rm inj} m_e c^2 \sigma v_A/(16 \sigma_T \Theta_{ss} L c)$ with $\Theta_{ss} = 75$ and $\eta_{\rm inj} = 1$.} and $\eta_{\rm inj} = 1$, is $\Theta_{ss} = 75$, which corresponds to $\gamma_{\rm ss} = 225$. As a result, $U_{\rm ph}$ varies with system size $L$, but in a way such that the dimensionless compactness parameter $\ell$ is fixed. Our choice of $U_{\rm ph}$ ensures that simulations quickly approach the equilibrium (as confirmed in Sec.~\ref{subsec:equil}).

We show the steady-state parameters and durations for the primary simulations employed in this study in Table~\ref{table:sims}. Our largest, fiducial case ($768^3$ cells) has $\langle\sigma\rangle = 0.90$ and $\langle\rho_e\rangle = 2.0 \Delta x$, giving an effective system size of~$L/2\pi\langle\rho_e\rangle = 61$. We also have three cases on $512^3$-cell lattices, covering $\langle\sigma\rangle \in \{ 0.2, 0.9, 3.4 \}$, constituting moderately large cases with modest variation in $\langle\sigma\rangle$. In the next tier, we have five cases on $384^3$-cell lattices, which consistute a broad scan in $\langle\sigma\rangle \in \{ 0.04, 0.2, 0.8, 3.3, 11.0 \}$. Since the resolution is fixed by the smallest kinetic scale ($\rho_e$ or $d_e$), the system size~$L/2\pi\rho_e$ does not necessarily correspond to the number of lattice cells: the low-$\langle\sigma\rangle$ cases have smaller~$L/2\pi\rho_e$ in order to properly resolve~$d_e$. In addition to these primary simulations, two of the simulations in the table are repeats of other cases (rM4 and rS1) with additional diagnostics: rM4* has a higher cadence of particle momentum distribution dumps (for Sec.~\ref{sec:ani}-Sec.~\ref{sec:sys}) and rS1* has a higher cadence of electric field and current density dumps (for Sec.~\ref{sec:trans}). The durations for the simulations typically range between $25 L/v_A$ and $35 L/v_A$, providing a long steady state for gathering statistics (from here onwards, we compute $v_A$ using the time-averaged magnetization $\langle\sigma\rangle$).

\begin{table}
\centering \caption{List of simulations and parameters \newline} \label{table:sims}
\begin{tabular}{|c|c|c|c|c|} 
	\hline
\hspace{0.5 mm} Case \hspace{0.5 mm}  & \hspace{1 mm} $N^3$ \hspace{1 mm}   & \hspace{1 mm} $L/2\pi\langle\rho_e\rangle$ \hspace{1 mm}  &   \hspace{1 mm} $\langle\sigma\rangle$ \hspace{1 mm} &   \hspace{1 mm} $t v_A/L$ \hspace{1 mm}  \\
	\hline
rL1 & $768^3$ & 60.4 & $0.90$ & $24.6$ \\
rM1d4 & $512^3$ & 29.6 & $0.20$ & $34.1$ \\
rM1 & $512^3$ & 39.4 & $0.86$ & $24.2$ \\
rM4 & $512^3$ & 38.9 & $3.4$ & $35.8$ \\
rM4* & $512^3$ & 39.1 & $3.4$ & $29.5$ \\
rS1d16 & $384^3$ & 10.4 & $0.041$ & $35.4$ \\
rS1d4 & $384^3$ & 21.4 & $0.19$ & $35.9$ \\
rS1 & $384^3$ & 28.3 & $0.82$ & $32.4$ \\
rS1* & $384^3$ & 28.3 & $0.83$ & $60.1$ \\
rS4 & $384^3$ & 28.1 & $3.3$ & $29.7$ \\
rS16 & $384^3$ & 24.9 & $11.0$ & $32.1$ \\
	\hline
\end{tabular}
\centering
\label{table-sims}
\end{table}

\section{Turbulence statistics} \label{sec3}

\subsection{Visuals} \label{sec:visuals}

\begin{figure}
 \includegraphics[width=\columnwidth]{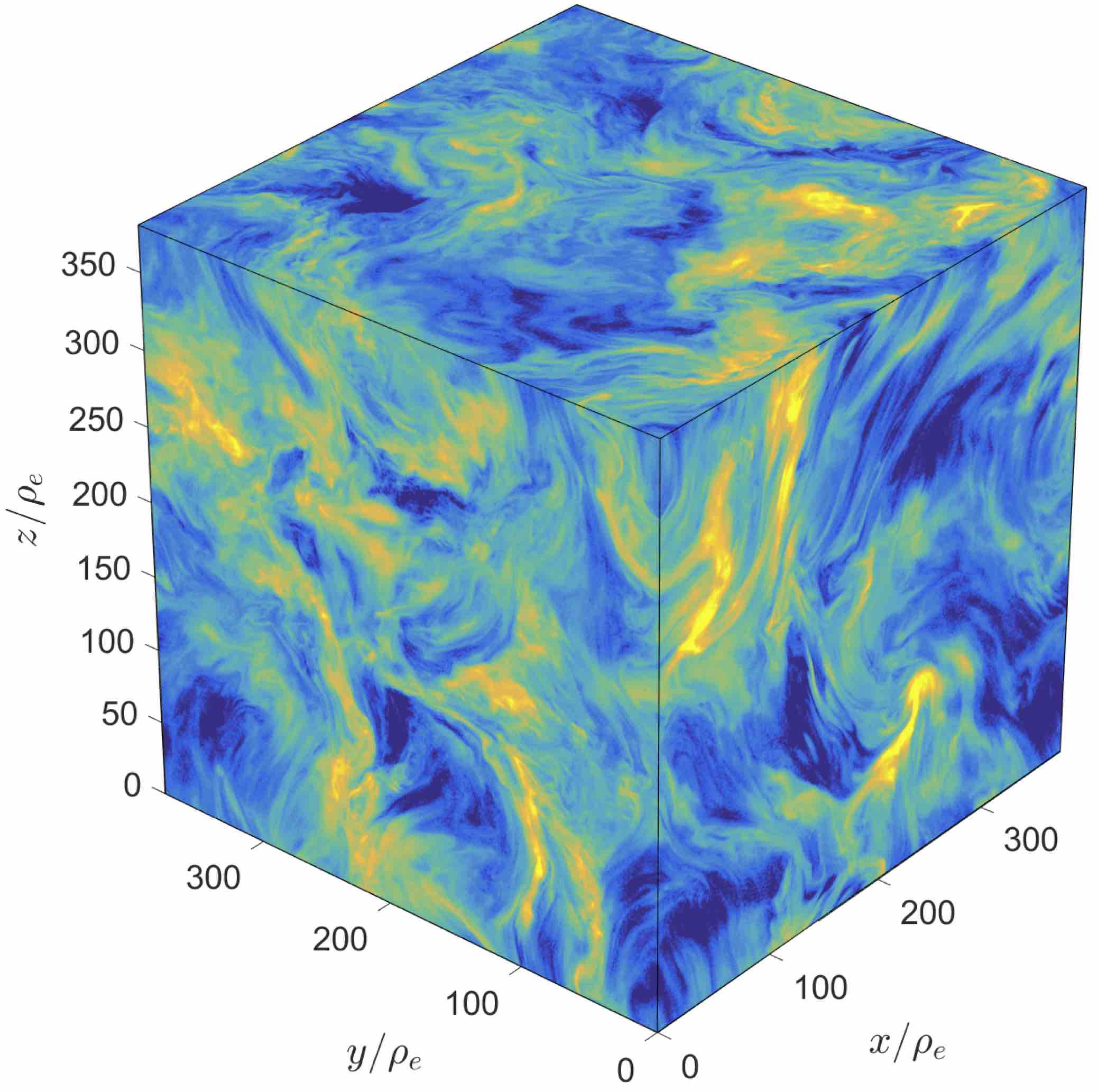}
  \centering
   \caption{\label{fig:emissivity_visual} Surface plot of the emissivity proxy $n \gamma_{\rm avg}^2$ for the $768^3$, $\langle\sigma\rangle = 0.9$ simulation.}
 \end{figure}

We first show visuals from our large case, the $768^3$ simulation with $\langle\sigma\rangle =0.9$. In Fig.~\ref{fig:emissivity_visual}, we show the surface visual for the isotropic emissivity proxy, given by $n \gamma_{\rm avg}^2$, where $n(\boldsymbol{x})$ is the plasma number density and $\gamma_{\rm avg}(\boldsymbol{x})$ is the local (cell-averaged) mean particle Lorentz factor.

To convey a qualitative picture of the steady-state turbulence at low and high magnetizations, we next show visuals for several quantities in the $512^3$ simulations with $\langle\sigma\rangle = 0.2$ ($L/2\pi\langle\rho_e\rangle = 30$) and $\langle\sigma\rangle = 3.4$ ($L/2\pi\langle\rho_e\rangle = 39$). For clarity, we show images of the data in an $xy$ slice, arbitrarily taken at $z = L/2$ in the final snapshots of the simulations.

\begin{figure}
 \includegraphics[width=\columnwidth]{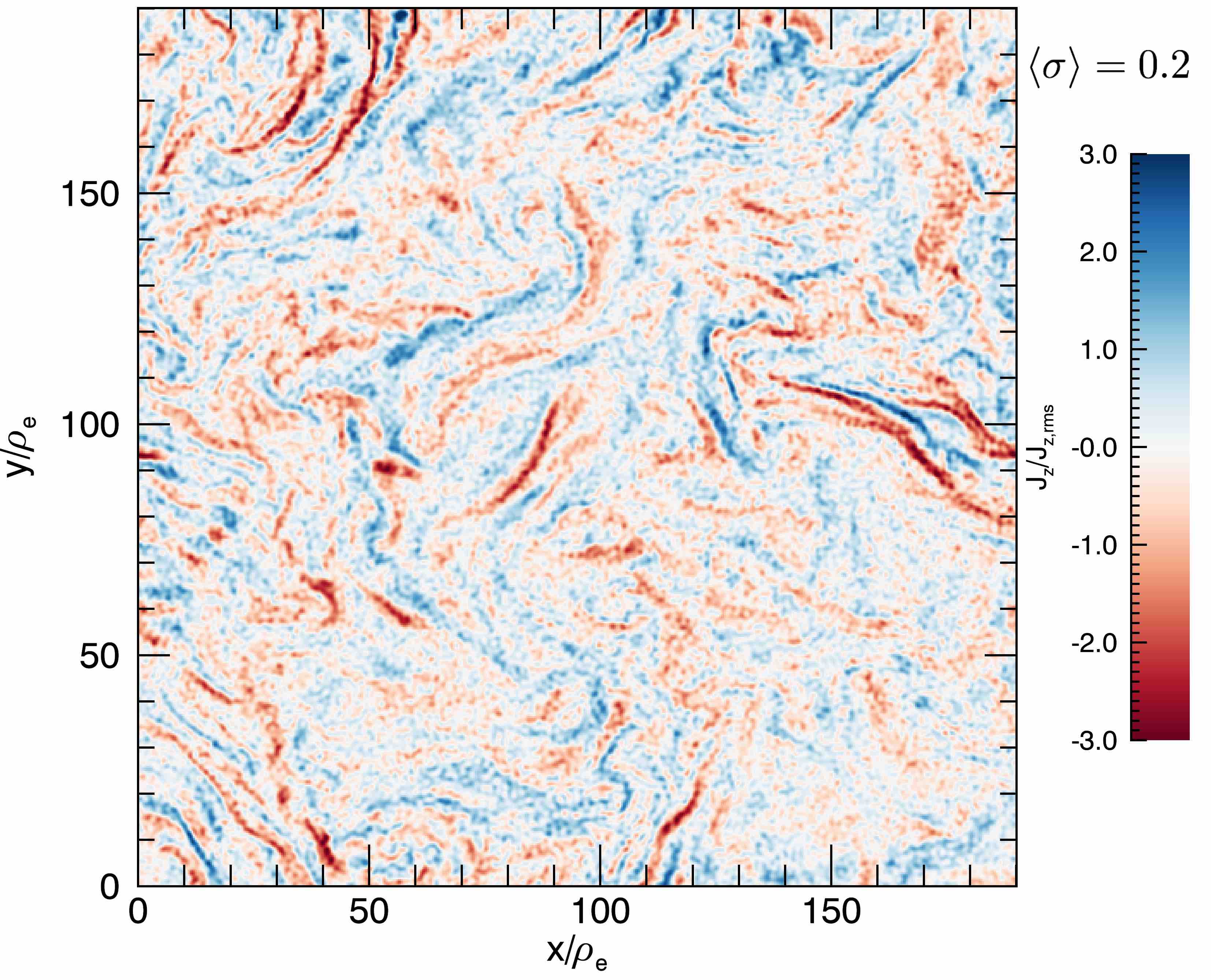}
  \includegraphics[width=\columnwidth]{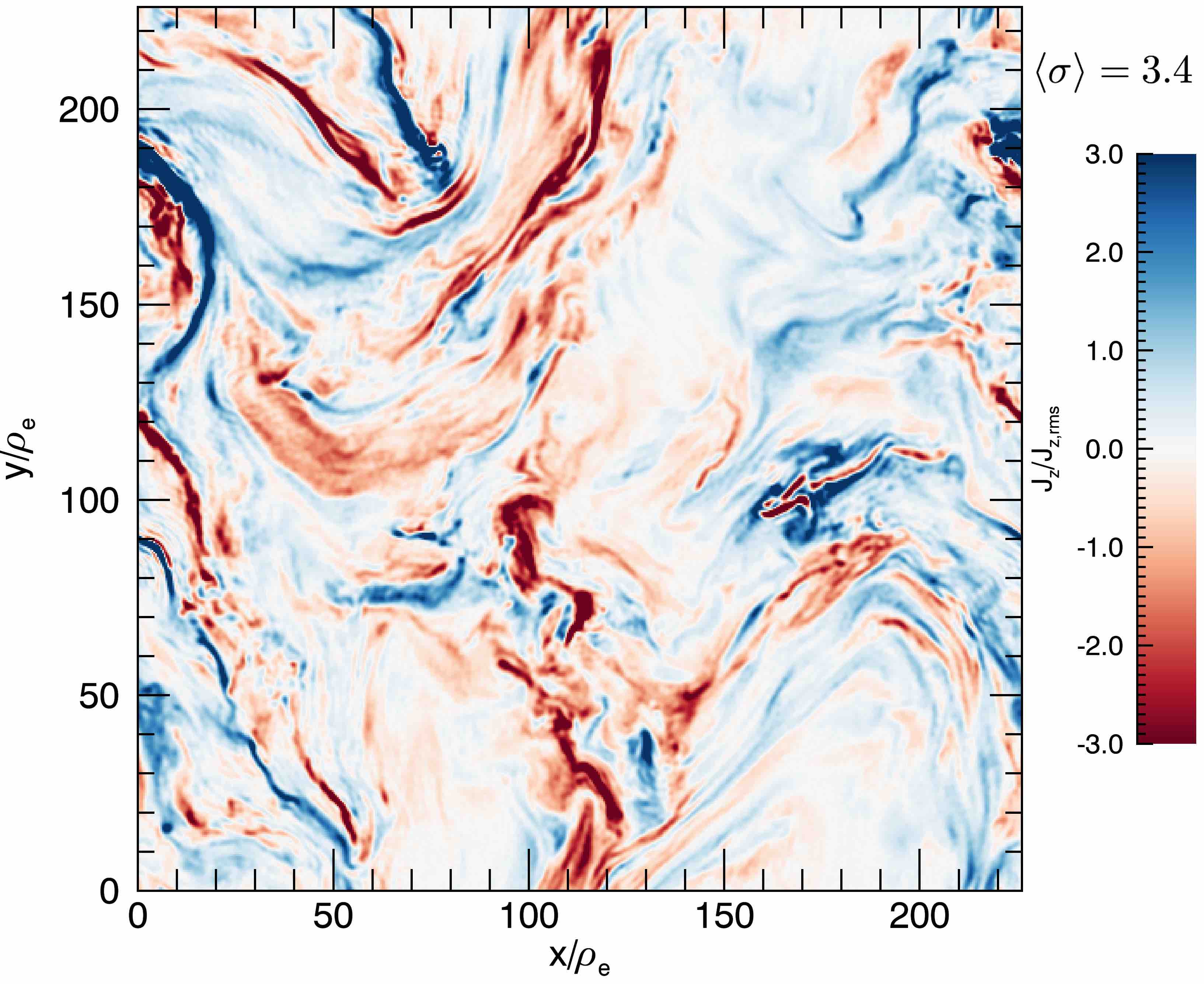}
  \centering
   \caption{\label{fig:jz_visual} Current density component along the mean field, $J_z$ (normalized to rms value $J_{z,\rm rms}$), for $\langle\sigma\rangle = 0.2$ (top) and $\langle\sigma\rangle = 3.4$ (bottom) in the $512^3$ simulations.}
 \end{figure}
 
 \begin{figure}
 \includegraphics[width=\columnwidth]{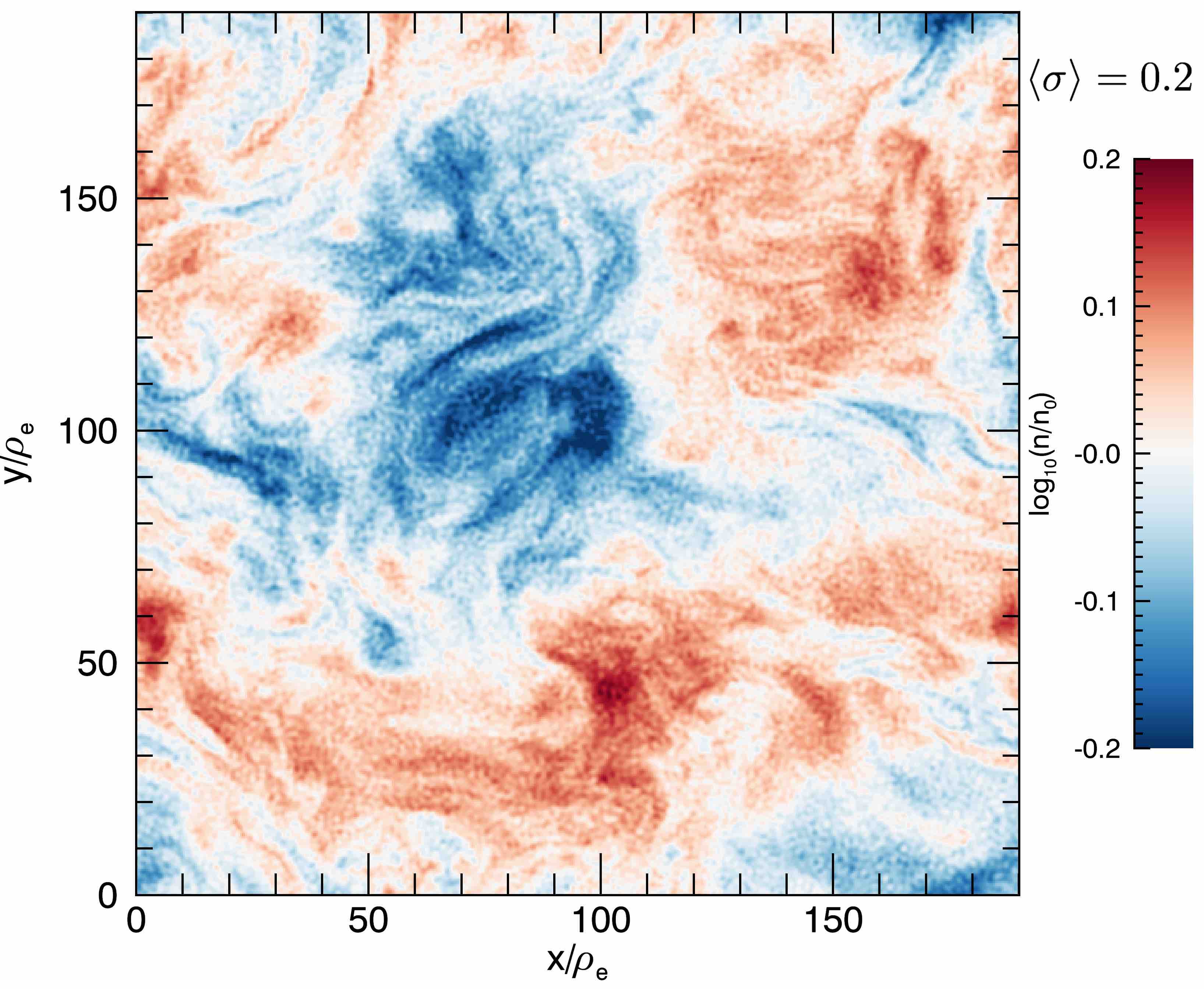}
  \includegraphics[width=\columnwidth]{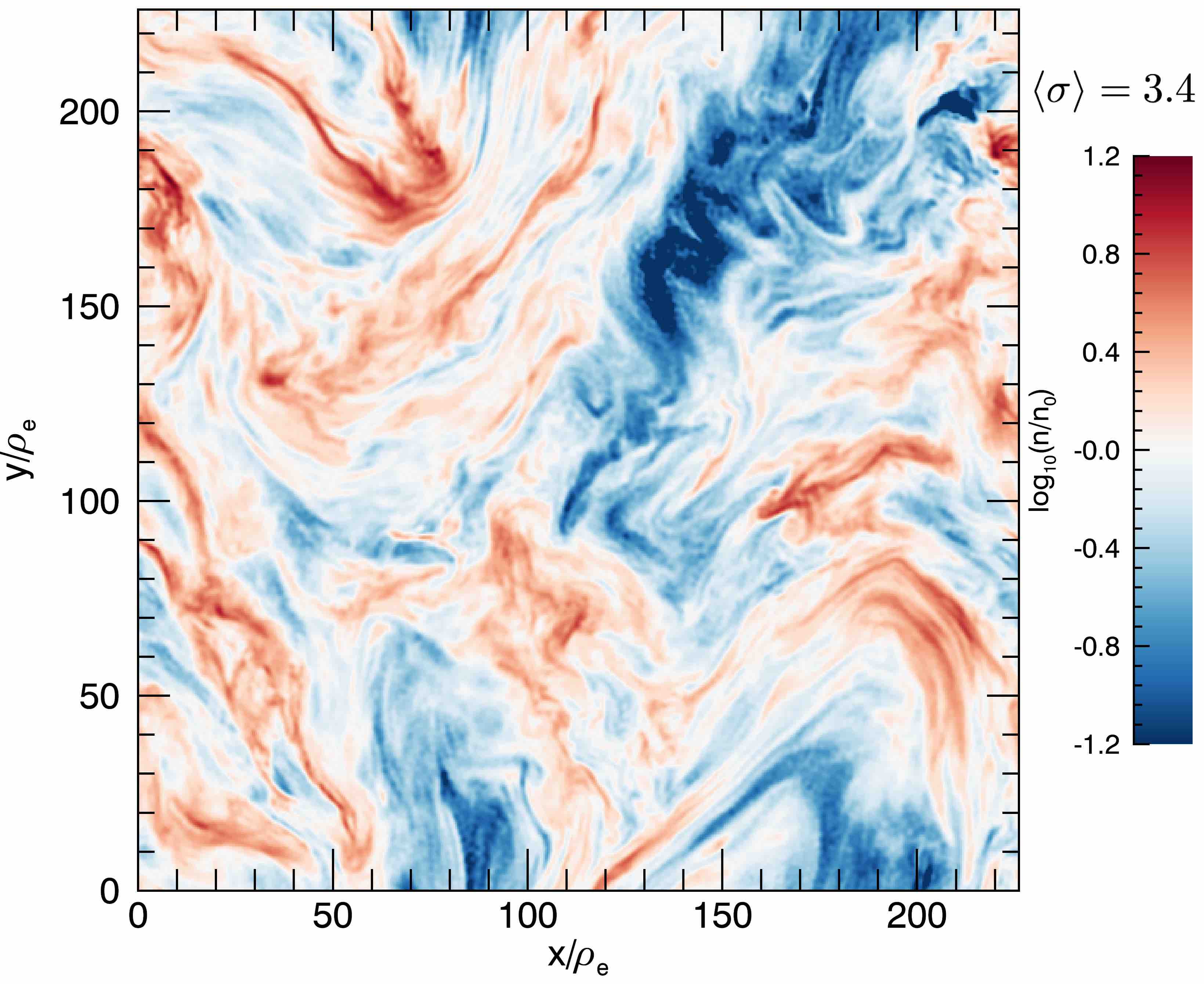}
  \centering
   \caption{\label{fig:density_visual} Particle number density $n$ (normalized to mean value $n_0$) for $\langle\sigma\rangle = 0.2$ (top) and $\langle\sigma\rangle = 3.4$ (bottom) in the $512^3$ simulations.}
 \end{figure}
 
 \begin{figure}
 \includegraphics[width=\columnwidth]{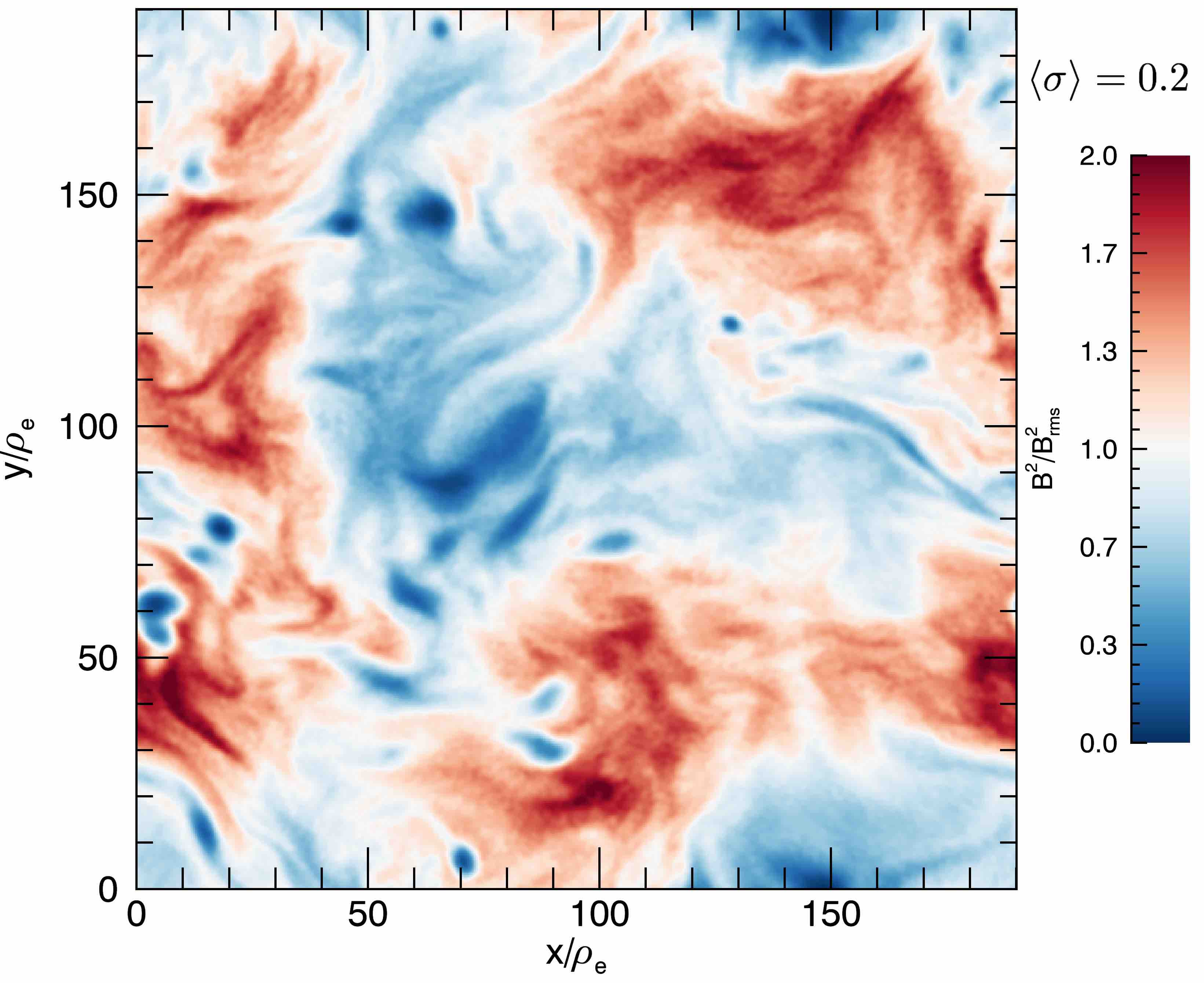}
  \includegraphics[width=\columnwidth]{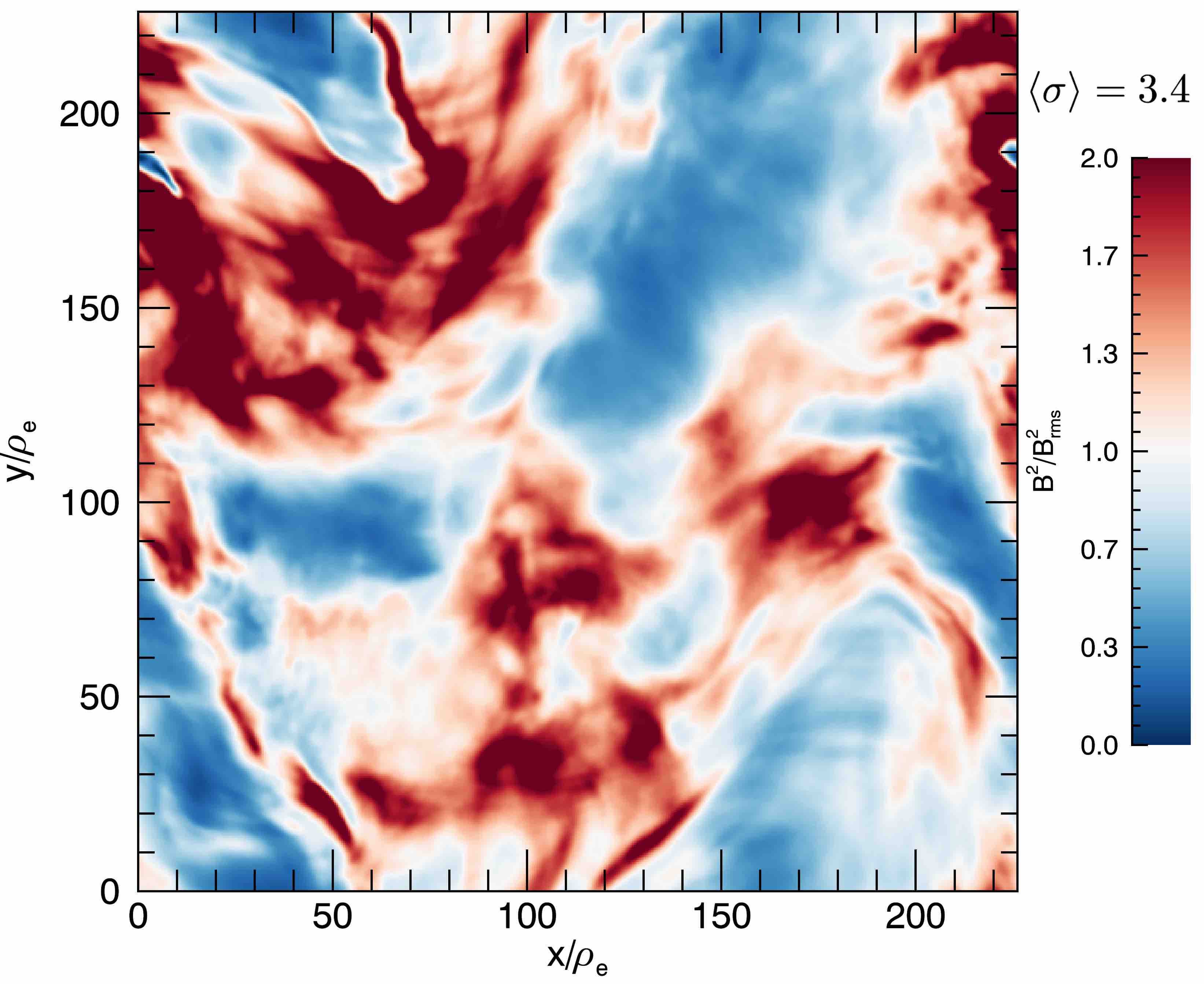}
  \centering
   \caption{\label{fig:mag_visual} Magnetic energy density $B^2/8\pi$ (normalized to mean value $B_{\rm rms}^2/8\pi$) for $\langle\sigma\rangle = 0.2$ (top) and $\langle\sigma\rangle = 3.4$ (bottom) in the $512^3$ simulations.}
 \end{figure}
 
  \begin{figure}
 \includegraphics[width=\columnwidth]{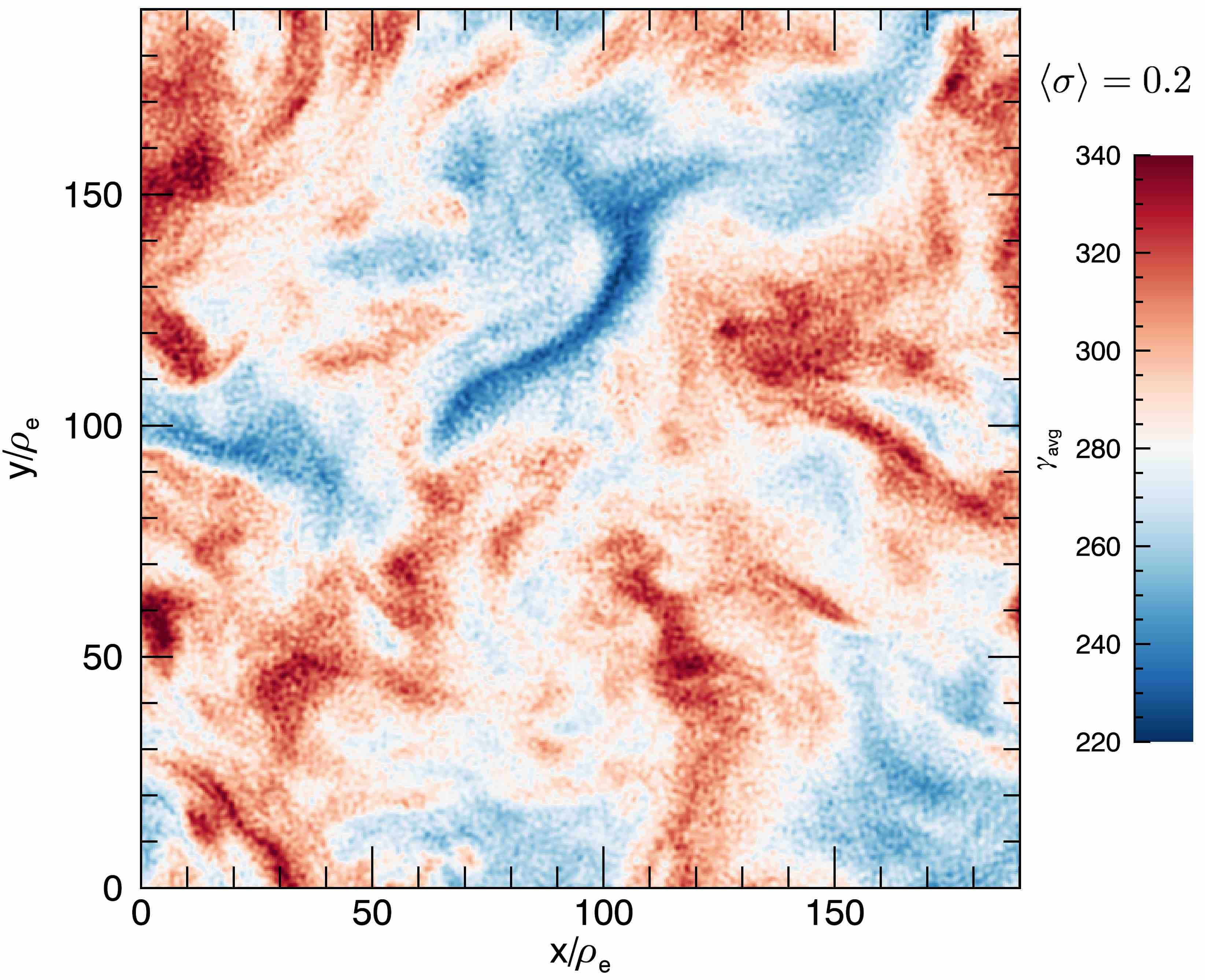}
  \includegraphics[width=\columnwidth]{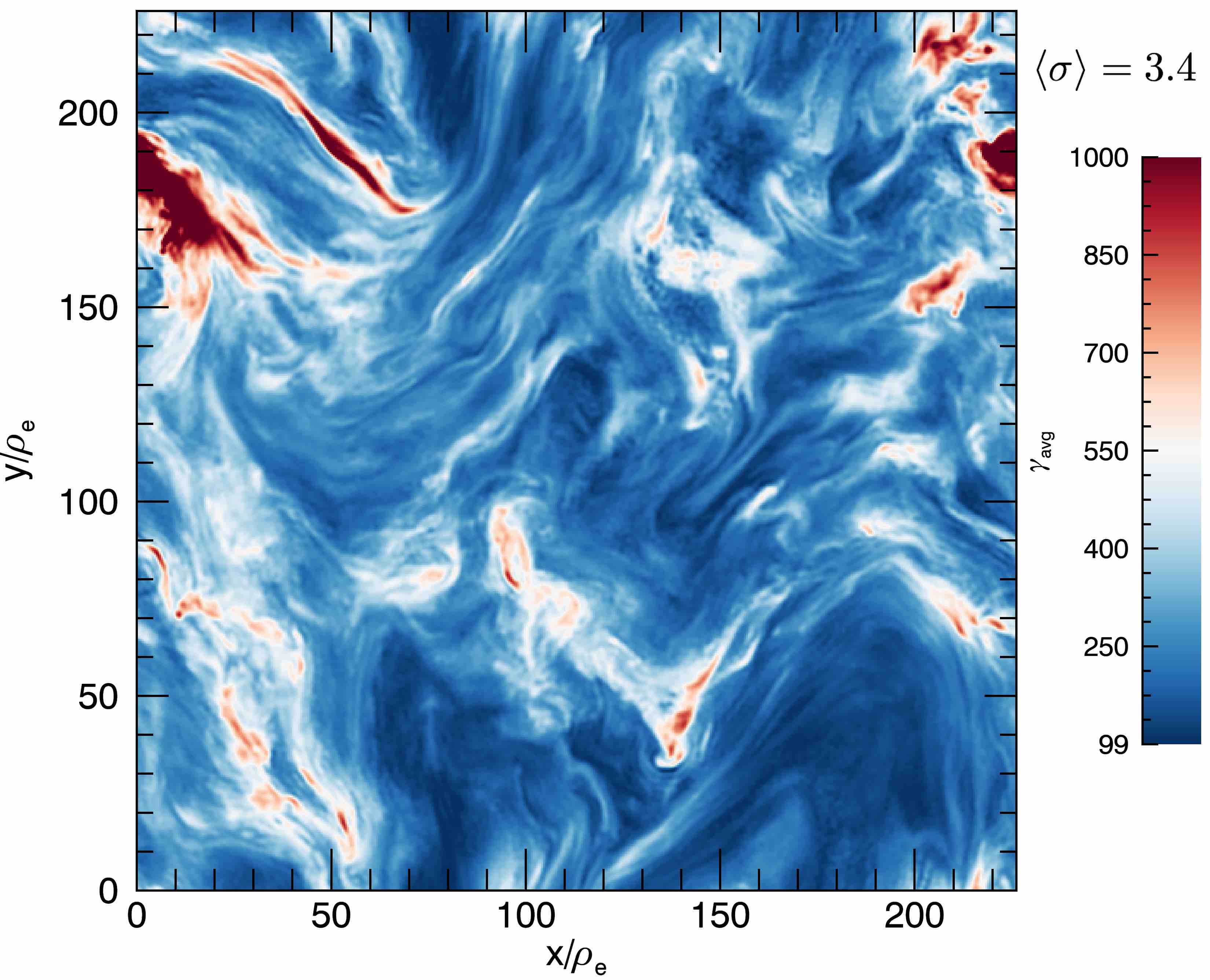}
  \centering
   \caption{\label{fig:gamma_visual} Local (cell) average particle Lorentz factor $\gamma_{\rm avg}$ for $\langle\sigma\rangle = 0.2$ (top) and $\langle\sigma\rangle = 3.4$ (bottom) in the $512^3$ simulations.}
 \end{figure}
 
In Fig.~\ref{fig:jz_visual}, we show the current density component along the guide field,~$J_z$. A large number of intermittent current sheets form in the turbulence. The high-$\sigma$ case exhibits current sheets that are more intense and more strongly clustered than in the low-$\sigma$ case, suggesting that energy dissipation may be more localized. In Fig.~\ref{fig:density_visual}, we show the particle number density~$n$ (on a logarithmic scale). Density fluctuations are much stronger in the high-$\sigma$ case, with $n/n_0 \gtrsim 10$ in some regions, concentrated in thin sheet-like structures. In Fig.~\ref{fig:mag_visual}, we show the magnetic energy density~$B^2/8\pi$. We find that the low-$\sigma$ case exhibits structures in the form of magnetic holes: circular coherent structures of size several~$\rho_e$ inside which the magnetic energy density drops to a small value, also seen in our previous non-radiative simulations \citep[][]{zhdankin_etal_2018a} and in non-relativistic kinetic turbulence \citep[e.g.,][]{roytershteyn_etal_2015}. These structures are correlated with high densities, consistent with local pressure equilibrium. Although they bear some resemblence to plasmoids resulting from the tearing instability in 2D MHD turbulence with high Reynolds number \citep{dong_etal_2018, walker_etal_2018}, these magnetic holes appear to have a different structure and origin in our simulations, which we defer to future study. In the high-$\sigma$ case, there are no magnetic holes; instead, the magnetic energy is more strongly concentrated in clumps, correlated with high density regions, indicating that the magnetic field may be compressed into thin sheets. Finally, in Fig.~\ref{fig:gamma_visual}, we show the local average particle Lorentz factor~$\gamma_{\rm avg}$, which coincides with regions where radiative cooling occurs (since the radiation backreaction force is proportional to~$\gamma^2$). The figure indicates that particle energization is more spatially localized in the high-$\sigma$ case, with intense hot spots that often coincide with current sheets.

 \subsection{Equilibrium temperature} \label{subsec:equil}
 
 \begin{figure}
 \includegraphics[width=\columnwidth]{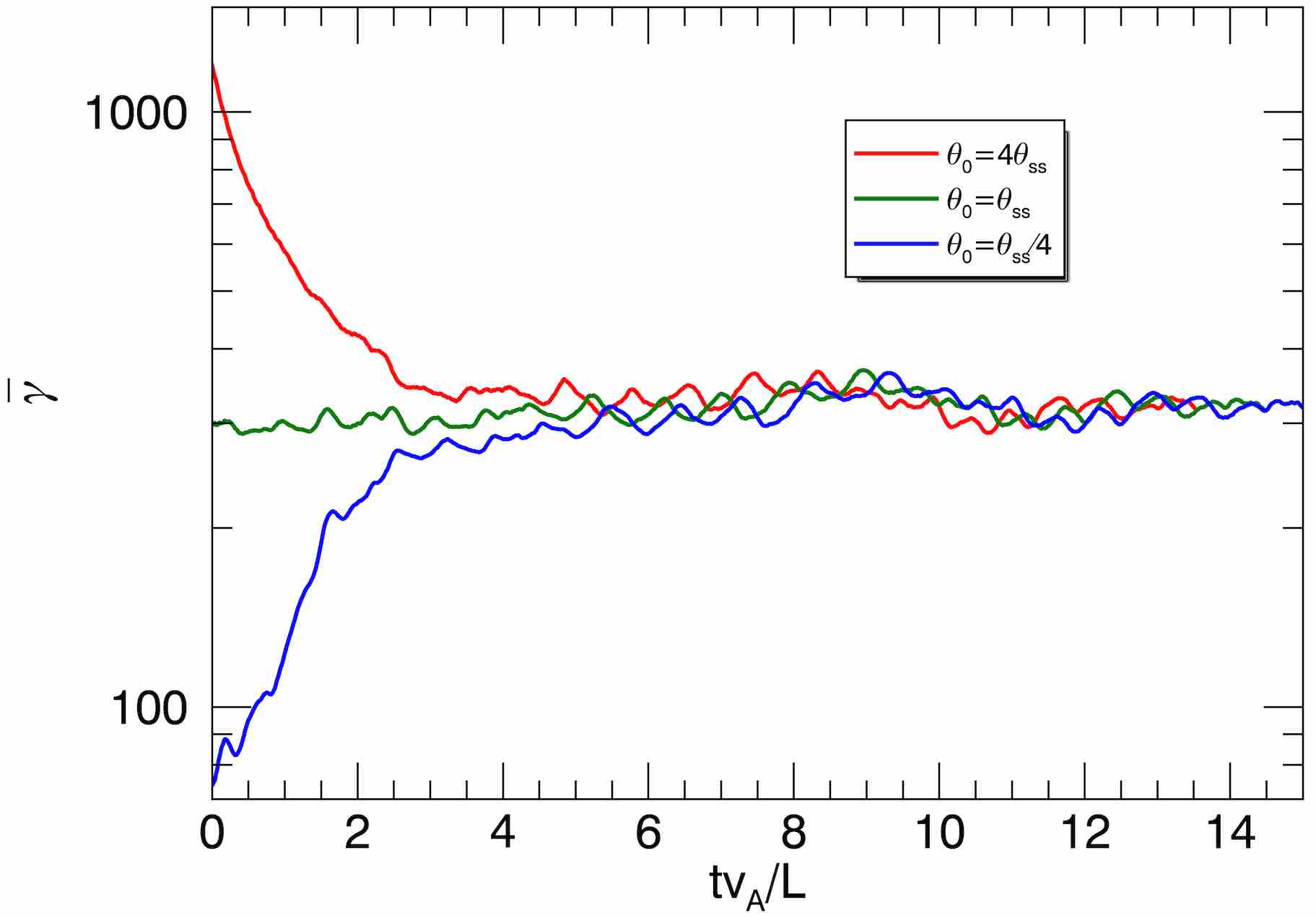}
 \includegraphics[width=\columnwidth]{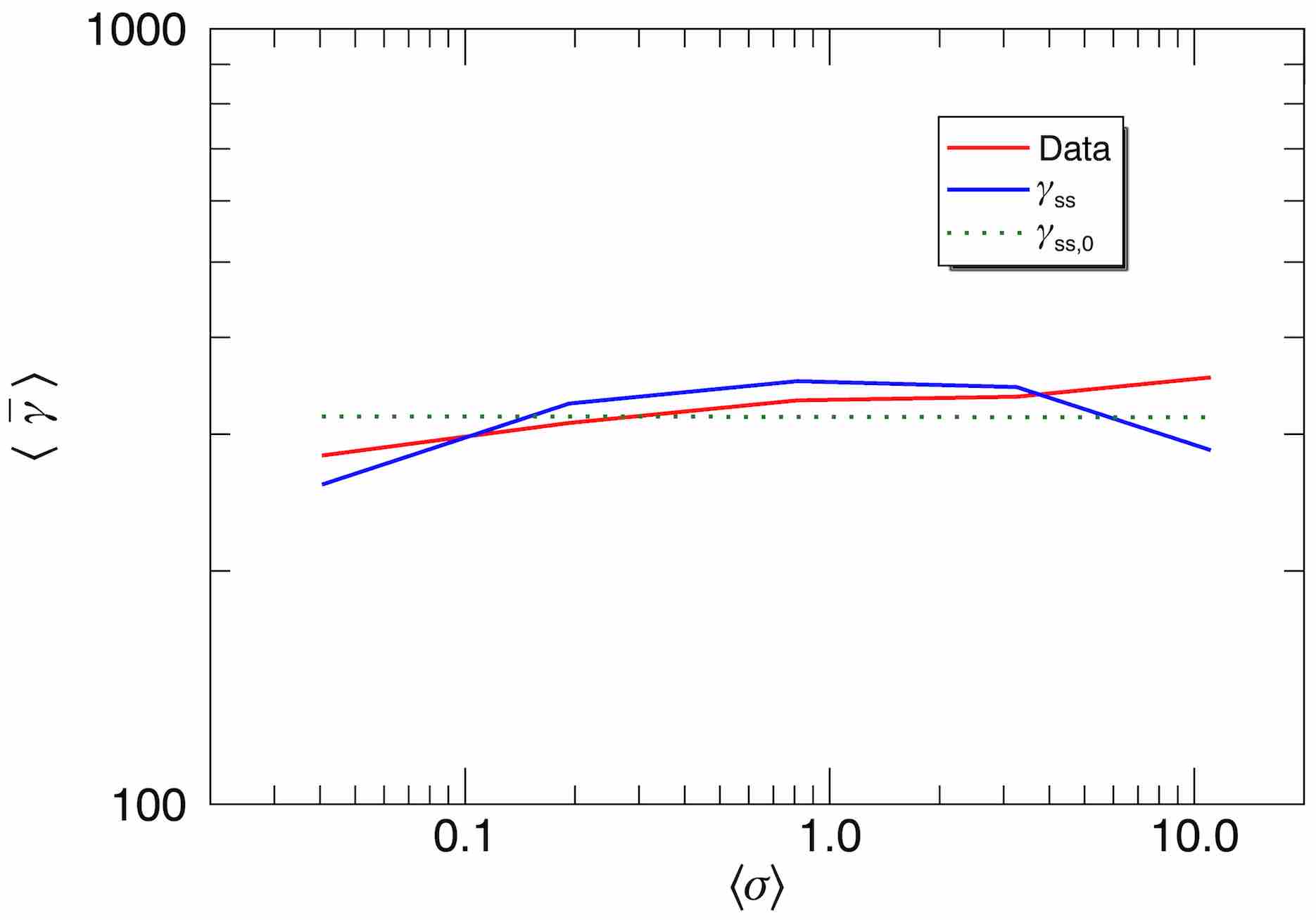}
  \centering
   \caption{\label{fig:theta} Top panel: Evolution of the mean particle energy~$\overline{\gamma}$ for a set of $256^3$ simulations with~$\langle\sigma\rangle \approx 1$ and different initial temperatures,~$\Theta_0 \in \{ \Theta_{ss}/4, \Theta_{ss}, 4 \Theta_{ss} \}$, demonstrating that all these cases are approach the specified steady state. Bottom panel: The time-averaged mean particle energy~$\langle\overline{\gamma}\rangle$ versus magnetization~$\langle\sigma\rangle$ for the $384^3$ simulation series (red). Also shown for reference are the predicted equilibrium values for the time-averaged parameters,~$\gamma_{ss}$ (blue), and for the initial parameters,~$\gamma_{ss,0}$ (green), assuming $\eta_{\rm inj} = 1.4$ in Eq.~\ref{eq:ss}.}
 \end{figure}
 
Before analyzing the turbulence statistics in detail, we confirm that the plasma attains the steady-state temperature expected for the given simulation parameters. In the top panel of Fig.~\ref{fig:theta}, we use a set of $256^3$ simulations with~$\langle\sigma\rangle \approx 1$ to demonstrate that simulations attain the same mean particle energy~$\overline{\gamma}$ at late times despite significantly different initial temperatures: $\Theta_0 \in \{ \Theta_{ss}/4, \Theta_{ss}, 4 \Theta_{ss} \}$. Recall that temperature and mean energy are related by~$\Theta = \overline{\gamma}/3$ for an ultra-relativistic Maxwell-J\"{u}ttner distribution. In the bottom panel of Fig.~\ref{fig:theta}, we show the steady-state mean particle energy~$\langle\overline{\gamma}\rangle$ versus magnetization~$\langle\sigma\rangle$ for the~$384^3$ simulation series. As expected from our numerical setup (in particular, tuning~$U_{\rm ph}$ to get fixed~$\Theta_{ss}$),~$\langle\overline{\gamma}\rangle$ does not vary significantly with~$\langle\sigma\rangle$, showing only a slight increase with~$\langle\sigma\rangle$. We compare $\langle\overline{\gamma}\rangle$ to the equilibrium values predicted from the initial parameters,~$\gamma_{ss,0}$, and predicted from the time-averaged parameters,~$\gamma_{ss}$, both calculated from Eq.~\ref{eq:ss} with~$\eta_{\rm inj} = 1.4$ chosen to get good agreement between the simulation measurement and the predictions. This value of~$\eta_{\rm inj}$ is close to the injection efficiency measured in our previous non-radiative simulations \citep{zhdankin_etal_2018a}. Hence, we conclude that Eq.~\ref{eq:ss} provides an accurate prediction for the steady-state temperature. Since the plasma is ultra-relativistically hot,~$\langle\overline{\gamma}\rangle \gg 1$, the absolute value of the temperature is irrelevant for the physics described in the remainder of the paper (as long as the relevant dimensionless plasma parameters~$\langle\sigma\rangle$ and $L/2\pi\langle\rho_e\rangle$ are held fixed).

\subsection{Energetics}

\begin{figure}
 \includegraphics[width=\columnwidth]{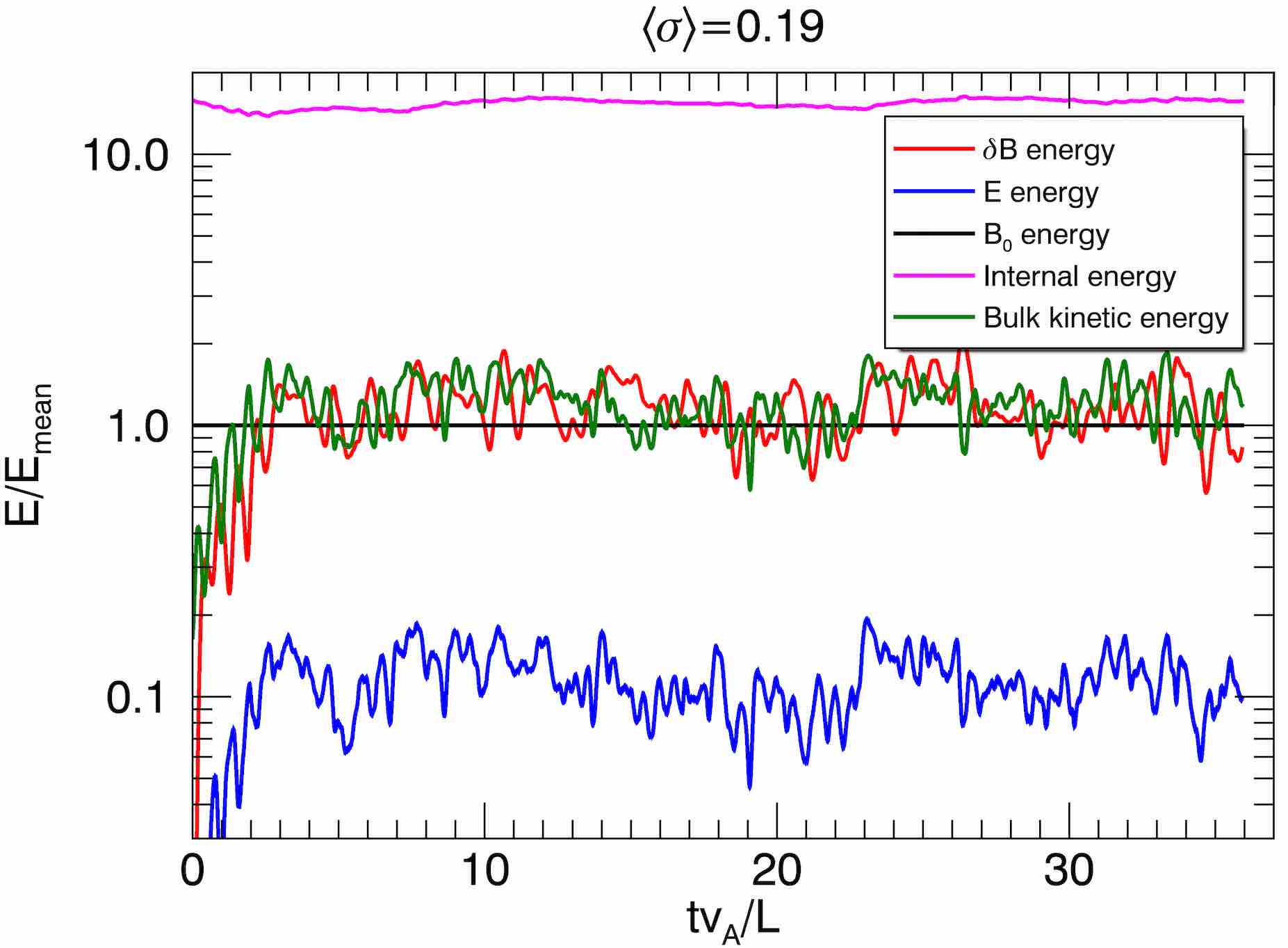}
  \includegraphics[width=\columnwidth]{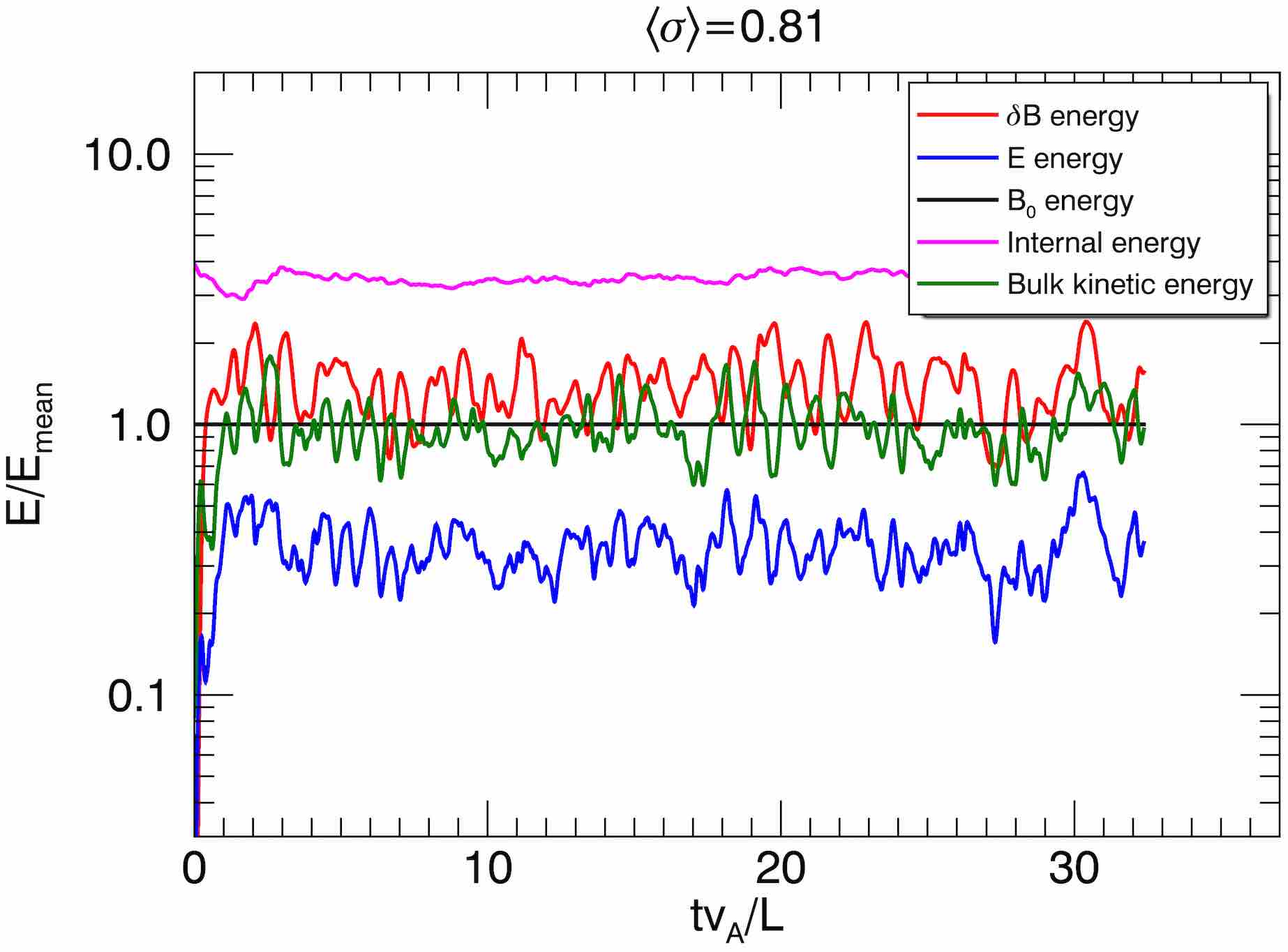}
 \includegraphics[width=\columnwidth]{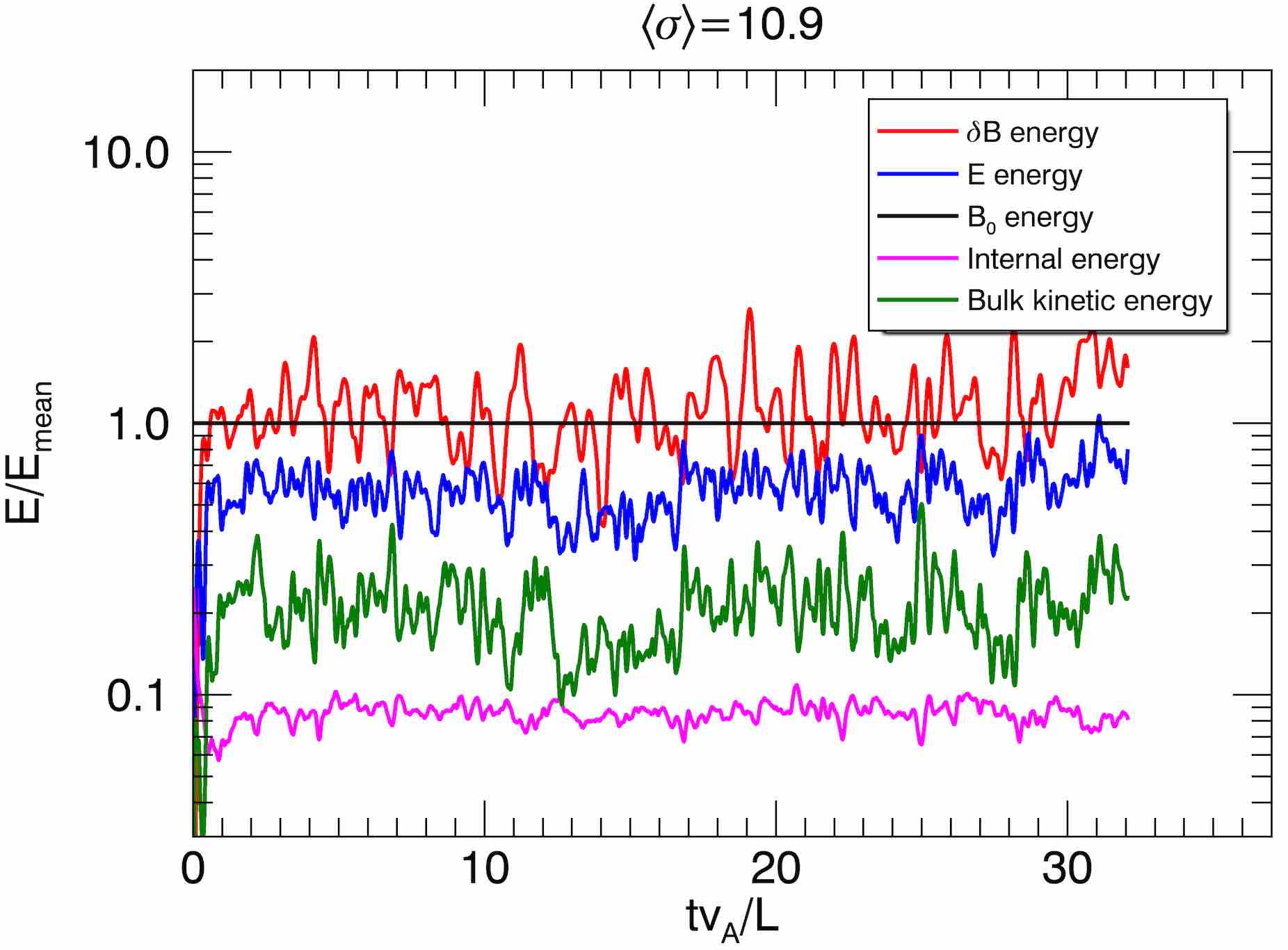}
  \centering
   \caption{\label{fig:turb_energy} Evolution of various energies at $\langle\sigma\rangle = 0.2$ (top), $\langle\sigma\rangle = 0.8$ (center), and at $\langle\sigma\rangle = 11.0$ (bottom), taken from $384^3$ simulations. Turbulent magnetic energy (red), electric energy (blue), internal energy (magenta), and bulk kinetic energy (green) are shown, all normalized to the energy of the mean magnetic field $E_{\rm mean}$ (black).}
 \end{figure}

We now consider the overall energetics of the simulations. In Fig.~\ref{fig:turb_energy}, we show the evolution of the different contributions to total energy in the system, for simulations in three magnetization regimes: low ($\langle\sigma\rangle = 0.2$), moderate ($\langle\sigma\rangle = 0.8$), and high ($\langle\sigma\rangle = 11$), for the $384^3$ simulation series. Specifically, we show magnetic, electric, internal, and bulk kinetic energy, with the latter two defined as in \cite{zhdankin_etal_2018a}\footnote{The internal and bulk energy density was computed separately for electrons and for positrons, then combined to obtain the total.}. For the given initial conditions, the energies arrive at a statistical steady state after a couple of large-scale Alfv\'{e}n crossing times ($L/v_A$).

 \begin{figure}
 \includegraphics[width=\columnwidth]{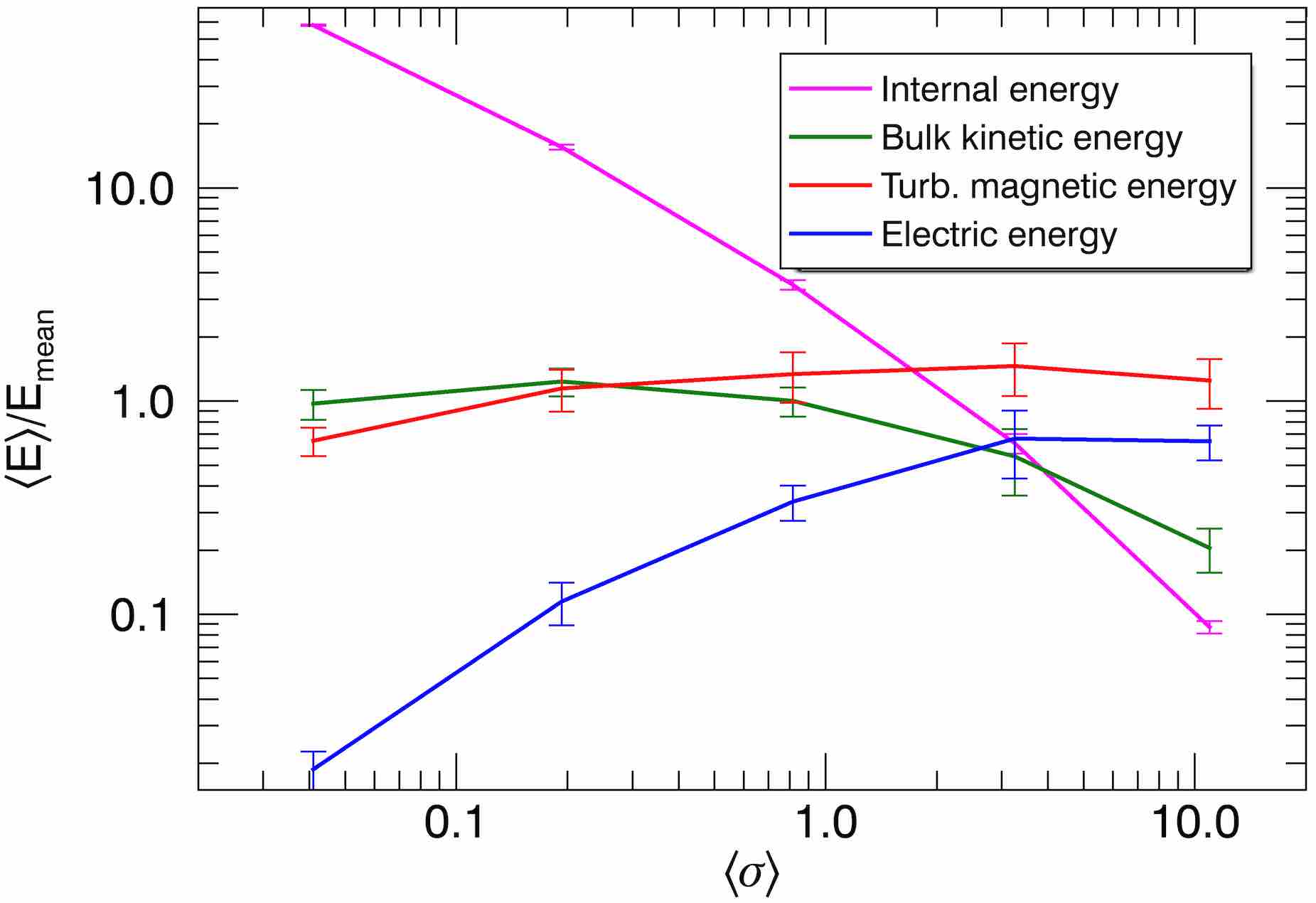}
  \centering
   \caption{\label{fig:energy_partition} Partitioning of the overall energy into time-averaged internal energy (magenta), turbulent magnetic energy (red), bulk kinetic energy (green), and electric energy (blue), versus magnetization $\langle\sigma\rangle$.}
 \end{figure}

Next, we consider the time-average of the different energies in steady state for varying $\sigma$ (for the $384^3$ series), shown in Fig.~\ref{fig:energy_partition}. For $\sigma \lesssim 1$, there are comparable amounts of turbulent energy in the bulk flows and in fluctuating magnetic fields (which are in turn comparable to the mean field energy), consistent with an Alfv\'{e}nic character. The electric energy is significantly smaller, roughly by a factor of $(v_A/c)^2$, consistent with it arising from the ideal MHD electric field component ($\boldsymbol{E} \approx \boldsymbol{v}_f/c \times \boldsymbol{B}$, where $\boldsymbol{v}_f$ is the bulk flow velocity). The internal energy is much larger than the turbulent energy, and thus the plasma has a high effective mass density and subsonic motions. For $\sigma \gg 1$, on the other hand, the total particle energy becomes small compared to the electric and magnetic energies. A large portion of the particle energy is contained in the bulk flows rather than the internal energy. The electric energy in this case is comparable to the magnetic energy. This is nominally the regime of relativistic force-free electrodynamics, as considered in previous works in the literature \citep{thompson_blaes_1998, cho_2005, cho_lazarian_2013, zrake_east_2016}.
 
\subsection{Turbulence spectrum}

 \begin{figure}
 \includegraphics[width=\columnwidth]{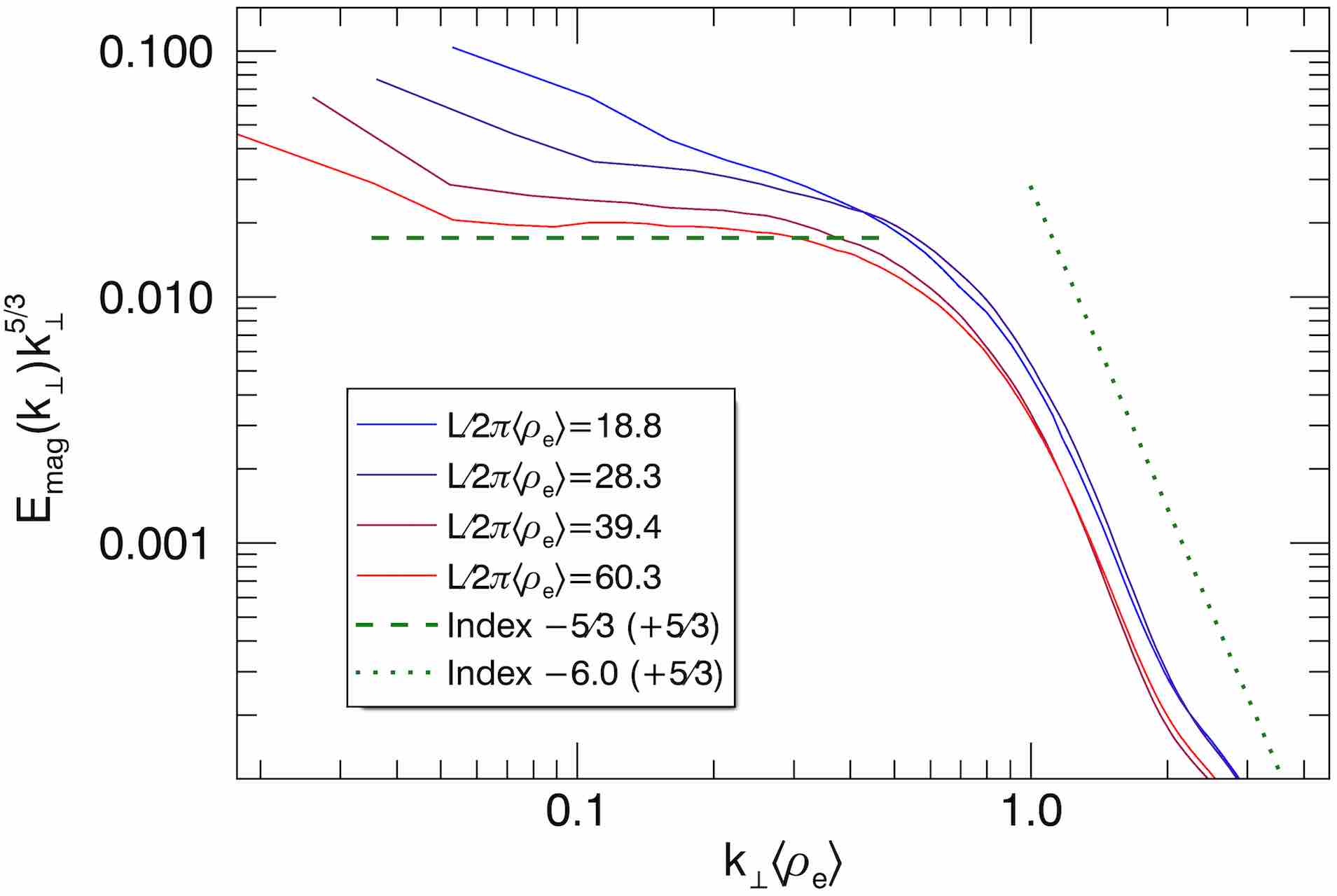}
   \includegraphics[width=\columnwidth]{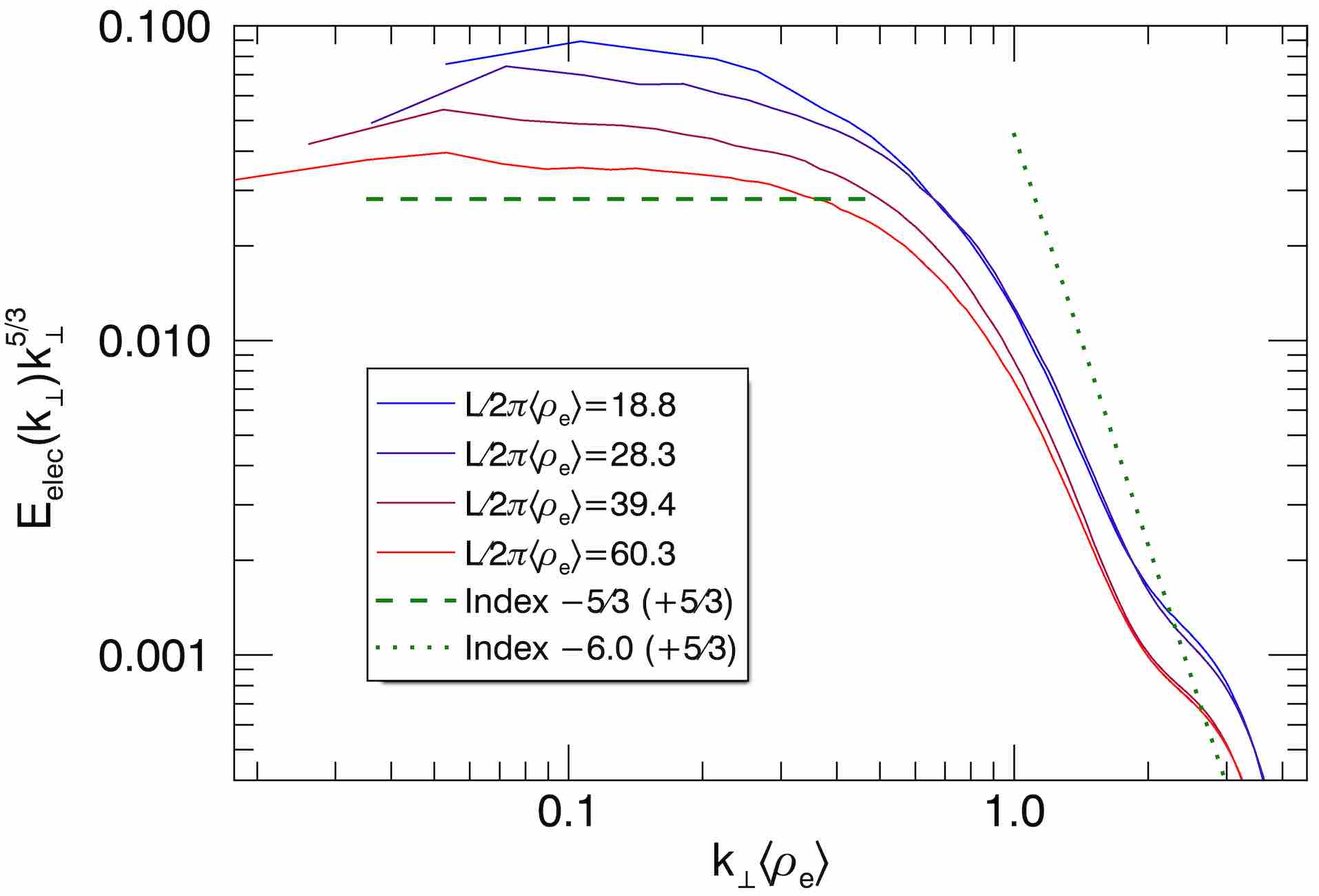}
 \includegraphics[width=\columnwidth]{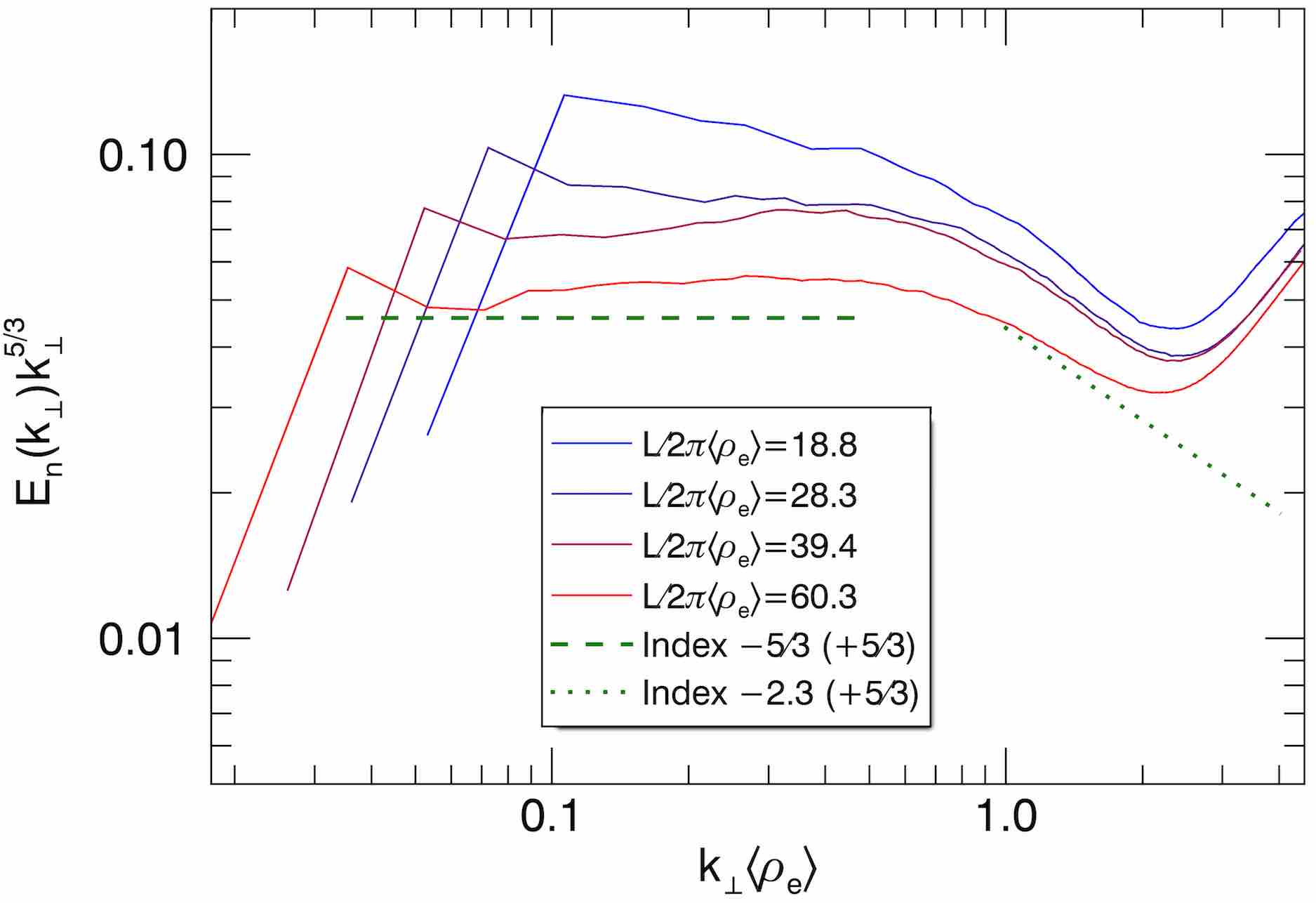}
  \centering
   \caption{\label{fig:turb_spec_size} Power spectra for magnetic fluctuations $E_{\rm mag}(k_\perp)$ (top), electric fluctuations $E_{\rm elec}(k_\perp)$ (center), and density fluctuations $E_n(k_\perp)$ (bottom), compensated by $k_\perp^{5/3}$, for varying system size and fixed magnetization $\langle\sigma\rangle \sim 1$. Inertial-range fits (prior to compensation) of $k_\perp^{-5/3}$ are shown for all cases (dashed lines). Kinetic-range scalings of $k_\perp^{-6}$ for $E_{\rm mag}(k_\perp)$ and $E_{\rm elec}(k_\perp)$, and $k_\perp^{-2.3}$ for $E_n(k_\perp)$, are also shown for reference (dotted lines).}
 \end{figure}
 
We proceed to describe the power spectra for various turbulent quantities. In the following, we integrate the power spectra across wavenumbers parallel to the global mean field $\boldsymbol{B}_0$ and angles around it, showing the resulting reduced spectra with respect to the perpendicular wavenumber $k_\perp = (k_x^2 + k_y^2)^{1/2}$. The motivation to focus on the perpendicular spectra comes from the anisotropy of MHD turbulence, which leads to different spectra perpendicular and parallel to the background magnetic field, as described by critical balance \citep{goldreich_sridhar_1995}. This anisotropy must be measured in coordinates with respect to the local magnetic field, which is generally tilted from the global guide field \citep{lazarian_vishniac_1999, cho_vishniac_2000}, and can be verified by applying structure functions or spectra along magnetic field lines \citep{cho_vishniac_2000, beresnyak_2015, zhdankin_etal_2018a}. For simplicity, in this work we only focus on the perpendicular scalings, which are accurately measured in the global coordinate system.

We show a series of power spectra (compensated by $k_\perp^{5/3}$) in Fig.~\ref{fig:turb_spec_size}. We find that the magnetic energy spectrum $E_{\rm mag}(k_\perp)$, electric energy spectrum $E_{\rm elec}(k_\perp)$, and particle density spectrum $E_n(k_\perp)$ all approach a scaling consistent with $k_\perp^{-5/3}$ in the inertial range ($k_\perp \rho_e \lesssim 1$) as the size is increased, for the cases with fixed $\langle\sigma\rangle \approx 1$. The largest case, on a $768^3$ lattice with $L/2\pi\langle\rho_e\rangle = 60.3$, has an inertial range roughly from $k_\perp \langle\rho_e\rangle \approx 0.05$ to $k_\perp \langle\rho_e\rangle \approx 0.3$. At small scales, $k_\perp\langle\rho_e\rangle > 1$, we find that the magnetic and electric energy spectra substantially steepen; although not a clear power law, these kinetic (sub-inertial) range spectra steepen to a scaling comparable to $k_\perp^{-6}$. The spectral indices in the kinetic range, however, are sensitive to numerical parameters such as filtering, resolution, and number of particles per cell; hence, we leave their asymptotic scaling to future work.

\begin{figure}
  \includegraphics[width=\columnwidth]{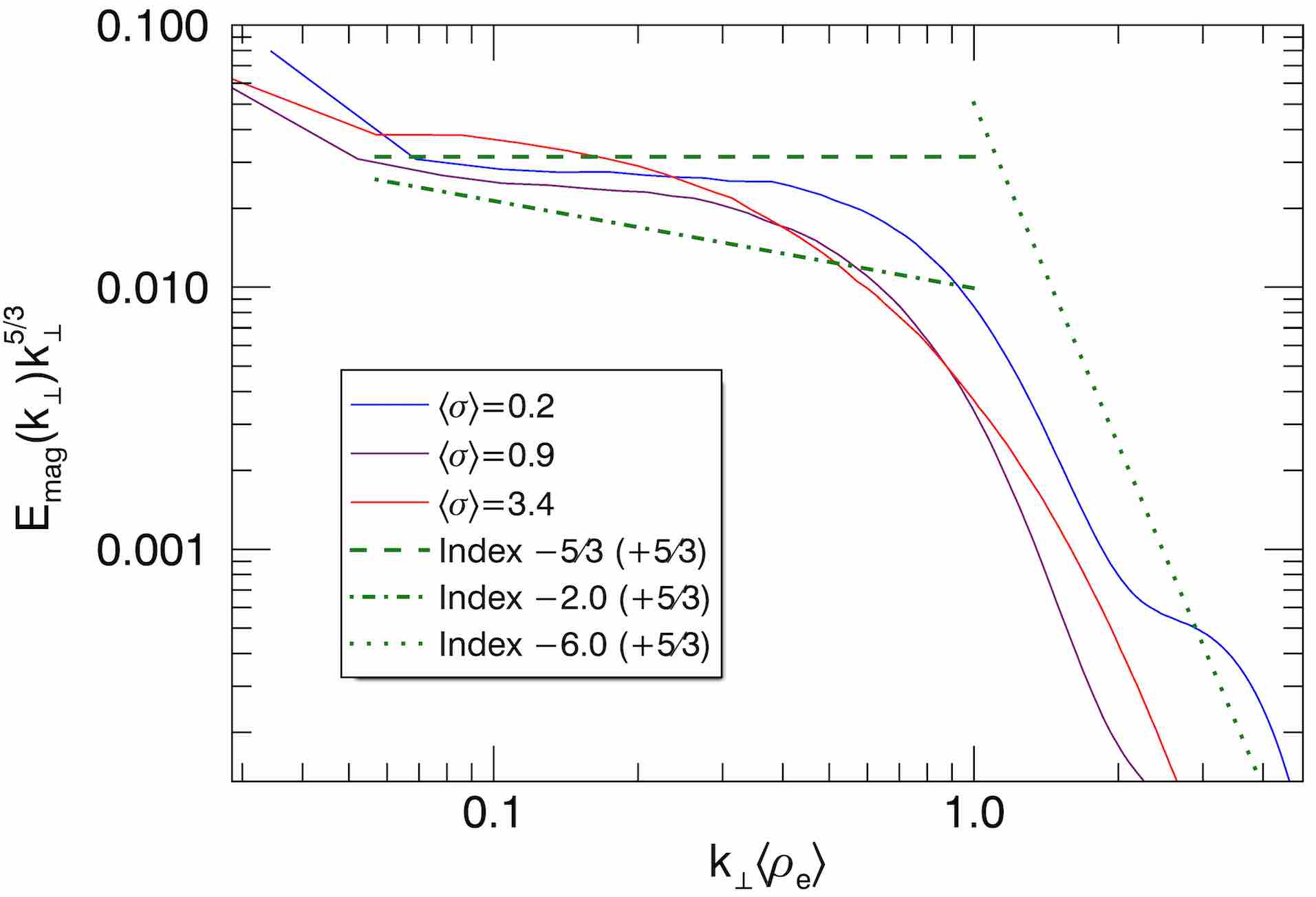}
  \includegraphics[width=\columnwidth]{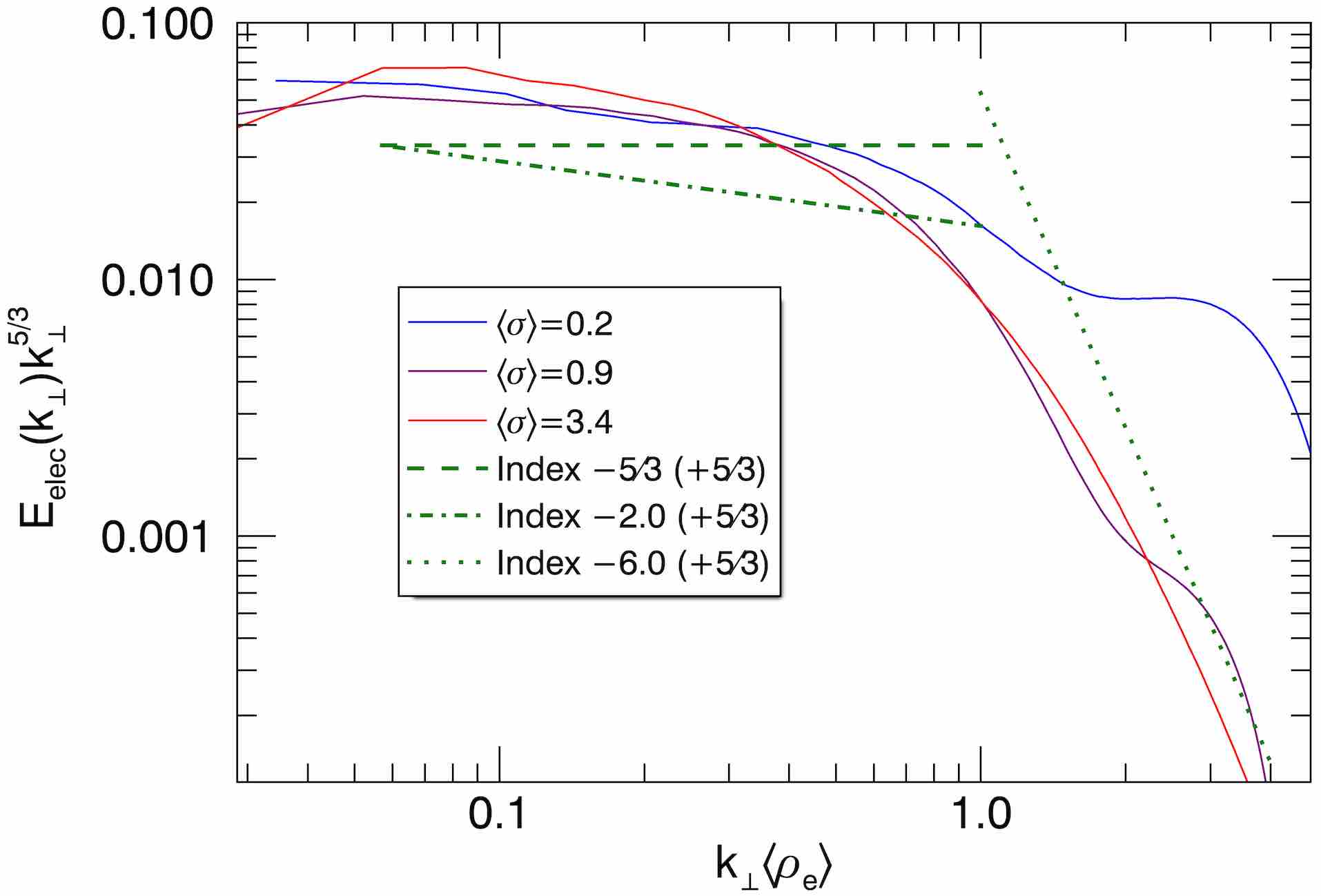}
  \includegraphics[width=\columnwidth]{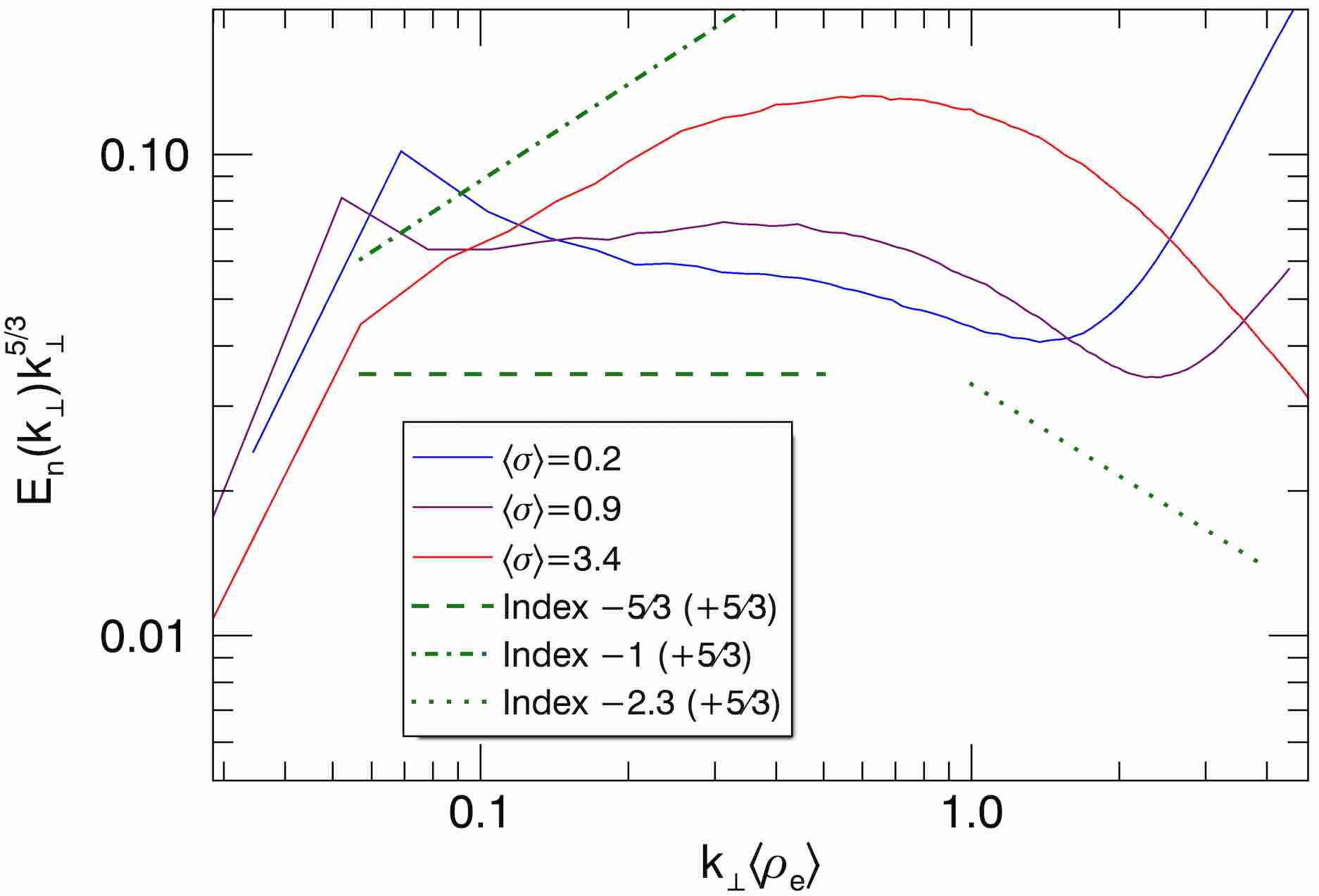}
  \centering
   \caption{\label{fig:turb_spec_sigma} Power spectra for magnetic fluctuations $E_{\rm mag}(k_\perp)$ (top), electric fluctuations $E_{\rm elec}(k_\perp)$ (center), and density fluctuations $E_n(k_\perp)$ (bottom), compensated by $k_\perp^{5/3}$, for fixed lattice size ($512^3$) and varying magnetization $\langle\sigma\rangle \in \{ 0.2, 0.9, 3.4\}$. Fits from Fig.~\ref{fig:turb_spec_size} are also shown, along with steeper $k_\perp^{-2}$ inertial-range scalings for $E_{\rm mag}(k_\perp)$ and $E_{\rm elec}(k_\perp)$, and a shallower $k_\perp^{-1}$ scaling for $E_n(k_\perp)$ (dash-dotted line).}
 \end{figure}
 
 In Fig.~\ref{fig:turb_spec_sigma}, we again show the three spectra $E_{\rm mag}(k_\perp)$, $E_{\rm elec}(k_\perp)$, and $E_n(k_\perp)$, compensated by $k_\perp^{5/3}$, this time for varying $\langle\sigma\rangle$ (taken from the $512^3$ simulations; recall that $L/\langle\rho_e\rangle$ is not strictly the same for these three cases). The magnetic energy spectrum is consistent with $k_\perp^{-5/3}$ for $\langle\sigma\rangle \lesssim 1$, but becomes steeper for the high-magnetization case, $\langle\sigma\rangle = 3.4$, with no clear power law. Evidently, the break near $k_\perp\langle\rho_e\rangle \sim 1$ becomes broader at high $\langle\sigma\rangle$, perhaps indicating that the transition to the kinetic range occurs at $d_e$ rather than $\rho_e$. The electric energy spectrum appears to be similar for all three cases (slightly steeper than $k_\perp^{-5/3}$). The $\langle\sigma\rangle \lesssim 1$ cases have density spectra close to $k_\perp^{-5/3}$, as expected for subsonic MHD turbulence; this becomes drastically shallower for the $\langle\sigma\rangle = 3.4$ case, with a scaling near $k_\perp^{-1}$. This shallow spectrum is qualitatively similar to the case of supersonic MHD turbulence, where the density spectrum becomes dominated by intense, small-scale compressive structures in the high Mach number regime \citep[e.g.,][]{beresnyak_etal_2005, kowal_etal_2007}.

\subsection{Energy transfer spectrum} \label{sec:trans}

\begin{figure}
 \includegraphics[width=\columnwidth]{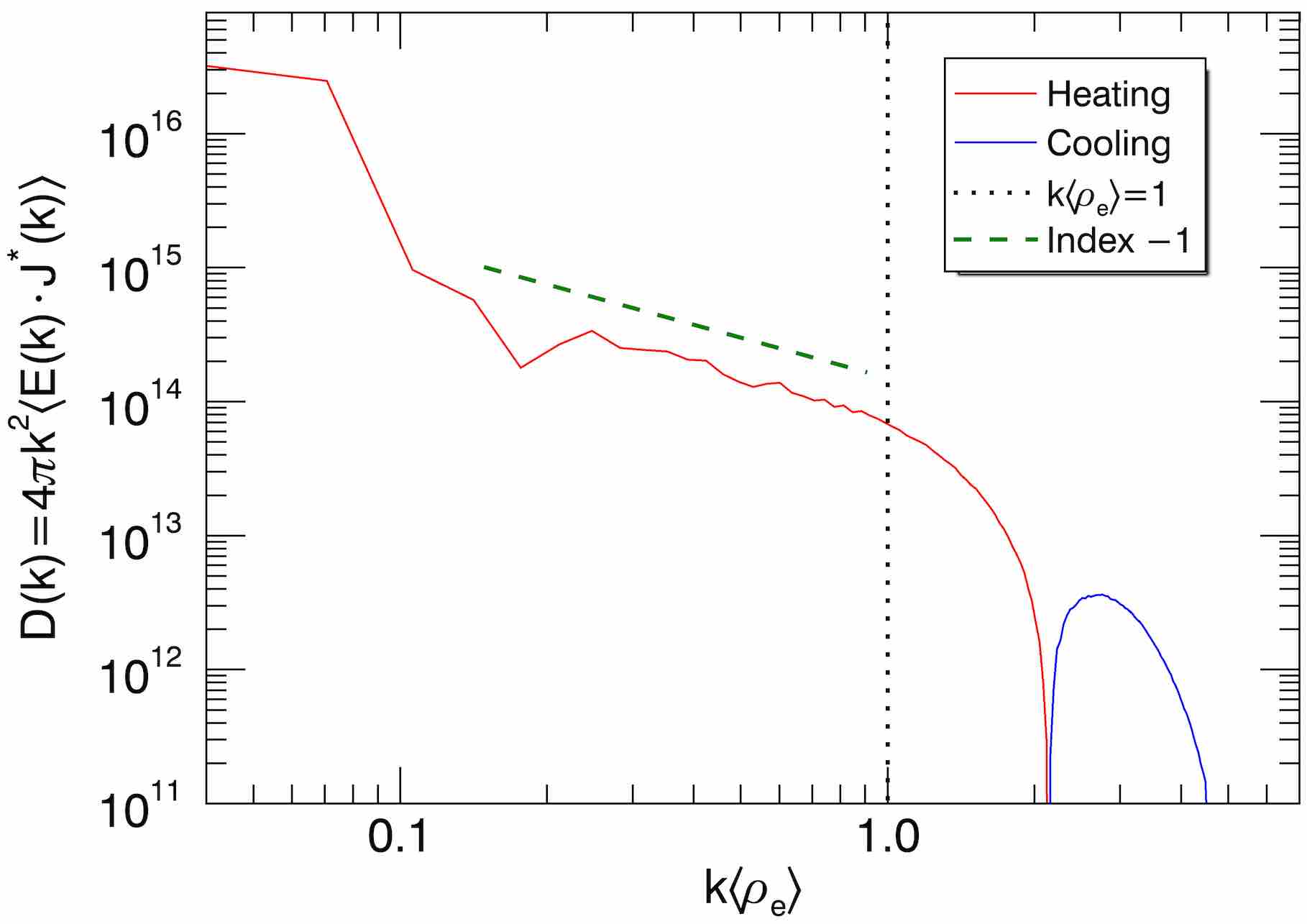}
  \includegraphics[width=\columnwidth]{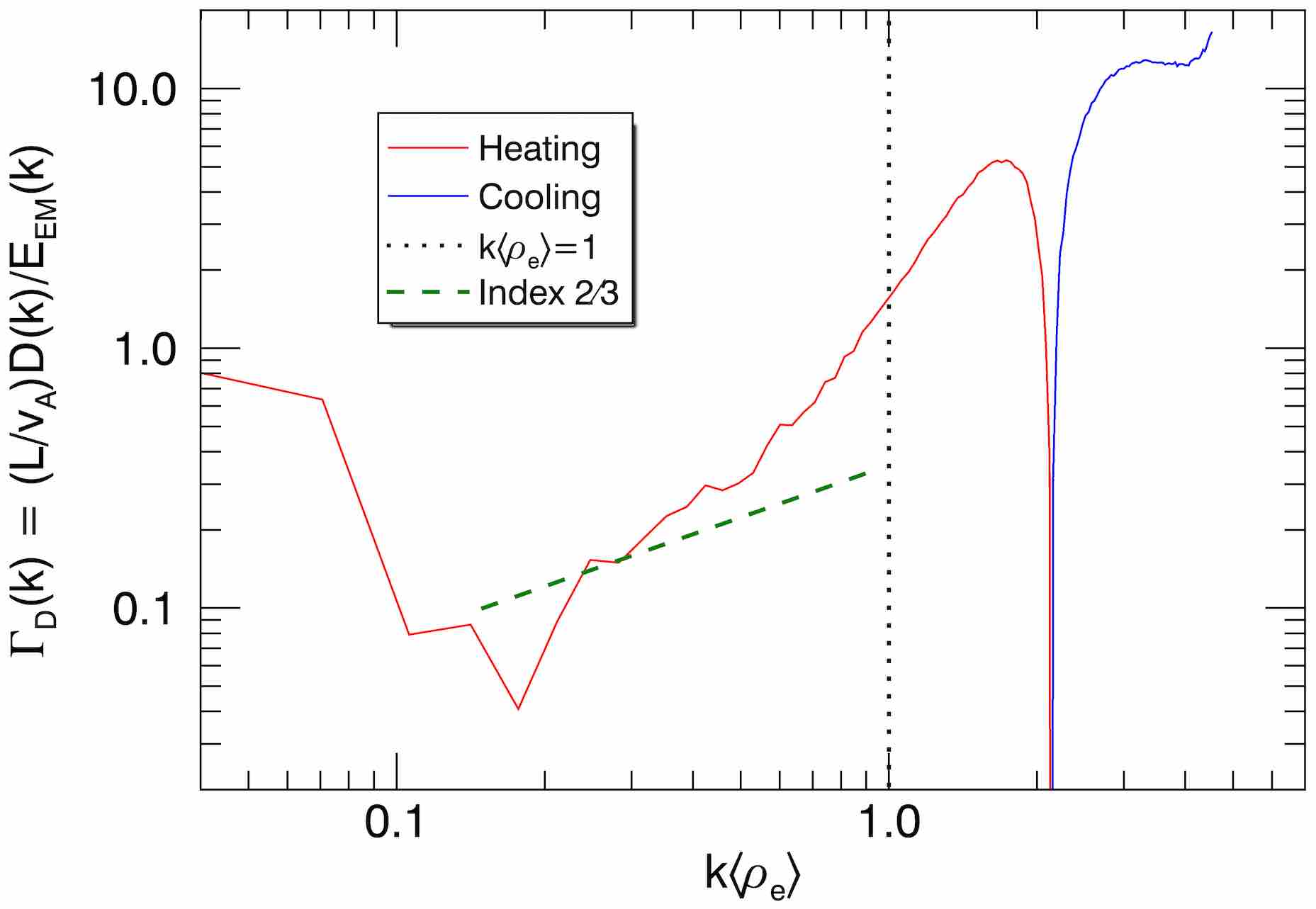}
  \centering
   \caption{\label{fig:spec_heating_prod} Top panel: The energy transfer spectrum ${\mathcal D}(k)$ in a $384^3$ simulation with $\langle\sigma\rangle \sim 1$, with positive values (red) and negative values (blue) both shown, and a $k^{-1}$ power-law fit (green, dashed). Bottom panel: The normalized energy transfer rate, $\Gamma_D(k) = (L/v_A) {\mathcal D}(k)/E_{\rm EM}(k)$, where $E_{\rm EM}(k)$ is the electromagnetic energy spectrum, with a $k^{2/3}$ power law for reference.}
 \end{figure}

A primary objective in studies of kinetic turbulence is to identify the nature of energy transfer from electromagnetic fields to internal energy \citep[e.g.,][]{matthaeus_etal_2015}. To take a first step in this direction, in this subsection, we measure the irreversible transfer of energy from the turbulent electromagnetic field to the plasma as a function of scale. To accomplish this, we implement the following diagnostic. We first expand the electric field and current density in terms of Fourier modes: $\boldsymbol{E}(\boldsymbol{x},t) = \int d^3k \tilde{\boldsymbol{E}}(\boldsymbol{k},t) e^{i \boldsymbol{k} \cdot \boldsymbol{x}}/(2\pi)^3$ and $\boldsymbol{J}(\boldsymbol{x},t) = \int d^3k \tilde{\boldsymbol{J}}(\boldsymbol{k},t) e^{i \boldsymbol{k} \cdot \boldsymbol{x}}/(2\pi)^3$, respectively. The rate of overall particle kinetic energy gain can then be written as
\begin{align}
\int d^3x \boldsymbol{E}(\boldsymbol{x},t) \cdot \boldsymbol{J}(\boldsymbol{x},t) &= \int \frac{d^3x d^3k d^3k'}{(2 \pi)^6} \tilde{\boldsymbol{E}}(\boldsymbol{k},t) \cdot \tilde{\boldsymbol{J}}(\boldsymbol{k}',t) e^{i (\boldsymbol{k} + \boldsymbol{k}') \cdot \boldsymbol{x}} \nonumber \\
&= \int \frac{d^3k d^3k'}{(2 \pi)^3} \tilde{\boldsymbol{E}}(\boldsymbol{k},t) \cdot \tilde{\boldsymbol{J}}(\boldsymbol{k}',t) \delta(\boldsymbol{k} + \boldsymbol{k}') \nonumber \\
&= \int \frac{d^3k}{(2 \pi)^3} \tilde{\boldsymbol{E}}(\boldsymbol{k},t) \cdot \tilde{\boldsymbol{J}}^*(\boldsymbol{k},t) \, .
\end{align}
Thus, the integrand $\tilde{\boldsymbol{E}}(\boldsymbol{k},t) \cdot \tilde{\boldsymbol{J}}^*(\boldsymbol{k},t)$ describes the rate of energy transfer from the electromagnetic field to the plasma (involving bulk flows, adiabatic compressions, heating, and nonthermal particle acceleration) for modes with the given wavenumber $\boldsymbol{k}$. Note that the energy transfer rate between the electric field and flows/compressions can be positive or negative, while irreversible energy dissipation (heating and nonthermal particle acceleration) has a net positive value. Thus, the signatures of flows and compressions are removed after integrating over directions of $\boldsymbol{k}$ and averaging over sufficiently long times. We are therefore led to define the {\it energy transfer spectrum\rm} by
\begin{align}
{\mathcal D}(k) = k^2 \int d\Omega \langle \tilde{\boldsymbol{E}}(\boldsymbol{k},t) \cdot \tilde{\boldsymbol{J}}^*(\boldsymbol{k},t) \rangle \, ,
\end{align}
where $d\Omega$ is the solid angle differential in $\boldsymbol{k}$ space (we do not take into account anisotropy with respect to $\boldsymbol{B}_0$ here). The integral of the energy transfer spectrum is proportional to the (average) rate of radiative energy loss, $\int dk {\mathcal D}(k) \propto \dot{E}_{\rm rad}$.

We measure ${\mathcal D}(k)$ for a simulation with $384^3$ cells and $\langle\sigma\rangle \sim 1$, run rS1* in Table~\ref{table:sims}, identical to run rS1 except with a longer duration and higher cadence of dumps for $\boldsymbol{E}$ and $\boldsymbol{J}$ fields (174 snapshots from $t v_A/L = 5.3$ to $t v_A/L = 40.5$). The result is shown in the first panel of Fig.~\ref{fig:spec_heating_prod}. We find that ${\mathcal D}(k)$ is very strongly peaked at driving scales, consistent with the secular injection of energy into the cascade by the electromagnetic driving. In the inertial range, ${\mathcal D} \propto k^{-1}$, indicating that the energy transfer is scale-invariant throughout this range, in the sense that $\int dk {\mathcal D}(k)$ is constant for any given logarithmic interval in wavenumbers. We interpret this as a signature of the energy cascade, which secularly transfers energy from bulk flows at low $k$ to bulk flows at high $k$ at a constant (scale-independent) rate, thus siphoning energy out of the electromagnetic field (which is continuously replenished from large-scale energy transfer). This result is consistent with the absence of dissipation mechanisms at MHD scales. Finally, ${\mathcal D}(k)$ strongly decreases at scales $k \langle\rho_e\rangle \gtrsim 1$, becoming negative (indicating net cooling) at large wavenumbers, $k \gtrsim 2/\langle\rho_e\rangle$. However, we find that this cooling region diminishes as the number of particles per cell is increased (not shown), indicating that it is likely a numerical artifact.

We next divide ${\mathcal D}(k)$ by the electromagnetic energy spectrum, $E_{\rm EM}(k) = E_{\rm mag}(k) + E_{\rm elec}(k)$, to obtain the rate of energy transfer (which we normalize to $v_A/L$), $\Gamma_D(k) = (L/v_A) D(k)/E_{\rm EM}(k)$. As shown in the second panel of Fig.~\ref{fig:spec_heating_prod}, $\Gamma_D \sim 1$ at driving scales, consistent with nonlinear energy transfer through Alfv\'e{n}ic motions. The energy transfer rate then drops just below the driving scales, and increases with wavenumber until $k\langle\rho_e\rangle \sim 1$, with a scaling somewhat steeper than $k^{2/3}$ (the scaling expected from a $k^{-5/3}$ energy spectrum; this is due to the simulation size being too small for a converged inertial-range spectral index). The rate then increases strongly at $k\langle\rho_e\rangle \gtrsim 1$, peaking near $k\langle\rho_e\rangle = 2$, indicative of damping at sub-inertial scales.

These results are broadly consistent with kinetic-scale dissipation mechanisms, such as gyroresonant particle acceleration, which is further discussed in the context of steady-state particle distributions in Sec.~\ref{sec:fp-fits}. We cannot preclude nonlocal contributions to the energy transfer (damping of compressive modes, magnetic reconnection, etc.), which are difficult to distinguish from the signatures of large-scale driving and the cascade.

\subsection{Magnetic field and density fluctuations} \label{sec:fluc}

\begin{figure}
 \includegraphics[width=\columnwidth]{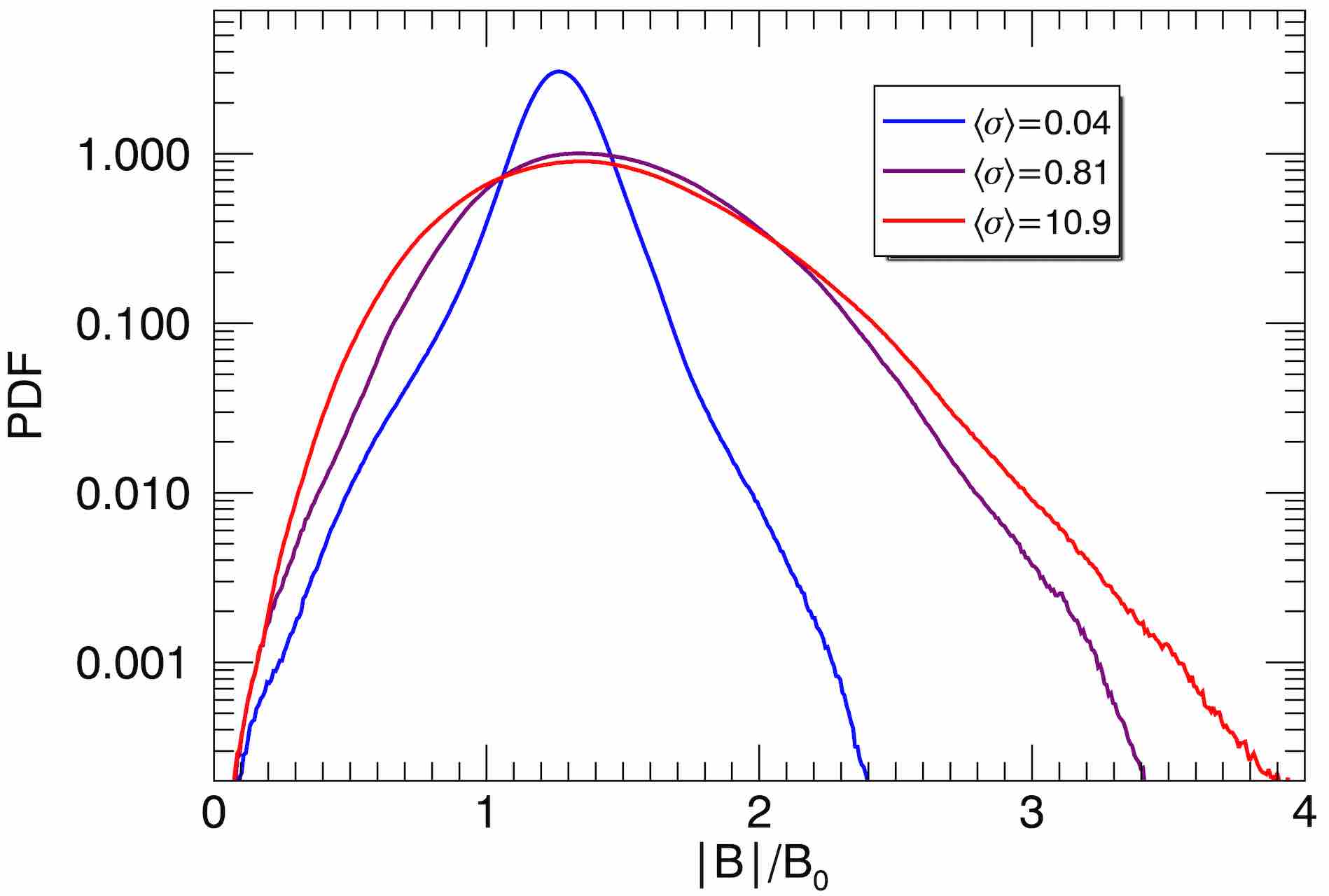}
  \includegraphics[width=\columnwidth]{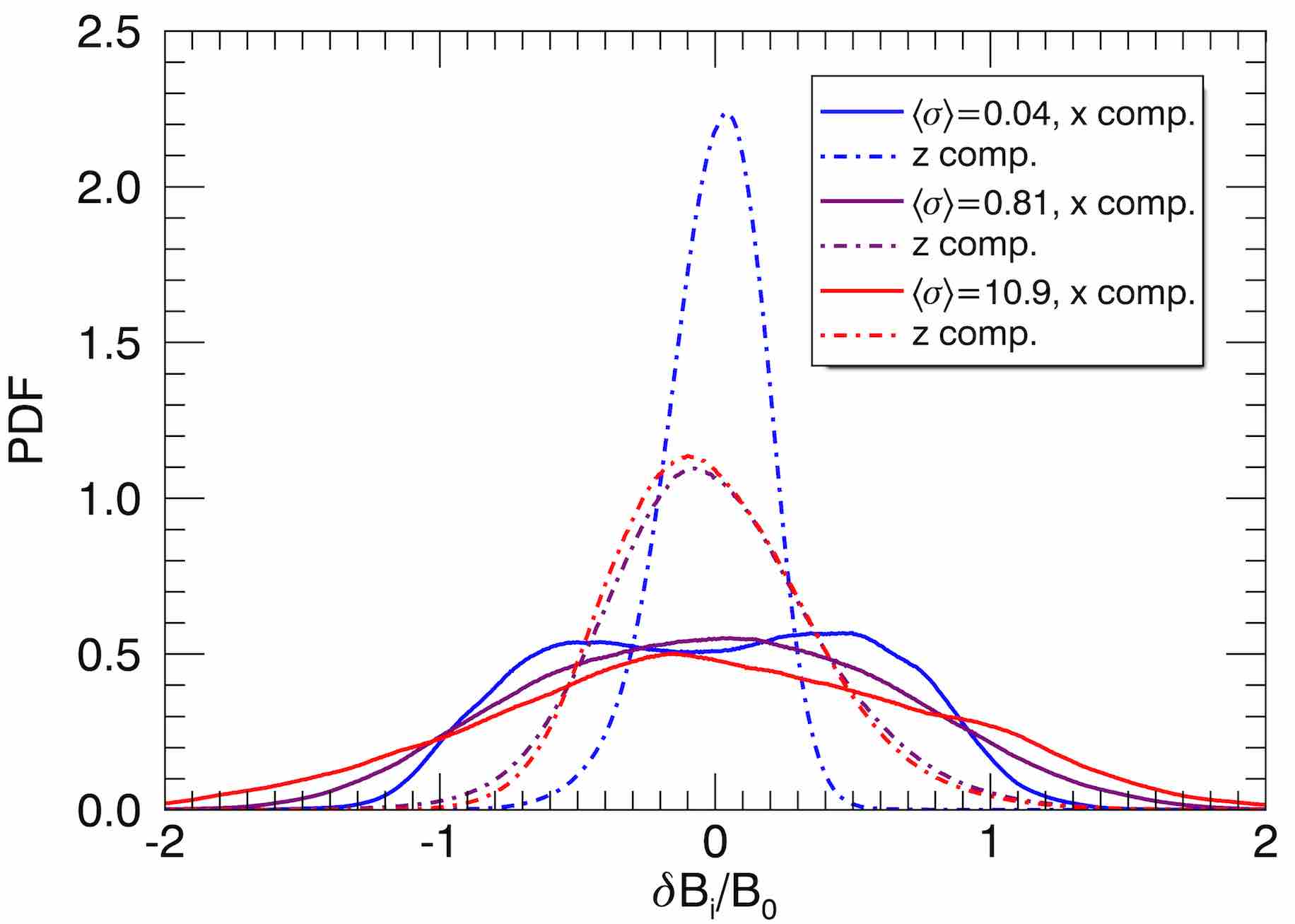}
  \centering
   \caption{\label{fig:pdf_mag} Top panel: probability distribution for the magnetic field magnitude $B$ for the $384^3$ simulations with $\langle \sigma \rangle \in \{ 0.04, 0.8, 11 \}$. Bottom panel: probability distribution for the fluctuating magnetic field components $\delta B_z$ (dashed) and $B_x$ (solid) for the same simulations.}
 \end{figure}
 
In this subsection, we comment on the statistics of the magnetic and density fluctuations. In the top panel of Fig.~\ref{fig:pdf_mag}, we show the probability distribution for the magnetic field magnitude $B$ in the $384^3$ series of simulations with $\langle \sigma \rangle \in \{ 0.04, 0.8, 11 \}$. We find that although all cases have distributions peaked near $B/B_0 \approx \sqrt{2}$ (as dictated by the driving amplitude), the low~$\sigma$ case has a much narrower distribution than the other two cases. In the bottom panel of Fig.~\ref{fig:pdf_mag}, we show the probability distribution for the magnetic field components $B_x$ and $\delta B_z = B_z - B_0$,  showing that while the perpendicular magnetic fluctuations have similar distributions for all of the simulations, the component along $\boldsymbol{B}_0$ has a narrower distribution in the low~$\sigma$ case. Together, these results indicate that the magnetic fluctuations become rotationally dominated at low~$\sigma$, which may be a consequence of kinetic instabilities or constraints from pressure anisotropy at high~$\beta$ \citep{tenerani_velli_2018, squire_etal_2019}. This appears to be a nontrivial signature of collisionless plasma physics affecting fluctuations at large scales,  which would not appear in an MHD simulation.

\begin{figure}
 \includegraphics[width=\columnwidth]{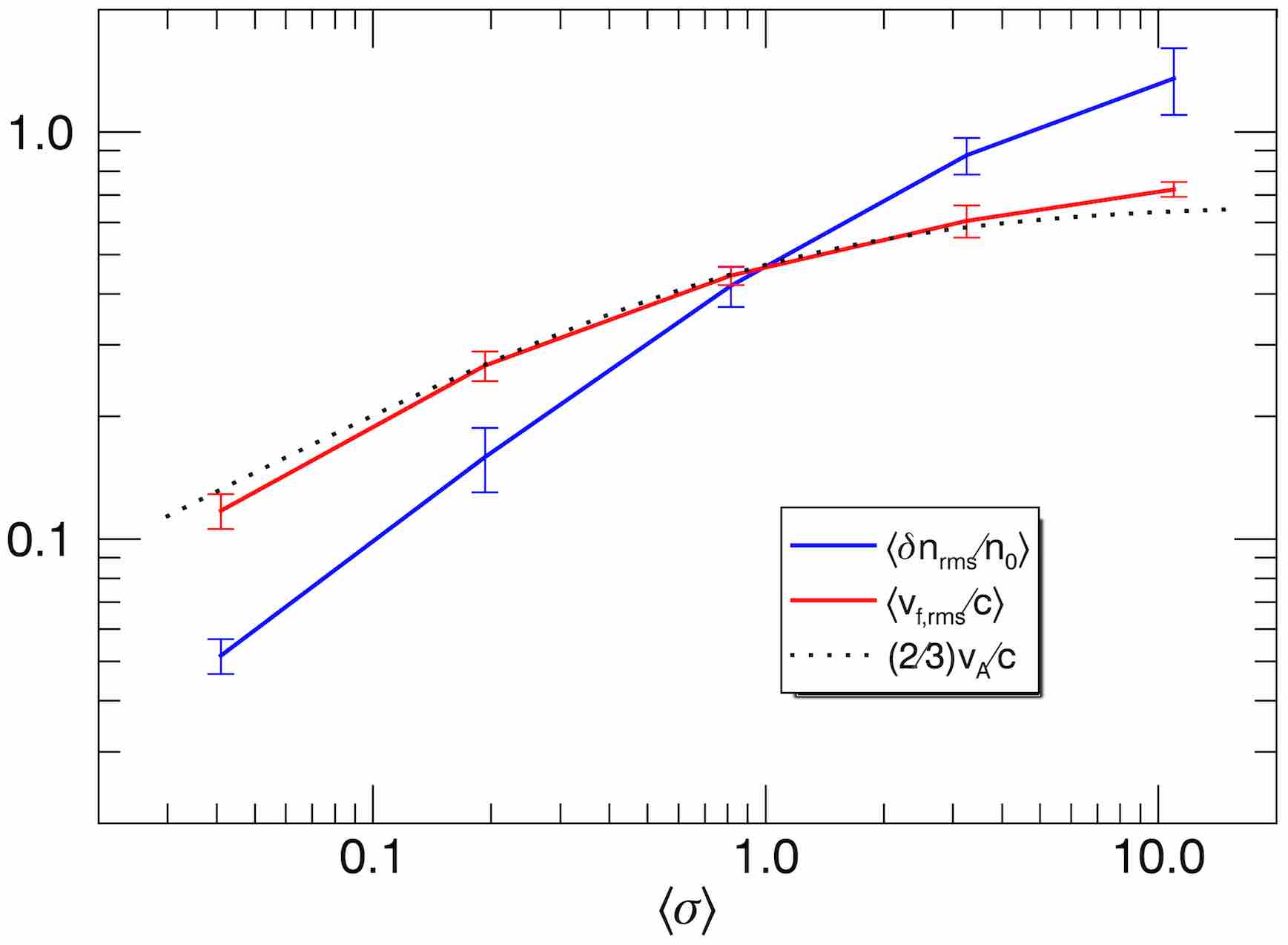}
  \includegraphics[width=\columnwidth]{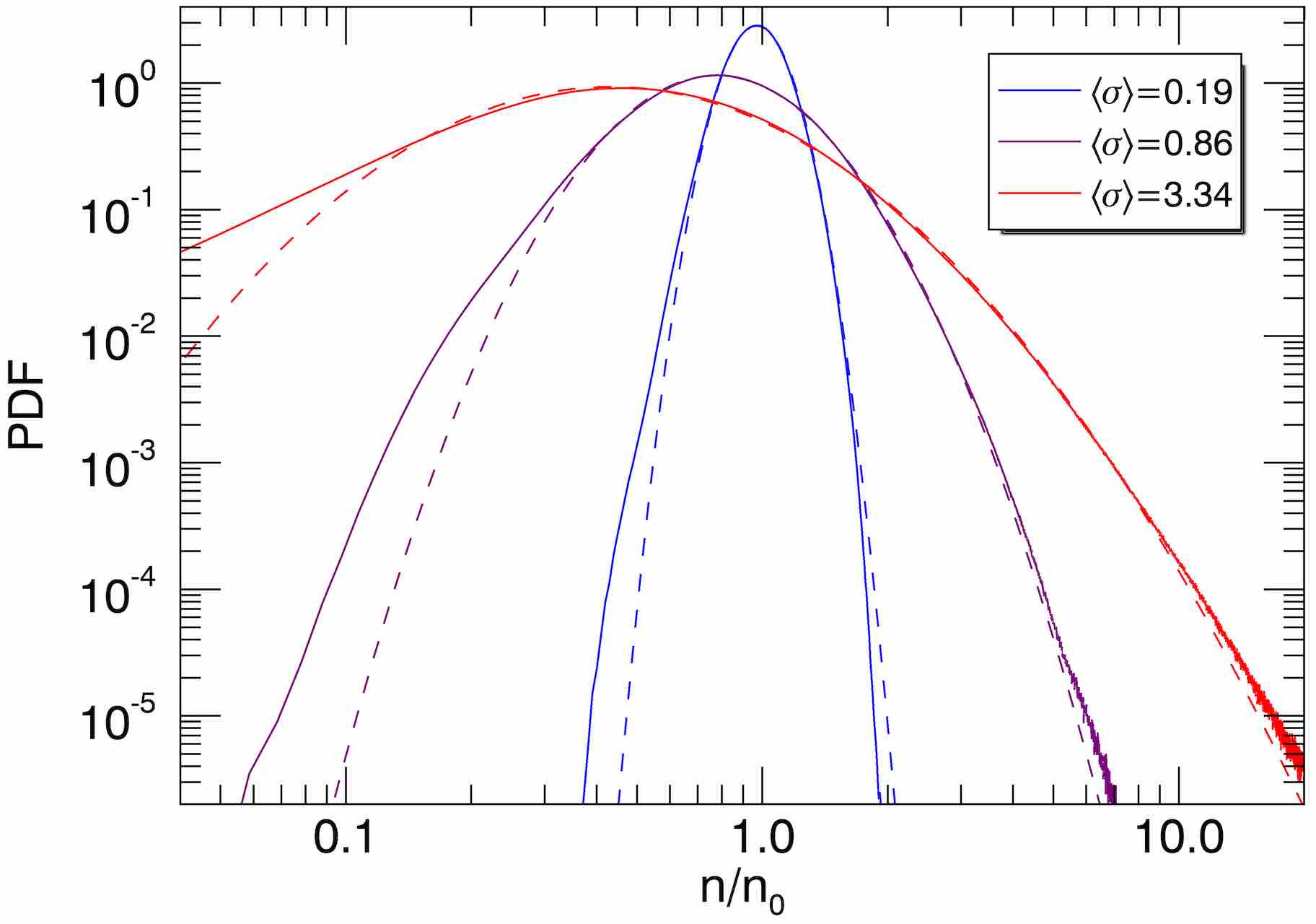}
  \centering
   \caption{\label{fig:density_velocity} Top panel: The time-averaged rms fluctuations for number density $\langle \delta n_{\rm rms} \rangle$ (blue) and flow velocity $\langle v_{f,\rm rms} \rangle$ (red) versus magnetization $\langle\sigma\rangle$ in the $384^3$ simulation series. For reference, the dotted line shows $(2/3) v_A/c$. Bottom panel: Probability distribution for the particle density $n$ (for the $512^3$ series), with log-normal fits (dashed).}
 \end{figure}
 
In the top panel of Fig.~\ref{fig:density_velocity}, we show the scaling of the rms fluctuations in number density, $\delta n_{\rm rms}$, and in velocity, $v_{f,\rm rms}$, with respect to $\langle\sigma\rangle$ (for the $384^3$ series). We find that the velocity scaling can be fit by $(2/3) v_A/c$, consistent with Alfv\'{e}nic fluctuations. We find that density fluctuations also increase with $\sigma$, with fluctuations becoming comparable to the mean ($\delta n_{\rm rms} \sim n_0$) at $\sigma \sim 3$. Since the speed of sound in a relativistic ideal gas is $c_s = c/\sqrt{3} \approx 0.58 c$ \citep[e.g.,][]{weinberg_1972}, shocks are formed for $\sigma \gtrsim 1$. In the bottom panel of Fig.~\ref{fig:density_velocity}, we show the time-averaged probability distribution for the particle density $n$, for the $512^3$ series of simulations. We find that the log-normal distribution provides a reasonable fit for the distributions (for the given range of $\langle \sigma \rangle$), apart from at low values of the density, where there is a noticable excess of particles compared to the fit. This excess diminishes when the number of particles per cell is increased (not shown), thus we consider it a numerical artifact due to PIC noise.

\subsection{Pressure anisotropy}

The pressure tensor in a collisionless plasma is generally anisotropic with respect to the local magnetic field. The establishment of a significant pressure anisotropy can affect the thermodynamics of the plasma and trigger kinetic (mirror and firehose) instabilities, as observed in, e.g., numerical simulations \citep{kunz_etal_2014, servidio_etal_2014b} and the solar wind \citep{bale_etal_2009}. In this subsection, we consider the statistics of the pressure anisotropy ratio $P_\perp/P_{||}$, with the pressure components parallel and perpendicular to the local magnetic field $\boldsymbol{B}$ defined by
\begin{align}
P_{||} &= \hat{\boldsymbol{B}} \hat{\boldsymbol{B}} : {\bf P} \nonumber \\
P_{\perp} &= \frac{1}{2} \left( {\bf I} - \hat{\boldsymbol{B}} \hat{\boldsymbol{B}} \right) : {\bf P} \, ,
\end{align}
where ${\bf P} = \int d^3p (\boldsymbol{p} \boldsymbol{p} c f)/\sqrt{m_e^2 c^2 + p^2}$ is the pressure tensor (with ram pressure terms from bulk flows included).

\begin{figure}
  \includegraphics[width=\columnwidth]{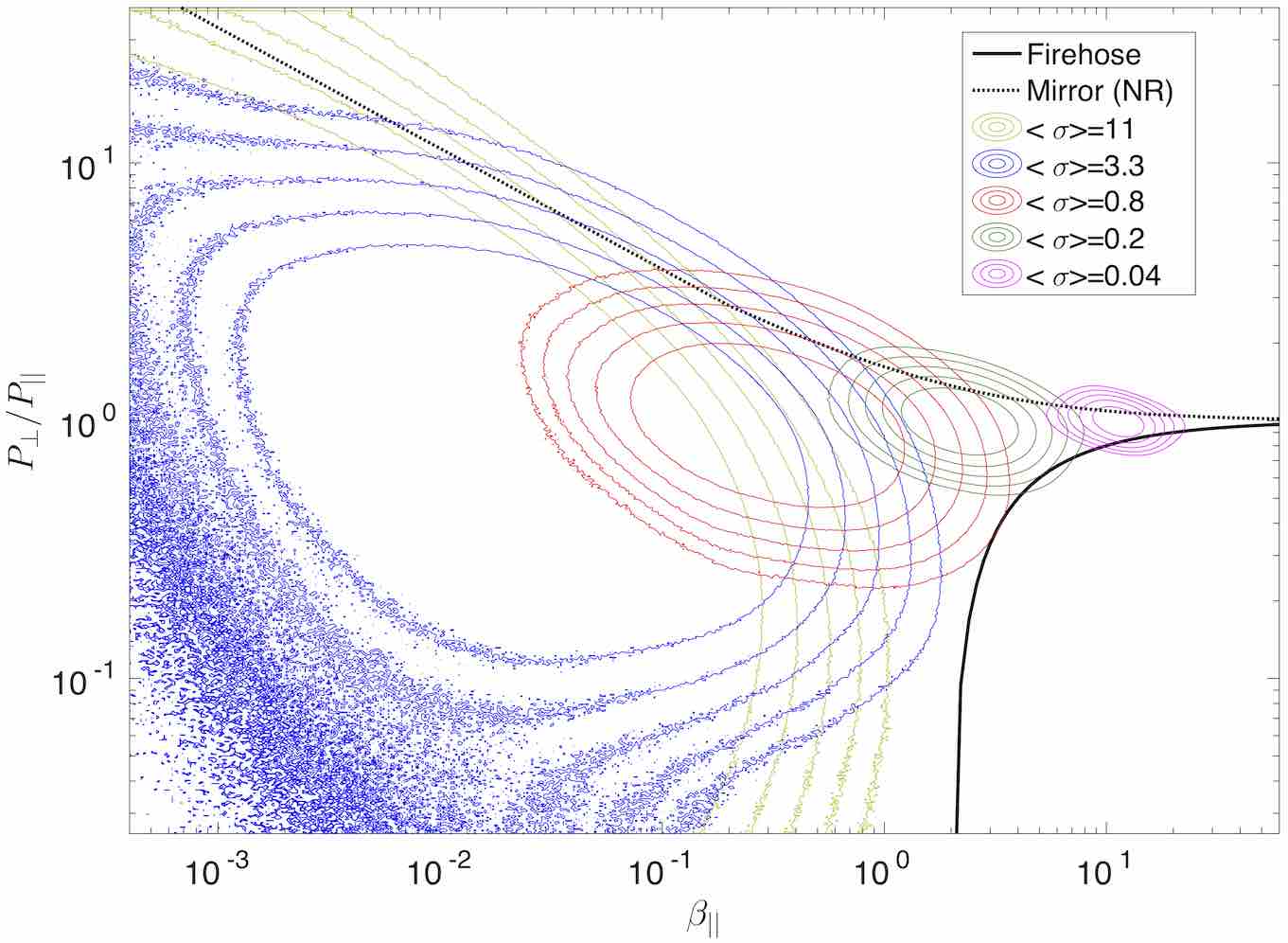}
  \centering
   \caption{\label{fig:dist_pressure} 2D probability distribution of pressure anisotropy ratio $P_\perp/P_{||}$ versus $\beta_{||}$ (for the $384^3$ series). Contours are shown at five values for each case: $\{ 1/2, 1/4, 1/8, 1/16, 1/32 \}$ of the maximum value of the $\langle\sigma\rangle=0.04$ distribution. Thresholds for the firehose instability (black, solid) and non-relativistic mirror instability (black, dashed) are shown for reference.}
 \end{figure}

In Fig.~\ref{fig:dist_pressure}, as per convention, we show the 2D distribution of pressure anisotropy ratio,~$P_\perp/P_{||}$, versus the plasma beta using the pressure component parallel to the magnetic field,~$\beta_{||} = 8 \pi P_{||} / B^2$, for the simulations in the $384^3$ series. We find that the distributions are peaked at isotropic pressure,~$P_\perp/P_{||} \sim 1$, and~$\beta_{||} \sim \beta \sim 1/(2\langle\sigma\rangle)$ (at least, for high~$\beta$). The low-$\beta$ (high-$\sigma$) cases have a broad spread in pressure anisotropy, ranging from $P_\perp/P_{||} \sim 0.1$ to $P_\perp/P_{||} \sim 10$, while the high-$\beta$ (low-$\sigma$) cases have much narrower distributions. We find that all of the simulations are within the marginal thresholds for the mirror and firehose instabilities \citep{kunz_etal_2014},
\begin{align}
\frac{P_\perp}{P_{||}} \Big|_{\rm Firehose} &\lesssim 1 - \frac{2}{\beta_{||}} \, , \nonumber \\
\frac{P_\perp}{P_{||}} \Big|_{\rm Mirror} &\gtrsim \frac{1}{2} \left( 1 + \sqrt{1 + \frac{4}{\beta_{||}}} \right) \, .
\end{align}
Note that both thresholds here are taken from the non-relativistic limit\footnote{For a discussion of the relativistic MHD mirror and firehose thresholds in the doubly polytropic approximation, which may not be strictly applicable for a collisionless plasma, see \cite{chou_hau_2004}. In this case, the firehose threshold remains unchanged for an ultra-relativistic plasma, but the mirror threshold shifts.}. These results suggest that mirror and firehose instabilities may become important in some regions of space for the high-$\beta$ cases, constraining the kinetic physics and possibly influencing aspects of the turbulence (such as the magnetic field fluctuations described in Sec.~\ref{sec:fluc}).

\section{Particle statistics} \label{sec4}

\subsection{Steady-state particle energy distributions}

\begin{figure}
 \includegraphics[width=\columnwidth]{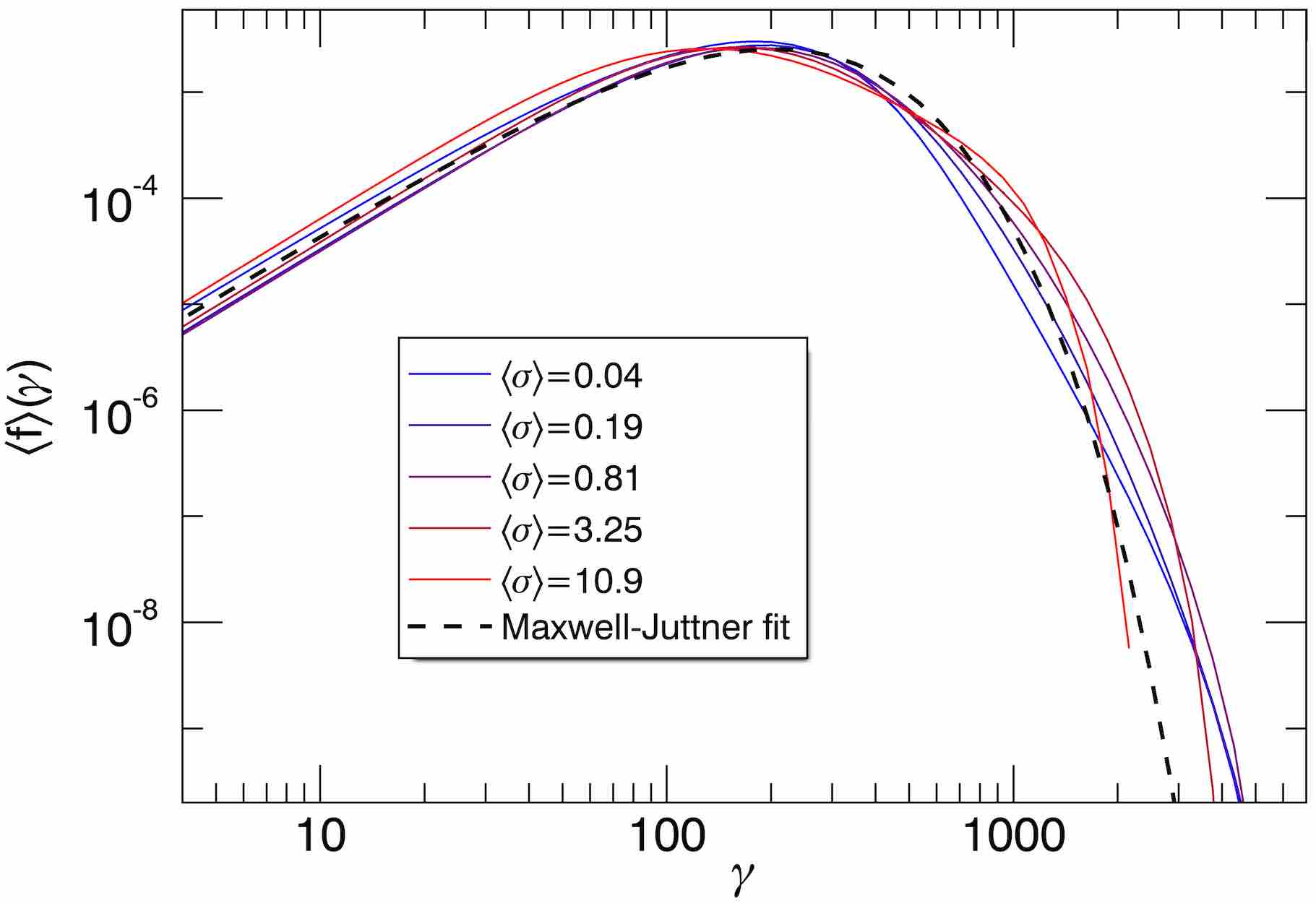}
 \includegraphics[width=\columnwidth]{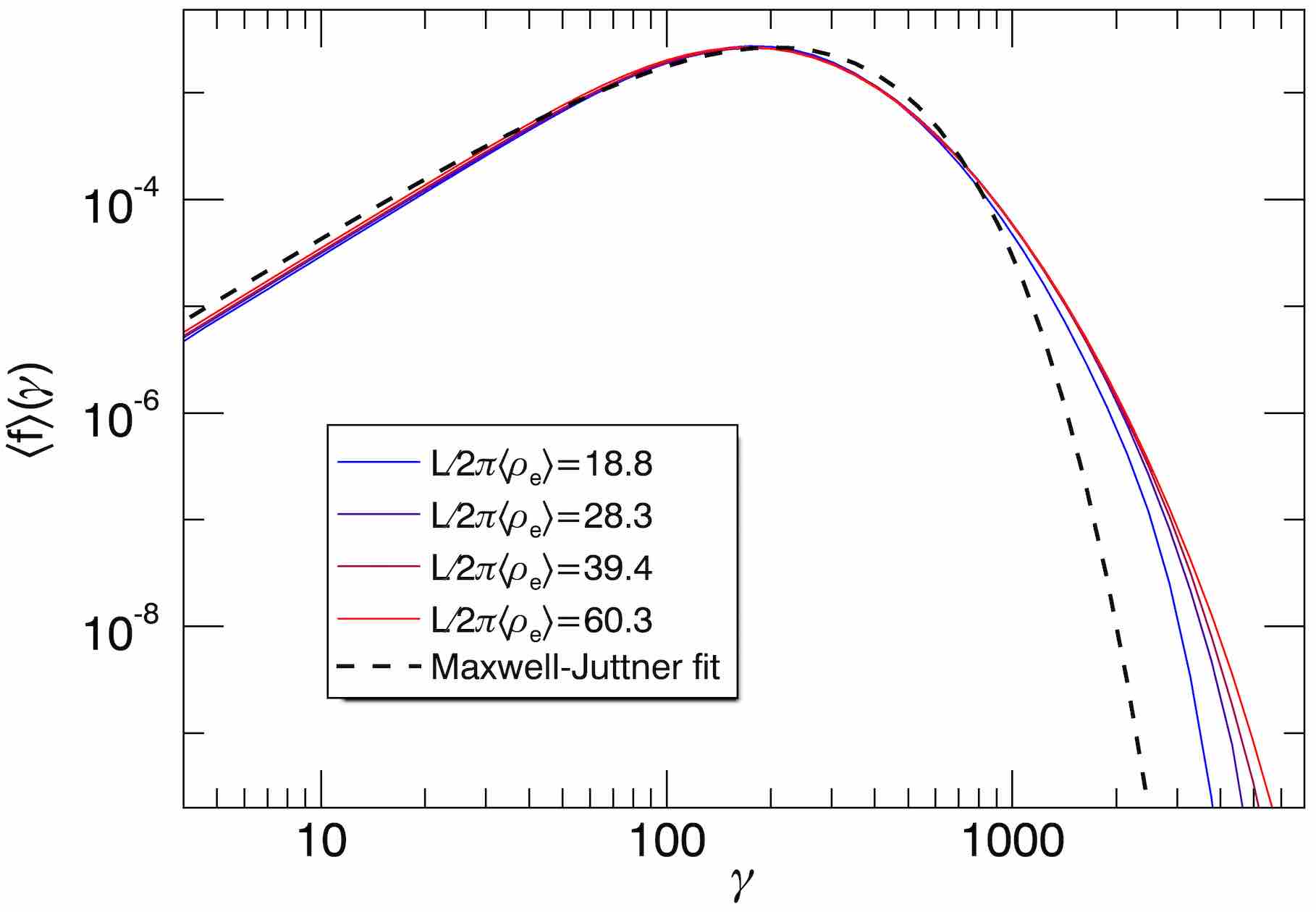}
  \centering
   \caption{\label{fig:dist_energy} Top panel: Time-averaged particle energy distributions $\langle f\rangle(\gamma)$ for varying $\langle\sigma\rangle$ (for the $384^3$ series). For reference, a Maxwell-J\"{u}ttner distribution with the same mean energy is also shown (black, dashed). Bottom panel: similar distributions for varying system size (at fixed $\langle\sigma\rangle \sim 1$).}
 \end{figure}

We now proceed to describe the particle statistics in our simulations, starting with the system-integrated energy distribution, $f(\gamma)$ (combined for electrons and positrons). Since $f(\gamma)$ is a global distribution, it in principle includes nonthermal signatures both from bulk flows and from particle acceleration. In the top panel of Fig.~\ref{fig:dist_energy}, we show the time-averaged particle energy distributions $\langle f \rangle(\gamma)$ for varying $\langle\sigma\rangle$ in the $384^3$ series of simulations. Our first main result is that the distributions are fairly close to the Maxwell-J\"{u}ttner distribution (Eq.~\ref{eq:mj}) for all $\sigma$. This is in contrast to the broad power-law distribution (extending up to the system-size limited energy $\gamma_{\rm max} = L e B_0/2 m_e c^2$) observed in simulations without radiative cooling \citep{zhdankin_etal_2017}, indicating that radiative cooling effectively thermalizes the plasma. In the bottom panel of Fig.~\ref{fig:dist_energy}, we similarly show $\langle f\rangle(\gamma)$ for varying system size $L/2\pi\langle\rho_e\rangle$ at fixed $\langle\sigma\rangle \sim 1$. We find that the system size weakly affects the far tail of the distribution, perhaps indicating that those high-energy particles are sensitive to fluctuations in the inertial range of turbulence. For $L/2\pi\langle\rho_e\rangle \gtrsim 25$, the dependence on the system size is negligible; this scale separation is achieved by most simulations in our study, apart from the $384^3$ low-$\langle\sigma\rangle$ cases. Note that the system-size limited energy $\gamma_{\rm max}$ is significantly higher than the energies obtained by any particles in these simulations: for example, the smallest case ($256^3$) in the system-size scan has $\gamma_{\rm max} = 1.28 \times 10^4 \approx 43 \overline{\gamma}$ while the largest case ($768^3$) has $\gamma_{\rm max} = 3.84 \times 10^4 \approx 128 \overline{\gamma}$.
 
 \begin{figure}[h]
 \includegraphics[width=\columnwidth]{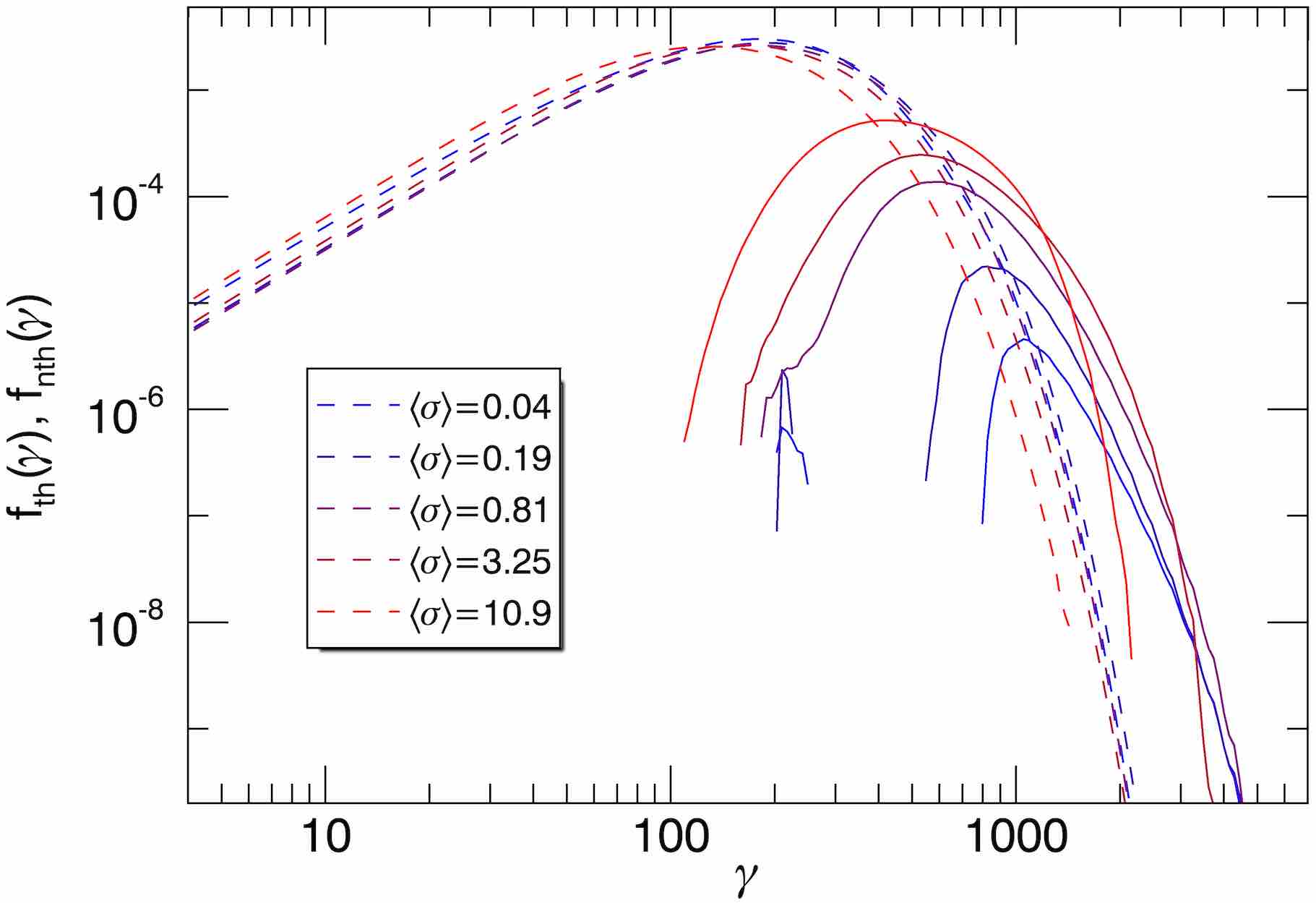}
 \includegraphics[width=\columnwidth]{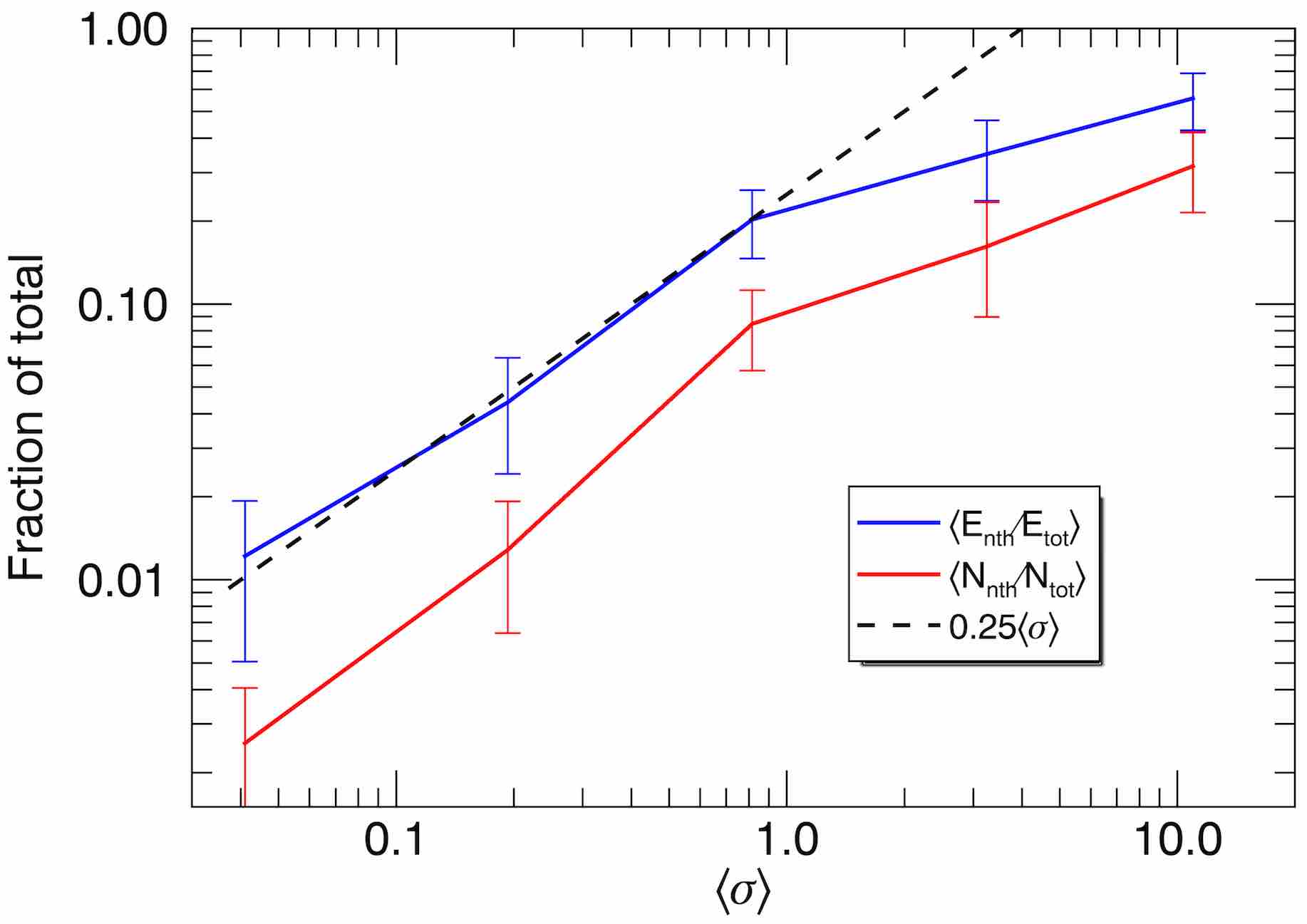}
\includegraphics[width=\columnwidth]{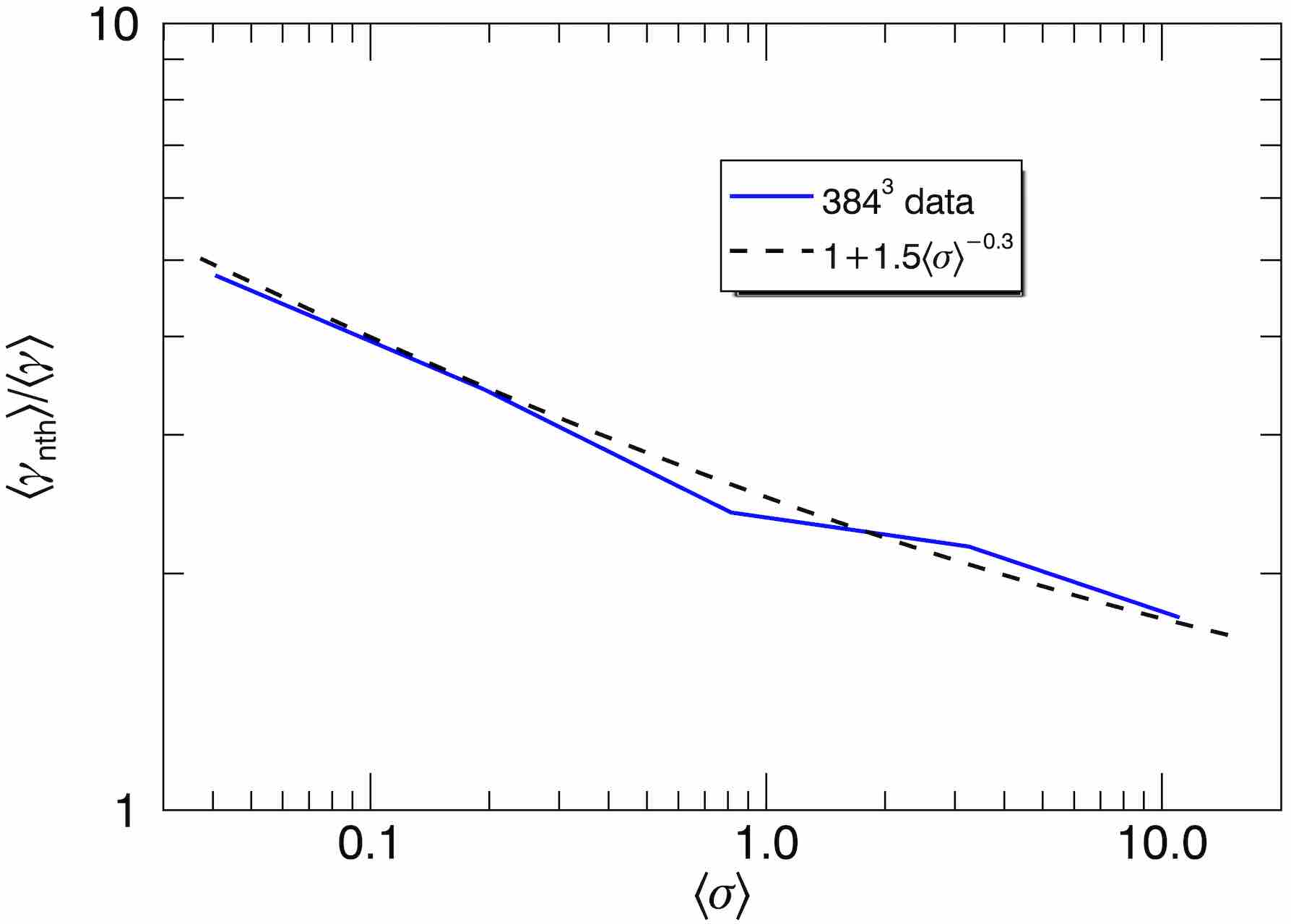}
  \centering
   \caption{\label{fig:dist_nonthermal} Top panel: Time-averaged thermal (dashed) and nonthermal (solid) components of the particle energy distribution $f(\gamma)$, for the $384^3$ simulation series. Center panel: Nonthermal particle energy fraction $E_{\rm nth}/E_{\rm tot}$ (blue) and number fraction $N_{\rm nth}/N_{\rm tot}$ (red) versus $\langle\sigma\rangle$ for same simulations, with a low-$\sigma$ fit of $\langle\sigma\rangle/4$. Bottom panel: Average particle energy of the nonthermal population $\langle\gamma_{\rm nth}\rangle$, relative to the overall mean energy $\langle\gamma\rangle$, versus $\langle\sigma\rangle$ for same simulations, with a fit of $1 + 1.5 \langle\sigma\rangle^{-0.3}$.}
 \end{figure}
 
Although the particle energy distributions are generally close to thermal, there are modest nonthermal deviations. We characterize these deviations by decomposing the distributions into thermal and nonthermal components, following the procedure in \cite{zhdankin_etal_2019}. We define the thermal part $f_{\rm th}(\gamma)$ to be a Maxwell-J\"{u}ttner distribution with temperature and normalization such that the peak coincides with the peak of the measured distribution. We also include all of the measured distribution at energies below the peak value in the thermal component. The nonthermal part $f_{\rm nth}(\gamma)$ is then defined to be the difference between the measured distribution and the thermal fit, $f_{\rm nth}(\gamma) \equiv f(\gamma) - f_{\rm th}(\gamma)$.
 
The top panel of Fig.~\ref{fig:dist_nonthermal} shows the thermal and nonthermal component distributions,~$f_{\rm th}(\gamma)$ and~$f_{\rm nth}(\gamma)$, for varying $\langle\sigma\rangle$ (time-averaged in the $384^3$ simulations). We find that all cases have a well-defined nonthermal population of high-energy particles, albeit without a power-law tail. The center panel of Fig.~\ref{fig:dist_nonthermal} shows the fraction of kinetic energy in the nonthermal population,~$E_{\rm nth}/E_{\rm tot}$, and the fraction of total particles in the nonthermal population,~$N_{\rm nth}/N_{\rm tot}$, as functions of~$\langle\sigma\rangle$. We find that both nonthermal fractions increase with~$\langle\sigma\rangle$, with a scaling that is close to linear for~$\langle\sigma\rangle \lesssim 1$, and weaker than linear for~$\langle\sigma\rangle \gtrsim 1$. The linear scaling at low $\sigma$ suggests that the efficiency of nonthermal particle acceleration is related to the dissipation of the available magnetic energy relative to the thermal kinetic energy. We note that~$\sim 25\%$ of the energy and~$\sim 10\%$ of the particles are in the nonthermal population at~$\langle\sigma\rangle \approx 1$. For the most extreme case,~$\langle\sigma\rangle = 11$, about~$\sim 50\%$ of energy and~$\sim 30\%$ of particles are in the nonthermal population, indicating that although the distribution does not span a broad range of energies, it does have a significantly nonthermal shape. The characteristic energy of the nonthermal population does not increase with $\langle\sigma\rangle$, as demonstrated in the bottom panel of Fig.~\ref{fig:dist_nonthermal}, which shows the average particle energy of the nonthermal population relative to the overall average energy, $\overline{\gamma}_{\rm nth}/\overline{\gamma} = (E_{\rm nth}/N_{\rm nth})/(E_{\rm tot}/N_{\rm tot})$, for varying $\langle\sigma\rangle$. Rather, the average energy of the nonthermal population is only twice the overall average energy in the highest-$\sigma$ case, while it is a factor of $\sim 5$ greater than the overall average in the lowest-$\sigma$ case, indicating that the nonthermal population moves further into the tail at low $\sigma$. We find that $\overline{\gamma}_{\rm nth}/\overline{\gamma}$ is well fit by $1 + 1.5 \langle \sigma \rangle^{-0.3}$.

\subsection{Fokker-Planck model fits} \label{sec:fp-fits}

To explain the nature of the quasi-thermal distributions observed in our simulations, we compare them to theoretical expectations from stochastic particle acceleration. Analytic models of nonthermal particle acceleration in a turbulent environment are generally formulated using the Fokker-Planck advection-diffusion equation in momentum space \citep[e.g.,][]{blandford_eichler_1987, schlickeiser_1989, chandran_2000, demidem_etal_2019}. In Appendix~\ref{app1}, we derive an analytic solution to the Fokker-Planck equation with a radiative cooling term, assuming that the advection coefficient is linear in momentum and the diffusion coefficient is quadratic in momentum:
\begin{align}
A_p &\propto \Gamma_h \gamma_0 + \Gamma_a \gamma \, , \nonumber \\
D_{pp} &\propto \Gamma_0 \gamma_0^2 + \Gamma_2 \gamma^2 \, . \label{eq:fpcoeffs}
\end{align}
These terms thus model first-order and second-order Fermi acceleration processes, respectively \citep{fermi_1954, fermi_1949}. We choose the characteristic energy $\gamma_0 = 300$, comparable to the mean particle energy in the simulations. The remaining four parameters describe the rates for the various terms in the Fokker-Planck equation: energy-independent diffusion $\Gamma_0$, second-order diffusive acceleration $\Gamma_2$, energy-independent advection (heating) $\Gamma_h$, and first-order acceleration $\Gamma_a$. Note that the second-order acceleration term ($\Gamma_2$) alone yields the Maxwell-J\"{u}ttner distribution as a solution, while other terms contribute to the modest nonthermal population (regardless of whether the first-order or second-order mechanism dominates). As a simplification, we choose $\Gamma_0 = \Gamma_2$. After these restrictions, the resulting steady-state solution (Eq.~\ref{eq:fpsolred}) that we fit to has three free parameters ($\Gamma_2$, $\Gamma_a$, and $\Gamma_h$) which may vary with $\langle\sigma\rangle$ and, in principle, $L/2\pi\langle\rho_e\rangle$.
 
 \begin{table}
\centering \caption{Fokker-Planck fit parameters \newline} \label{table:fit}
\begin{tabular}{|c|c|c|c|c|} 
	\hline
	\hspace{0.5 mm} Case \hspace{0.5 mm}  & \hspace{1 mm} $\langle\sigma\rangle$ \hspace{1 mm}   & \hspace{1 mm} $\Gamma_0=\Gamma_2$ \hspace{1 mm}  &   \hspace{1 mm} $\Gamma_h$ \hspace{1 mm} &   \hspace{1 mm} $\Gamma_a$ \hspace{1 mm}  \\
	\hline
rM1d4 & $0.20$ & 3.0 & $-1.0$ & $-20.0$ \\
rL1 & $0.90$ & 1.8 & $-3.0$ & $-8.0$ \\
rM4 & $3.4$ & 1.4 & $-5.0$ & $-3.5$ \\
	\hline
\end{tabular}
\centering
\end{table}

\begin{figure}
    \includegraphics[width=\columnwidth]{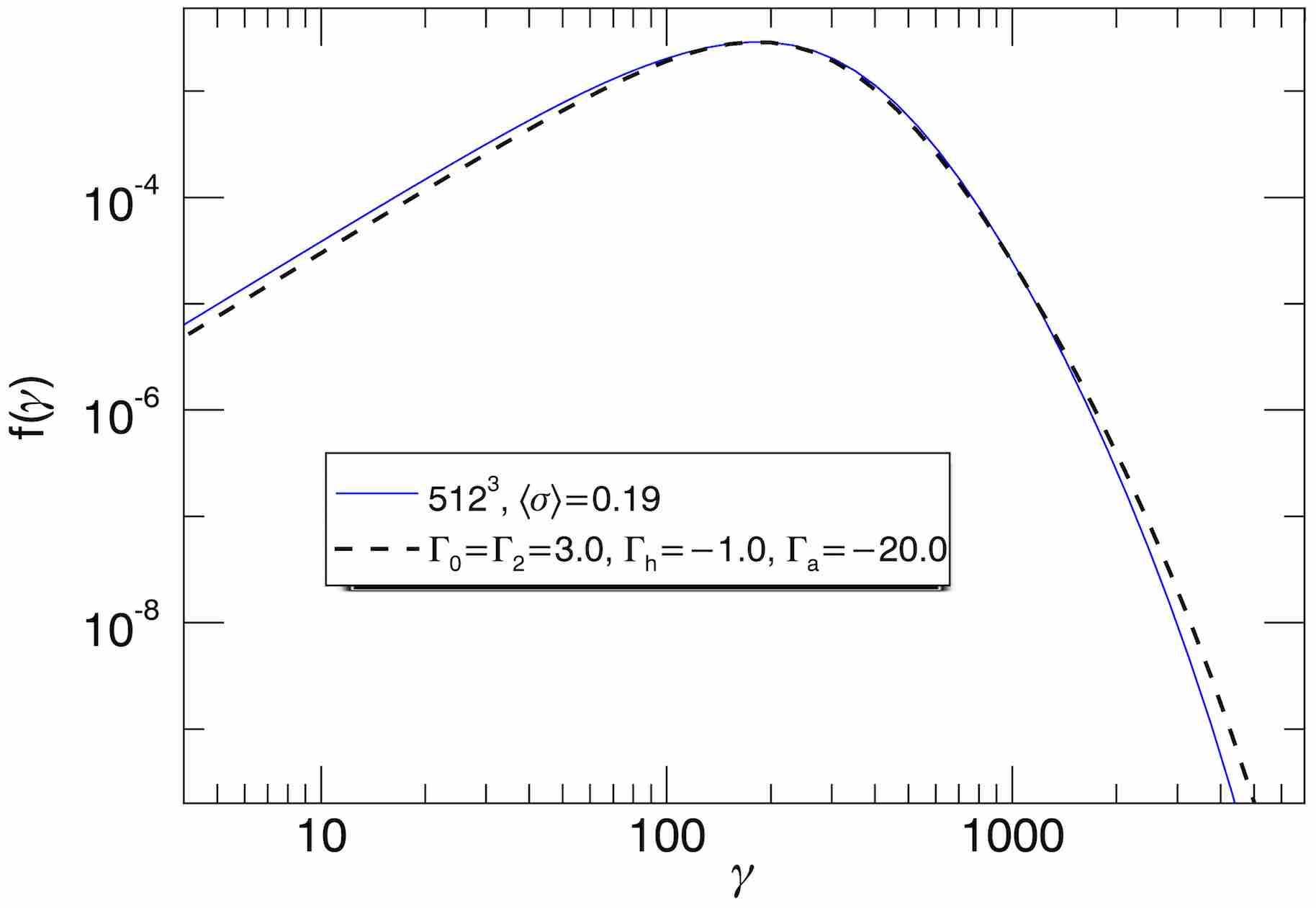}
   \includegraphics[width=\columnwidth]{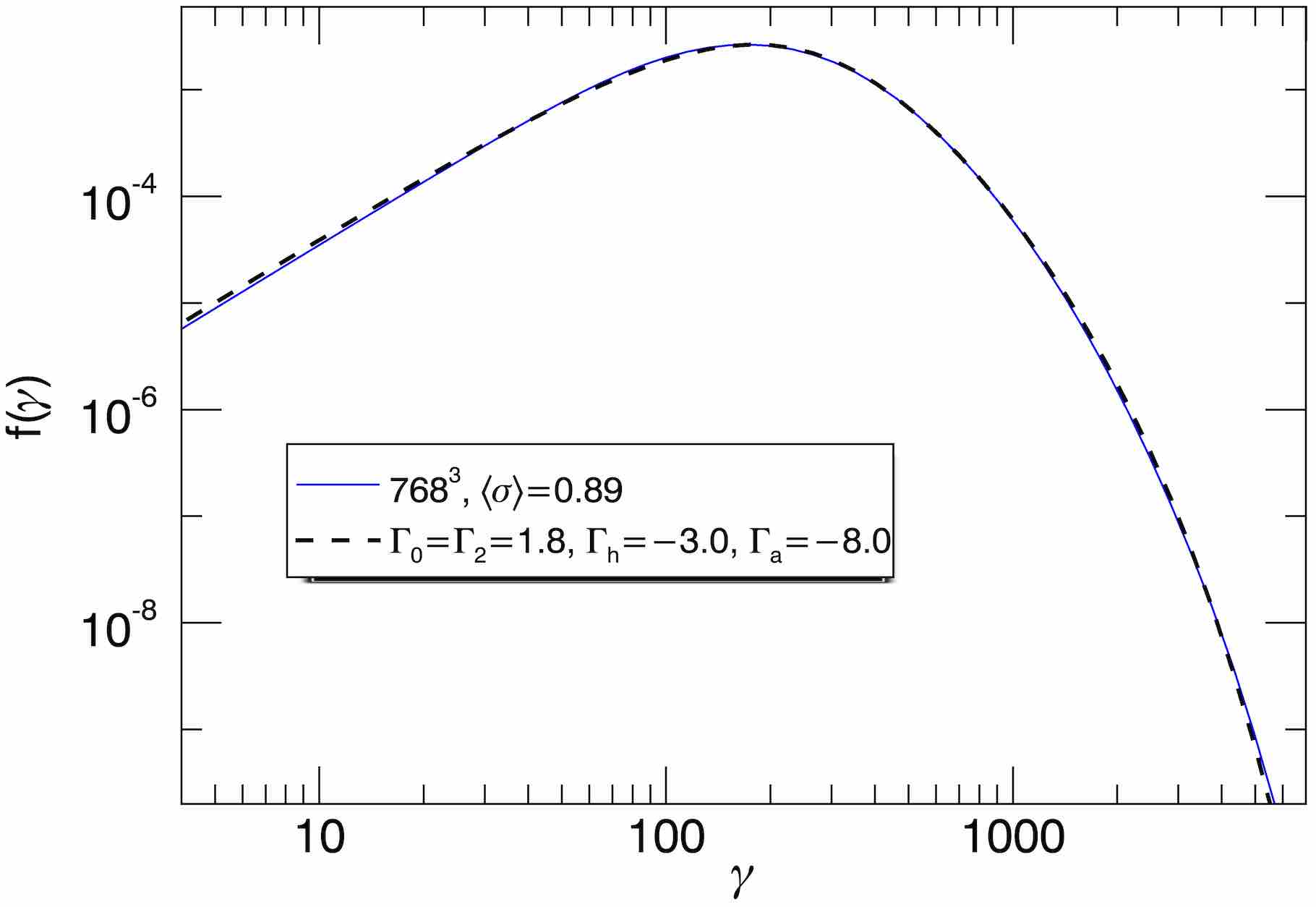}
    \includegraphics[width=\columnwidth]{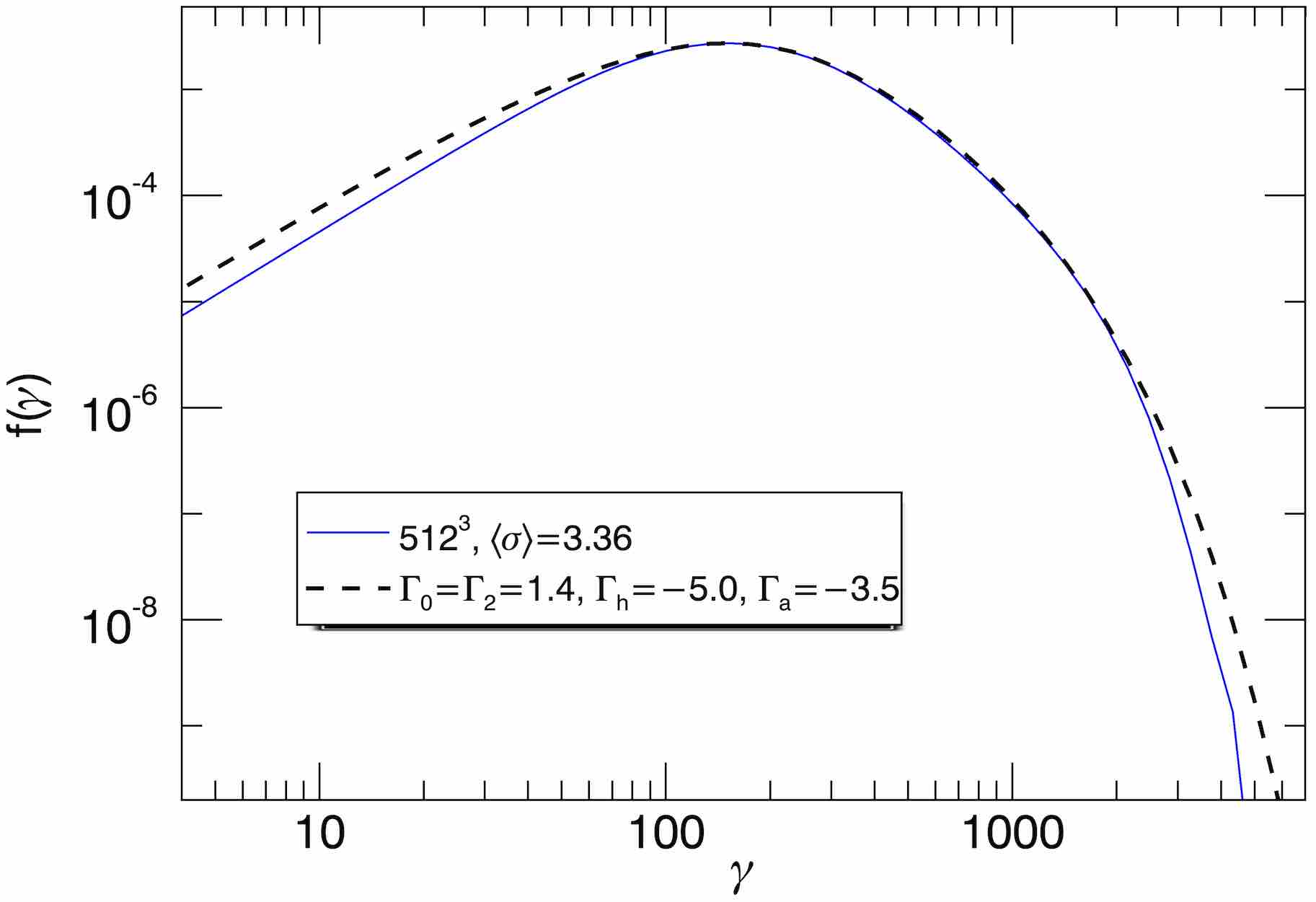}
  \centering
   \caption{\label{fig:dist_energy_fits} Top panel: Steady-state particle energy distribution $f(\gamma)$ for the $512^3$, $\langle\sigma\rangle = 0.2$ case (blue) with fit from the Fokker-Planck model (black, dashed). Center panel: similar for the $768^3$, $\langle\sigma\rangle = 0.9$ case. Bottom panel: similar for the $512^3$, $\sigma = 3.4$ case.}
 \end{figure}

In Fig.~\ref{fig:dist_energy_fits}, we fit the particle energy distributions from the largest simulations (cases rM1d4, rL1, and rM4) with the solution from the Fokker-Planck model. We are able to obtain very good fits to all three simulations, with the fitting parameters depending on $\langle\sigma\rangle$ as shown in Table~\ref{table:fit}. In these fits, we limited ourselves to $\Gamma_2 \propto c/v_A$, motivated by second-order Fermi acceleration or gyroresonant interactions with Alfv\'{e}n waves (Eq.~\ref{eq:gamma2}) \citep{miller_etal_1990}. Curiously, the advection terms are negative, implying that there is a first-order cooling (or deceleration) process, consistent with measurements from tracked particles in previous non-radiative turbulence work \citep{wong_etal_2019}. The first-order acceleration term ($\Gamma_a$) becomes less negative as $\langle\sigma\rangle$ increases, indicating that there may be a competing first-order acceleration process that becomes important in the high-$\sigma$ regime. Given the admittedly considerable amount of freedom in fitting to the Fokker-Planck solution and interpreting the parameters, we limit this work to a proof of concept. Thus, we defer a more complete analysis of the Fokker-Planck equation (using tracked particles to directly measure the diffusion and advection coefficients) to future work.

\subsection{Temporal variability of energy distribution} \label{sec:variability}
 
   \begin{figure}
 \includegraphics[width=\columnwidth]{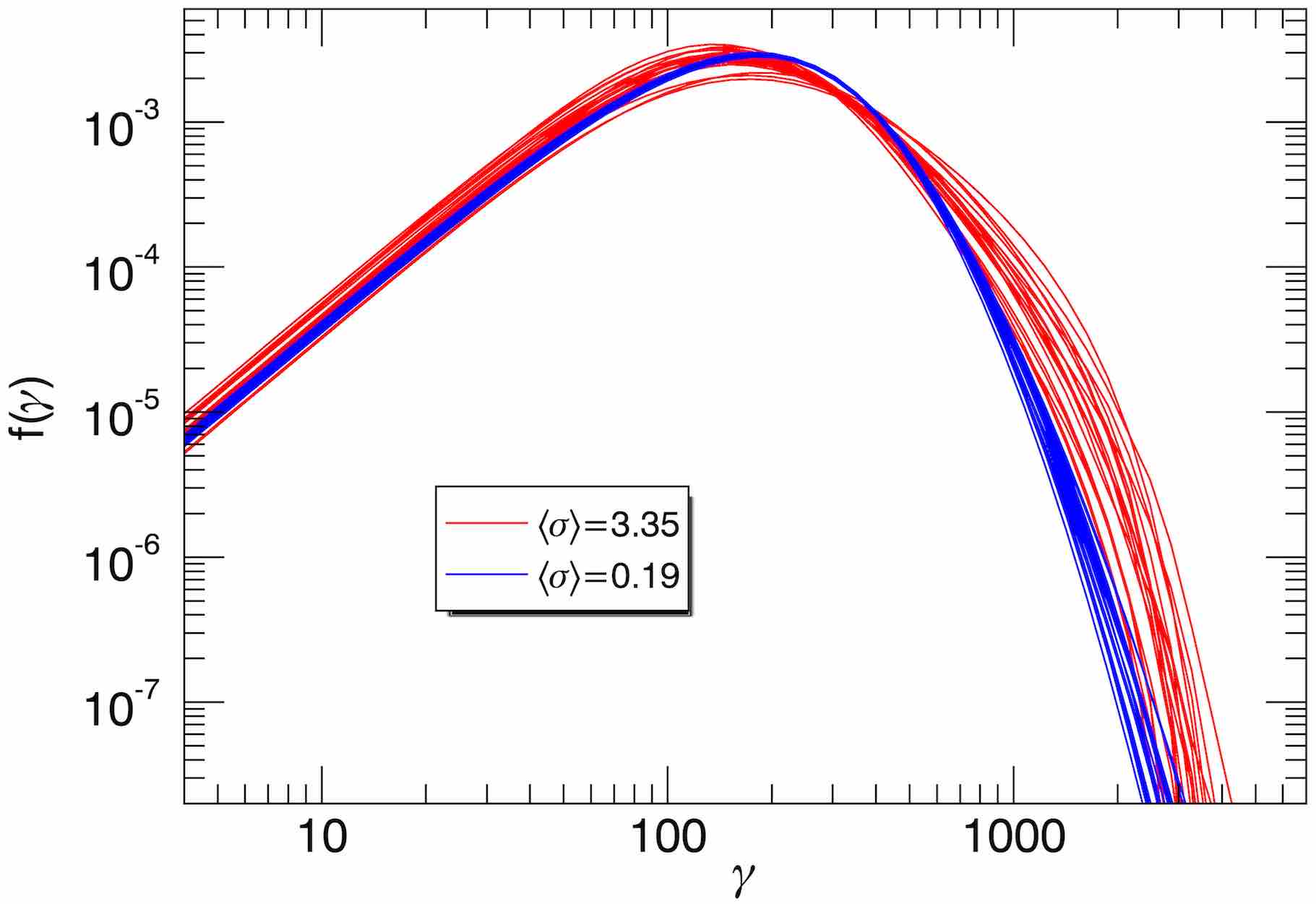}
  \centering
   \caption{\label{fig:dist_time} Particle energy distribution $f(\gamma)$ at 20 different times (output at constant cadence) for the $\langle\sigma\rangle = 3.4$, $512^3$ case (red; times $7.3 < t v_A/L < 35.3$) and for the $\langle\sigma\rangle = 0.2$, $512^3$ case (blue; times $6.8 < t v_A/L < 32.7$).}
 \end{figure}
 
Having completed our characterization of the steady-state particle energy distribution, we next turn to the temporal variability of the distribution. To illustrate the overall variability of the energy distributions, in Fig.~\ref{fig:dist_time} we overlay $f(\gamma)$ at 20 different times (during statistical steady state) for two simulations: a high-magnetization case ($\langle\sigma\rangle = 3.4$, $512^3$) and a low-magnetization case ($\langle\sigma\rangle = 0.2$, $512^3$). We find that the high-magnetization case exhibits a significant amount of variability in the high-energy tail, with the tail often shifting by a factor of $\sim 2$ in energy. The low-magnetization case, on the other hand, is very stable in time.
 
  \begin{figure}
 \includegraphics[width=\columnwidth]{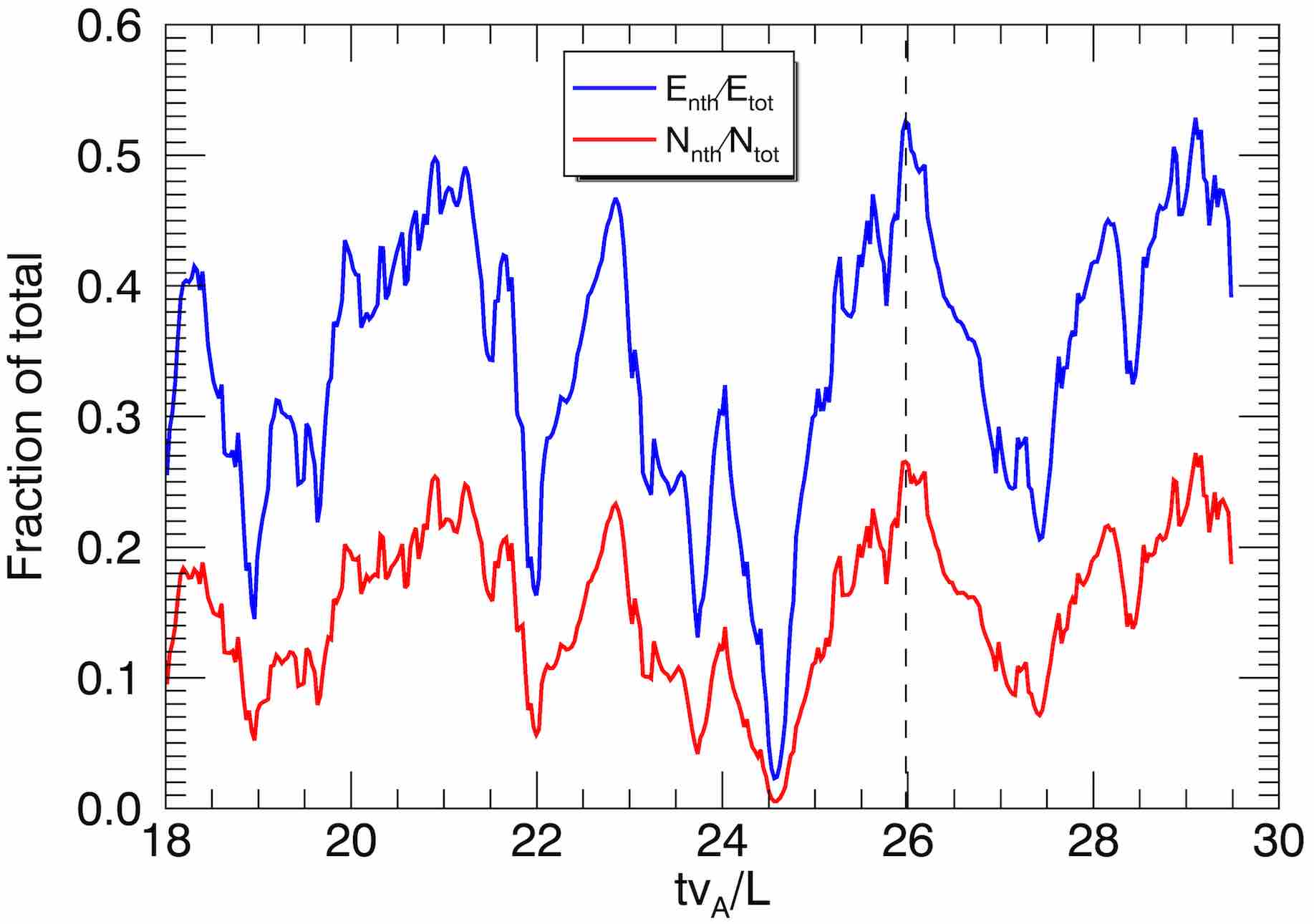}
\includegraphics[width=\columnwidth]{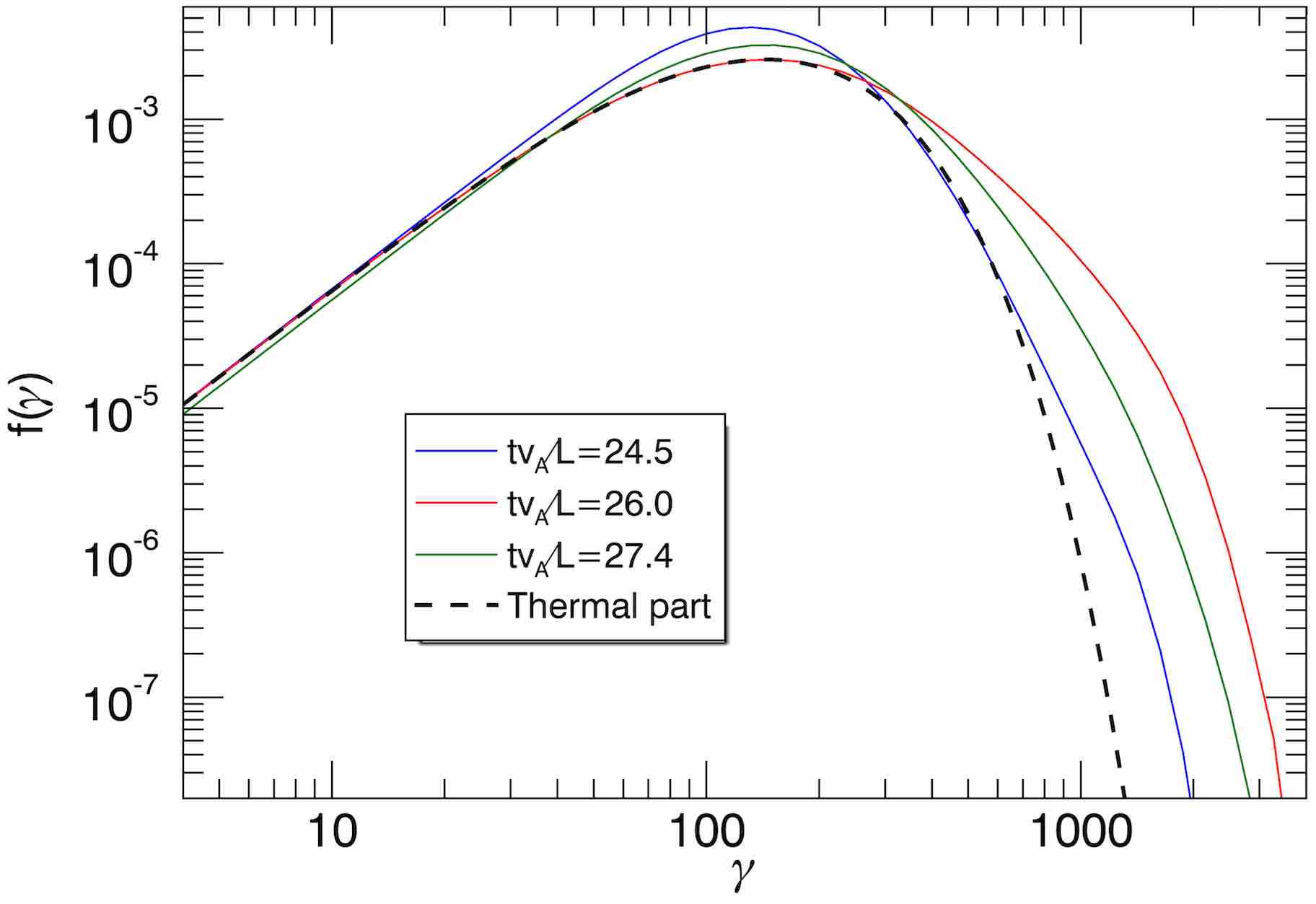}
 \includegraphics[width=\columnwidth]{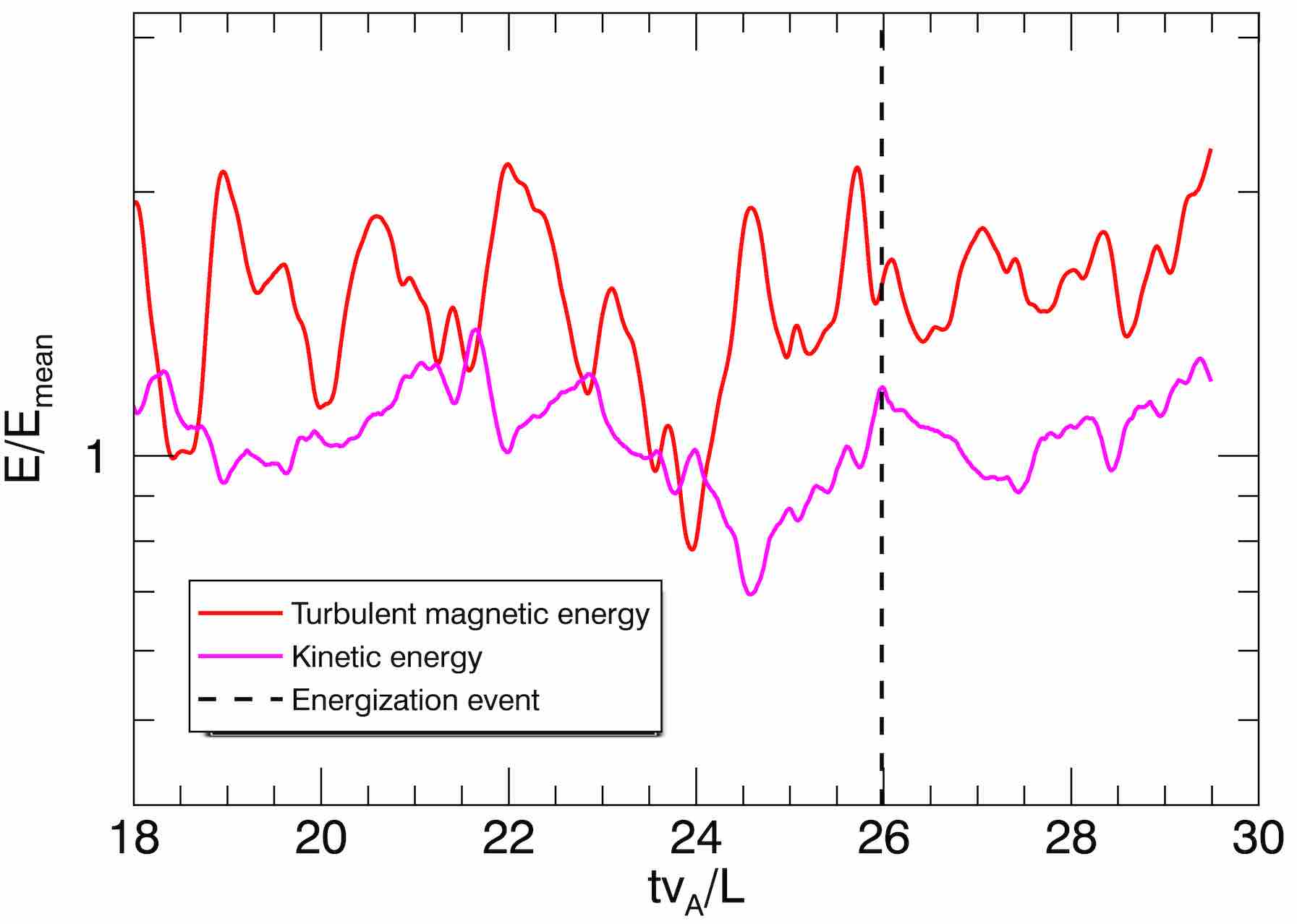}
  \centering
   \caption{\label{fig:variability_nonthermal} Top panel: Evolution of nonthermal energy fraction (blue) and number fraction (red) for the $\langle\sigma\rangle = 3.4$, $512^3$ case. The fiducial nonthermal energization event at $t v_A/L = 26.0$ is marked with the vertical dashed line. Center panel: Particle energy distributions $f(\gamma)$ at times $t v_A/L \in \{ 24.5, 26.0, 27.4 \}$, showing hardening of the distribution during the fiducial event. The thermal part at $t v_A/L = 26.0$ is also shown (black, dashed). Bottom panel: Evolution of turbulent magnetic energy (red) and particle kinetic energy (magenta) relative to the energy of the mean magnetic field $E_{\rm mean}$.}
 \end{figure}

We now consider the temporal variability of the $\langle\sigma\rangle = 3.4$, $512^3$ case in more detail. In the top panel of Fig.~\ref{fig:variability_nonthermal}, we show the time-evolution of the nonthermal energy and particle number fractions, $E_{\rm nth}/E_{\rm tot}$ and $N_{\rm nth}/N_{\rm tot}$. We find that $E_{\rm nth}/E_{\rm tot}$ and $N_{\rm nth}/N_{\rm tot}$ both typically vary by a factor of $\sim 2$ on timescales of $\sim L/v_A$ (with values ranging from $\sim 0.1$ to $\sim 0.5$ for the former and $\sim 0.05$ to $\sim 0.25$ for the latter). There are several nonthermal energization events indicated by spikes in the energy and number of nonthermal particles. We focus on one particular energization event, at time $tv_A/L \sim 26.0$, which we refer to as the {\it fiducial event} in the remainder of the paper. We show the particle distribution corresponding to the fiducial event, along with an earlier time $t v_A/L = 24.5$ and later time $t v_A/L = 27.4$, in the center panel of Fig.~\ref{fig:variability_nonthermal}. Whereas $f(\gamma)$ is close to thermal at an earlier time ($t v_A/L = 24.5$), it becomes significantly harder and extends to higher energies during the event. This indicates that the system exhibits significant kinetic variability on global scales. In the bottom panel of Fig.~\ref{fig:variability_nonthermal}, we show the evolution of turbulent magnetic energy and particle kinetic energy (bulk and internal combined) for the simulation, with the fiducial event marked by a dashed line. We find that the magnetic energy has a noticeable drop (by $\sim 50\%$) just before the event; the particle energy grows to a local maximum just after the event, although its increase is small compared to the decrease in magnetic energy (indicating that a significant fraction of the energy was lost to radiation). These results together suggest that magnetic reconnection of the large-scale fields may play a role in the energization, which we revisit in Sec.~\ref{sec:nhe}. We will investigate the fiducial event in more detail in the following subsections.

\subsection{Momentum anisotropy distributions} \label{sec:ani}

The next topic that we tackle is the anisotropy of the global particle momentum distribution $F(\boldsymbol{p})$. Since radiation is emitted in the direction of particle motion, the momentum anisotropy has implications for observable radiative signatures. In particular, phenomena such as the kinetic beaming of high-energy particles may cause rapid time variability in the overall momentum anisotropy, as discussed by \cite{cerutti_etal_2012, cerutti_etal_2013, mehlhaff_etal_inprep} in the context of relativistic magnetic reconnection.

To characterize the anisotropy of the momentum distribution, we condition $F(\boldsymbol{p})$ on particles in given energy ranges. We define the particle momentum anisotropy distribution for energies between a lower threshold $\gamma_{{\rm thr},1}$ and upper threshold $\gamma_{{\rm thr},2}$ by
\begin{align}
f(\theta, \phi|\gamma_{{\rm thr},1}<\gamma<\gamma_{{\rm thr},2}) \equiv \int_{\gamma_{{\rm thr},1}}^{\gamma_{{\rm thr},2}} d\gamma f(\theta, \phi, \gamma)  \, ,
\end{align}
where $f(\theta, \phi, \gamma) = m_e c p^2 \sin{\theta} F(\boldsymbol{p})$, with $F(\boldsymbol{p})$ being the (spatially averaged) three-dimensional particle momentum distribution (for positrons and electrons combined) and $\boldsymbol{p}/m_e c = (\gamma \cos{\phi} \sin{\theta}, \gamma \sin{\phi} \sin{\theta}, \gamma \cos{\theta})$. We assumed ultra-relativistic particles. Thus,~$\theta$ and~$\phi$ are the polar and azimuthal angles (respectively) of the momentum vector with respect to the global mean field $\boldsymbol{B}_0$. In the following, we use the Mollweide projection to display the particle anisotropy distributions in an area-preserving way. We normalize $f(\theta, \phi| \gamma)$ to the time- and direction-averaged value for the given range of~$\gamma$.

\begin{figure}
 \includegraphics[width=\columnwidth]{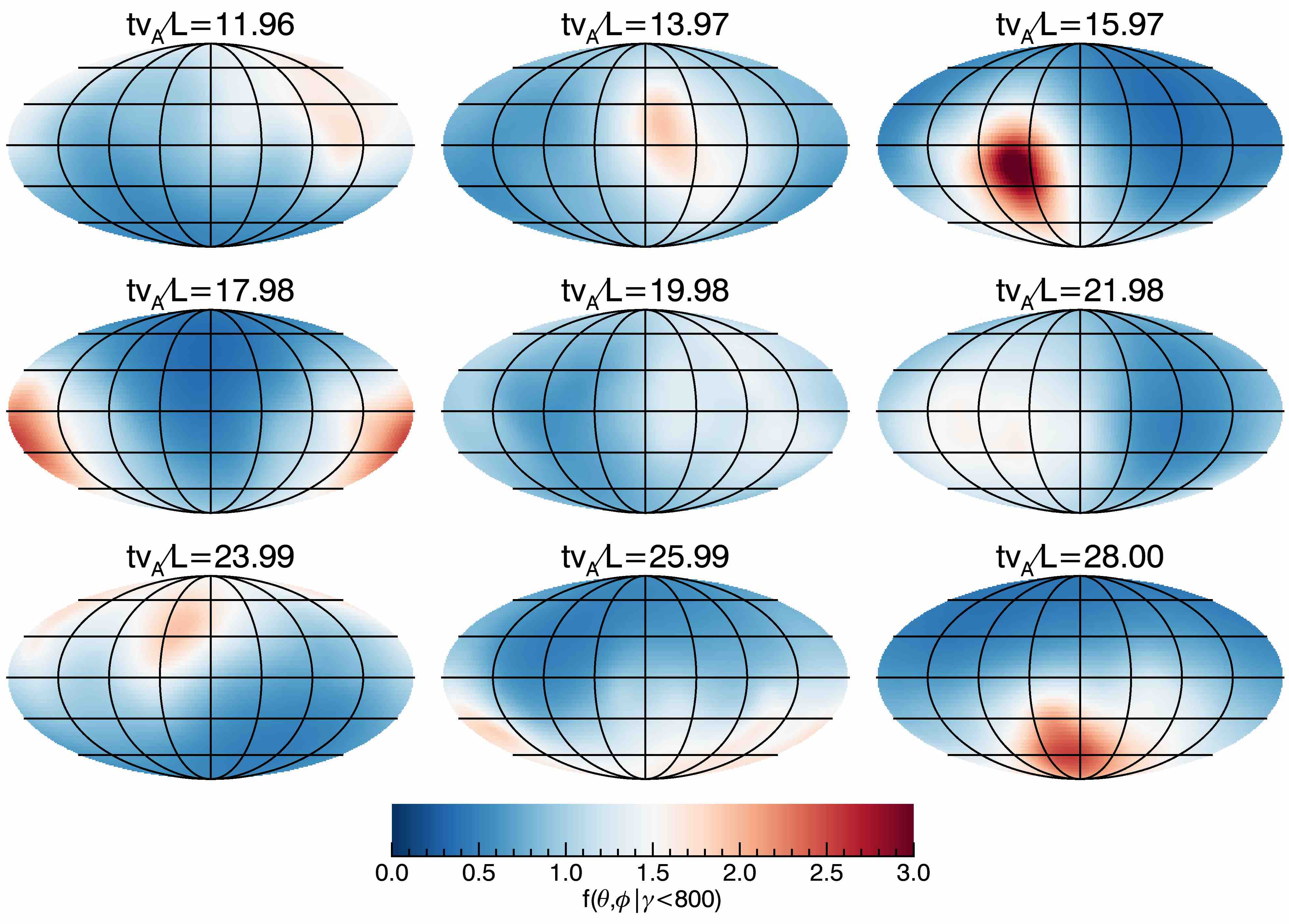}
  \centering
   \caption{\label{fig:dist_angular} Momentum anisotropy distributions for low-energy particles, $f(\theta,\phi|\gamma<800)$, at nine different times for the $512^3$, $\langle\sigma\rangle = 3.4$ case.}
 \end{figure}
 
We first consider the momentum anisotropy distributions for the low-energy particles, defined as those with~$\gamma < 800$. In Fig.~\ref{fig:dist_angular}, we show the low-energy anisotropy distributions, $f(\theta,\phi|\gamma<800)$, in nine snapshots of the $512^3$, $\langle\sigma\rangle = 3.4$ case (spanning from $tv_A/L = 12.0$ to $tv_A/L = 28.0$, shown every $\approx 2 L/v_A$). We find that the anisotropy distribution typically exhibits wide-angle particle beams, which are signatures of large-scale bulk flows. An example of a particularly conspicuous bulk flow is at time $tv_A/L = 16.0$, where a beam with amplitude $\sim 3$ times the mean isotropic value dominates the anisotropy map.

\begin{figure}
\includegraphics[width=\columnwidth]{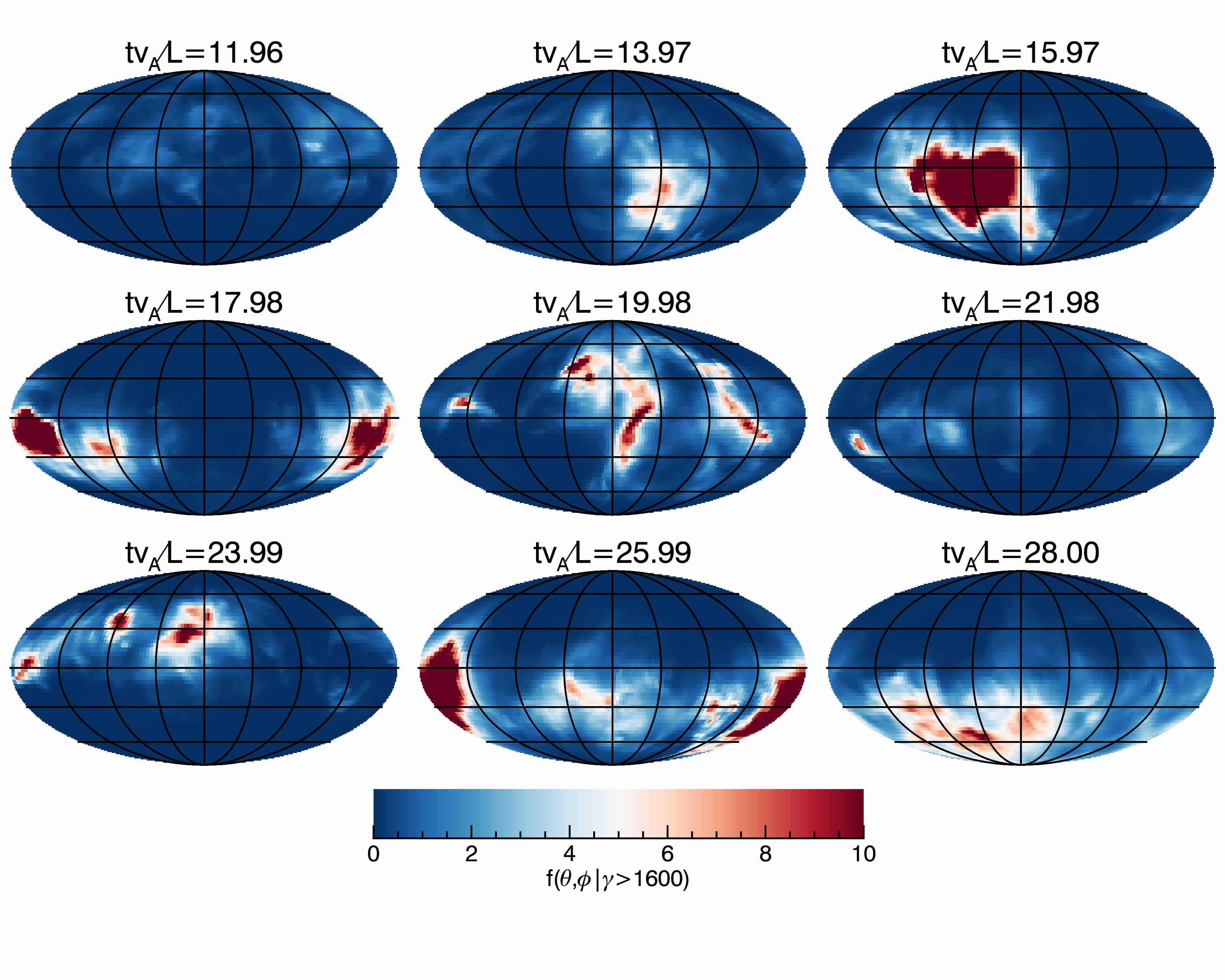}
  \centering
   \caption{\label{fig:dist_angular_he} Momentum anisotropy distributions for high-energy particles, $f(\theta,\phi|\gamma>1600)$, at nine different times for the $512^3$, $\langle\sigma\rangle = 3.4$ case.}
 \end{figure}
 
Next, in Fig.~\ref{fig:dist_angular_he}, we show similar momentum anisotropy distributions for high-energy particles ($\gamma > 1600$) in the same nine snapshots of the $512^3$, $\langle\sigma\rangle = 3.4$ case. In contrast to the low-energy particles, we find that the energetic particle populations are characterized by finer angular structure, organized in an ensemble of narrow beams. These beams have random orientations, but are most commonly directed perpendicular to $\boldsymbol{B}_0$. We find that the intense high-energy particle beams are often correlated with the low-energy bulk flows (e.g., at time $tv_A/L = 16.0$), indicating that the nonthermal population is boosted by the flow, not unlike minijet models \citep{lyutikov_2006, giannios_etal_2009, narayan_kumar_2009, kumar_narayan_2009}. However, we also find cases of multiple narrow high-energy beams in the absence of significant low-energy bulk flows (e.g., at times $tv_A/L = 20.0$ and $tv_A/L = 24.0$), suggesting inherently kinetic beaming phenomena. We also find some times where beams are almost entirely absent (e.g., at times $tv_A/L = 12.0$ and $tv_A/L = 22.0$). An animation of the momentum anisotropy distribution is available as supplementary material.

\begin{figure}
\includegraphics[width=\columnwidth]{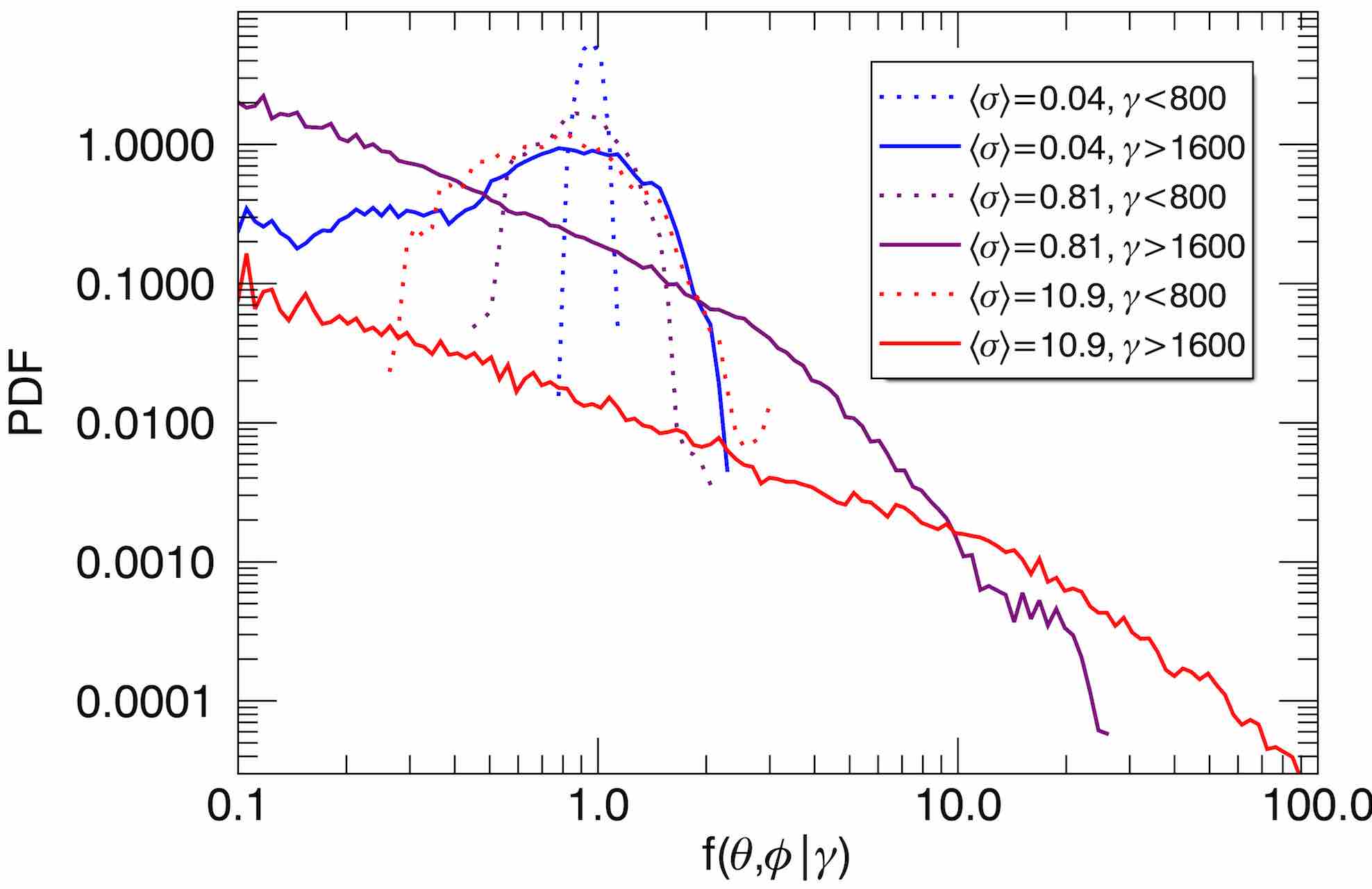}
  \centering
  \caption{\label{fig:dist_angular_sigma} The probability distribution function for the momentum anisotropy distribution $f(\theta, \phi | \gamma)$ bin values, for low-energy particles ($\gamma < 800$, dashed) and high-energy particles ($\gamma > 1600$, solid), in $384^3$ simulations with $\sigma \in \{ 0.04, 0.8, 11 \}$.}
 \end{figure}
 
In contrast to the high magnetization cases, we find that the low magnetization cases ($\langle\sigma\rangle \lesssim 1$) do not exhibit significant kinetic beaming. To demonstrate this, we show the probability distribution function (PDF) for the momentum anisotropy distribution bin values. This PDF highlights the probability for $f(\theta,\phi|\gamma)$ to reach extreme values. The measured PDF is shown in Fig.~\ref{fig:dist_angular_sigma} for $384^3$ simulations with $\sigma \in \{ 0.04, 0.8, 11 \}$, both for low-energy particles ($\gamma < 800$) and for high-energy particles ($\gamma>1600$). The PDF for high-energy particles has broader tails when $\sigma$ is higher, indicating more extreme beaming events. The low-energy particles exhibit a much narrower PDF, which however also broadens at high $\sigma$, indicating more intense bulk flows.

\begin{figure}
\includegraphics[width=\columnwidth]{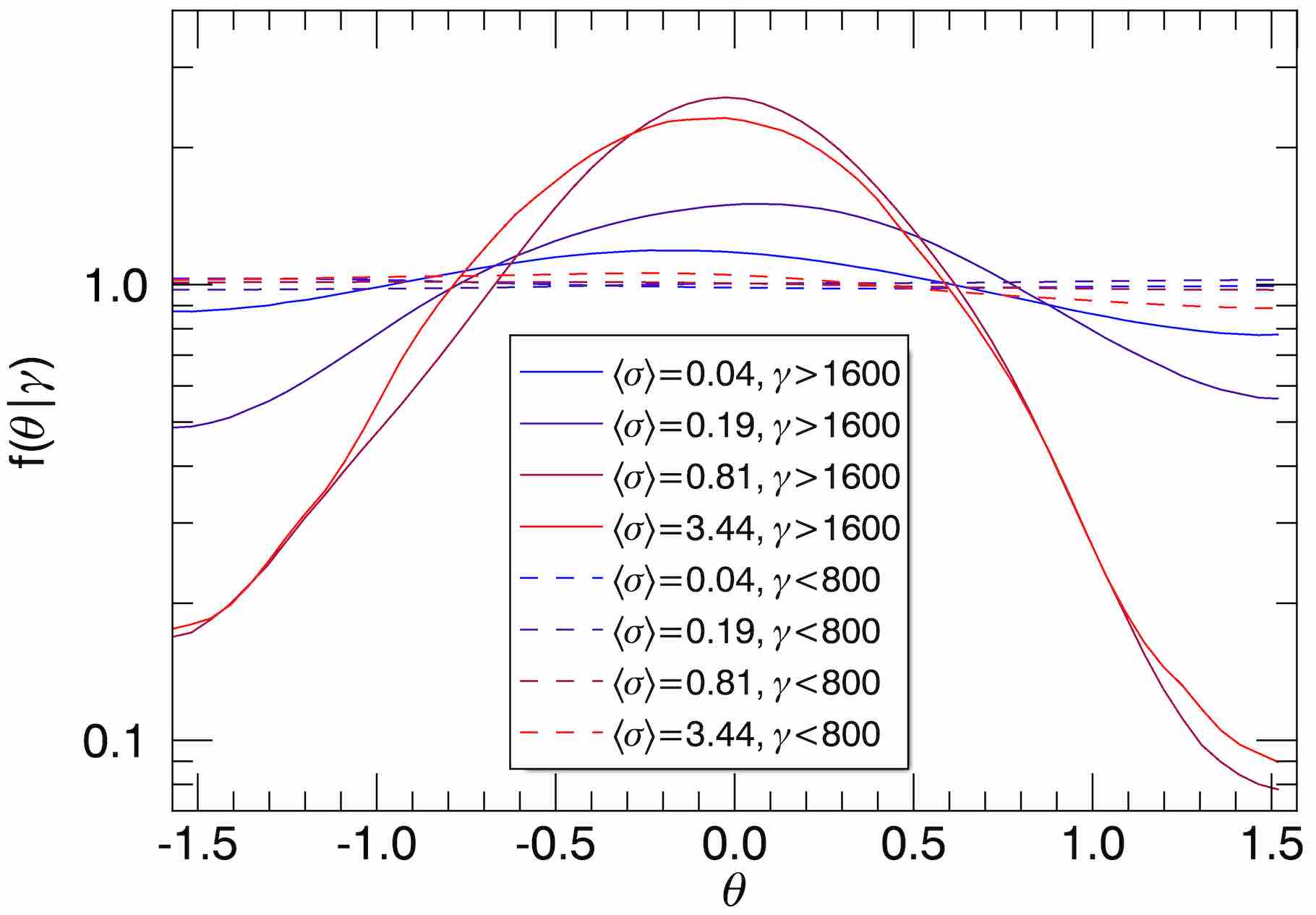}
  \centering
   \caption{\label{fig:dist_angular_theta} Momentum anisotropy distributions averaged over the azimuthal angle $\phi$ for low-energy particles, $f(\theta | \gamma < 800)$ (dashed), and for high-energy particles, $f(\theta|\gamma>1600)$ (solid). Simulations with varying $\langle\sigma\rangle$ from the $384^3$ series are shown in different colors.}
 \end{figure}
 
To characterize the statistical anisotropy of the momentum distribution with respect to $\boldsymbol{B}_0$, we integrate the distribution over azimuthal angles~$\phi$, defining the reduced distribution $f(\theta | \gamma) = \int d\phi f(\theta, \phi|\gamma)$. We show the resulting time-averaged distribution for low-energy particles and for high-energy particles, for simulations with varying $\langle\sigma\rangle$, in Fig.~\ref{fig:dist_angular_theta}. We find that the low-energy population is essentially isotropic with respect to $\boldsymbol{B}_0$, indicating that the particles contributing to the bulk flows are not significantly influenced by the global mean magnetic field. The high-energy population, on the other hand, has momentum vectors that are preferentially oriented perpendicular to $\boldsymbol{B}_0$, with the anisotropy being strongest for cases with $\langle\sigma\rangle \gtrsim 1$. This result supports dissipation mechanisms that are associated with the perpendicular (ideal) electric field, which we further explore in Sec.~\ref{sec:track}.

\subsection{Kinetic beam evolution}

\begin{figure*}
\includegraphics[width=2\columnwidth]{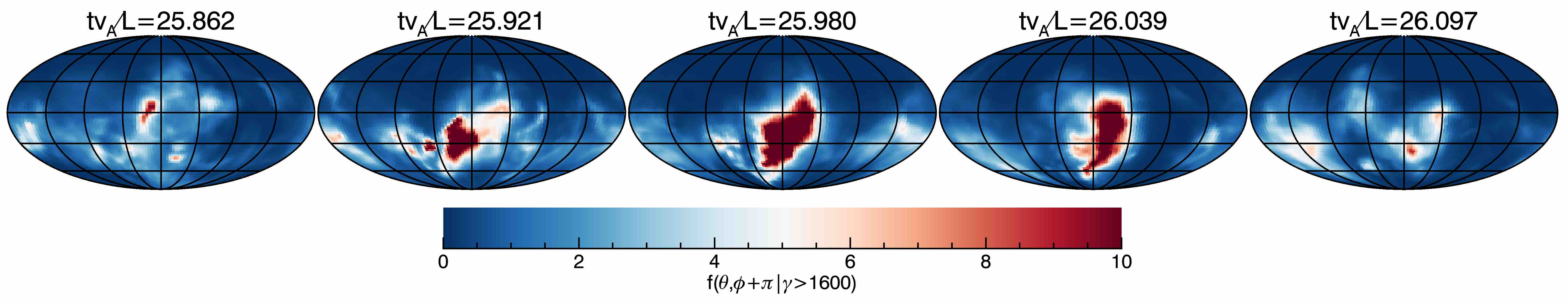}
  \centering
   \caption{\label{fig:dist_angular_event} Momentum anisotropy distributions for high-energy particles, $f(\theta,\phi+\pi|\gamma>1600)$, over a duration of $0.24 L/v_A$ covering the fiducial beaming event at $\theta = 122^\circ$ and $\phi = 174^\circ$, from the $512^3$, $\langle\sigma\rangle = 3.4$ case. Note that the coordinate system has been shifted by $180^\circ$ in $\phi$ for clarity.}
 \end{figure*}
 
To understand kinetic beaming in more detail, we next focus on a particularly strong beaming event as a case study, which coincides with the fiducial nonthermal energization event described in Sec.~\ref{sec:variability}. This beam reaches a peak (of $\sim 40$ times the average bin value) at $t v_A/L \approx 26.0$ in the $\langle\sigma\rangle = 3.4$, $512^3$ simulation (thus appearing in the bottom center panel of Fig.~\ref{fig:dist_angular_he}). The evolution of the particle anisotropy distribution during five snapshots (over a duration of $\sim 0.24 L/v_A$) that cover the fiducial event is shown in Fig.~\ref{fig:dist_angular_event}. The peak of the beam is located at $\theta = 122^{\circ}$ and $\phi = 174^{\circ}$; note that the $\phi$ coordinates in the figure have been shifted by $180^\circ$ to show the beam more clearly. The beam is visible in three of the five times shown, indicating a duration of roughly $\sim 0.12 L/v_A$, thus comparable to the timescale of the energy-containing scale ($L/2\pi v_A \approx 0.16 L/v_A$). The beam covers an angle of roughly $90^\circ$ in both $\theta$ and $\phi$. In comparison, smaller beams generally have a faster evolution. A systematic study of the statistical properties of beams (intensities, sizes, durations, and motion) is left to future work.

\begin{figure}
 \includegraphics[width=\columnwidth]{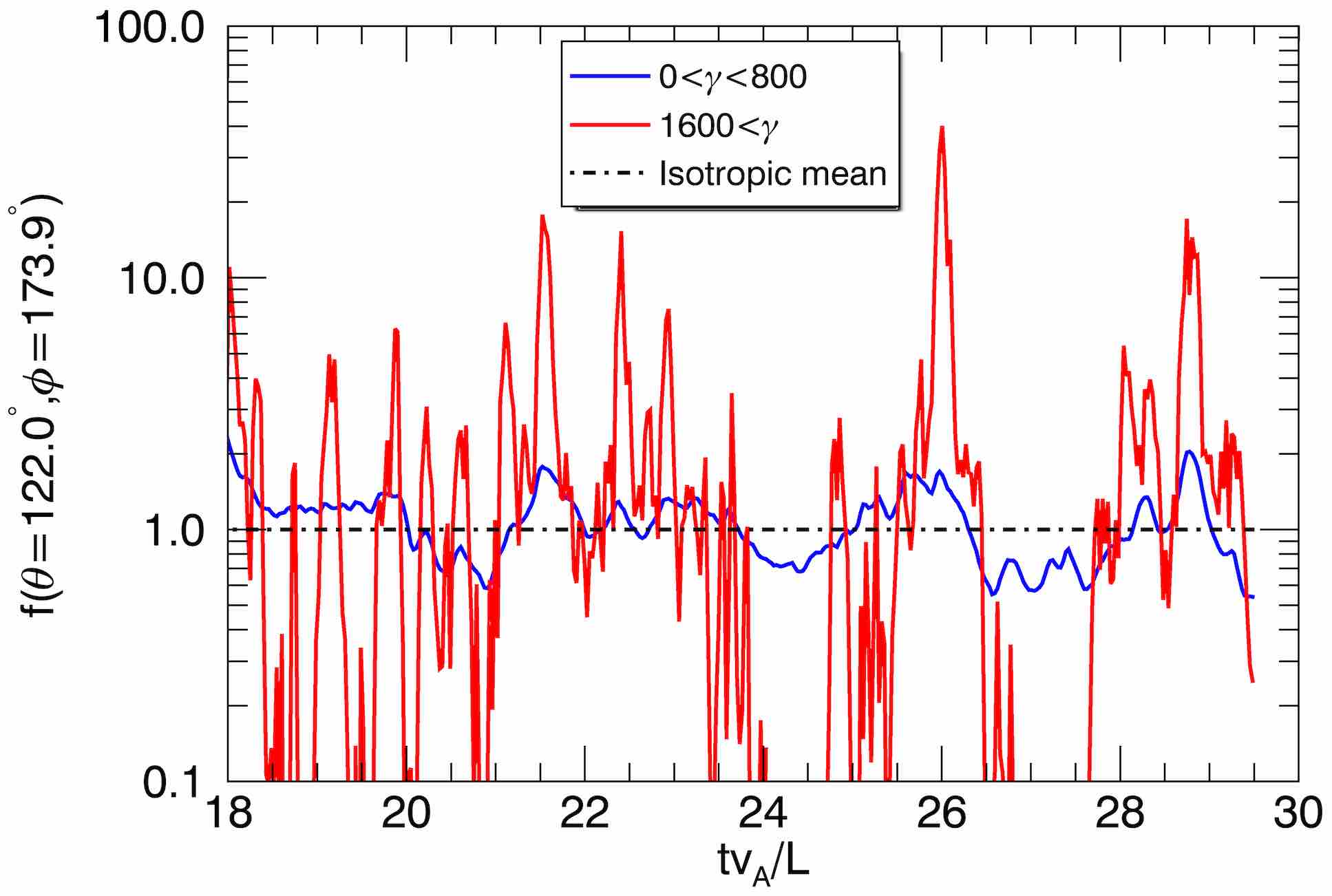}
 \includegraphics[width=\columnwidth]{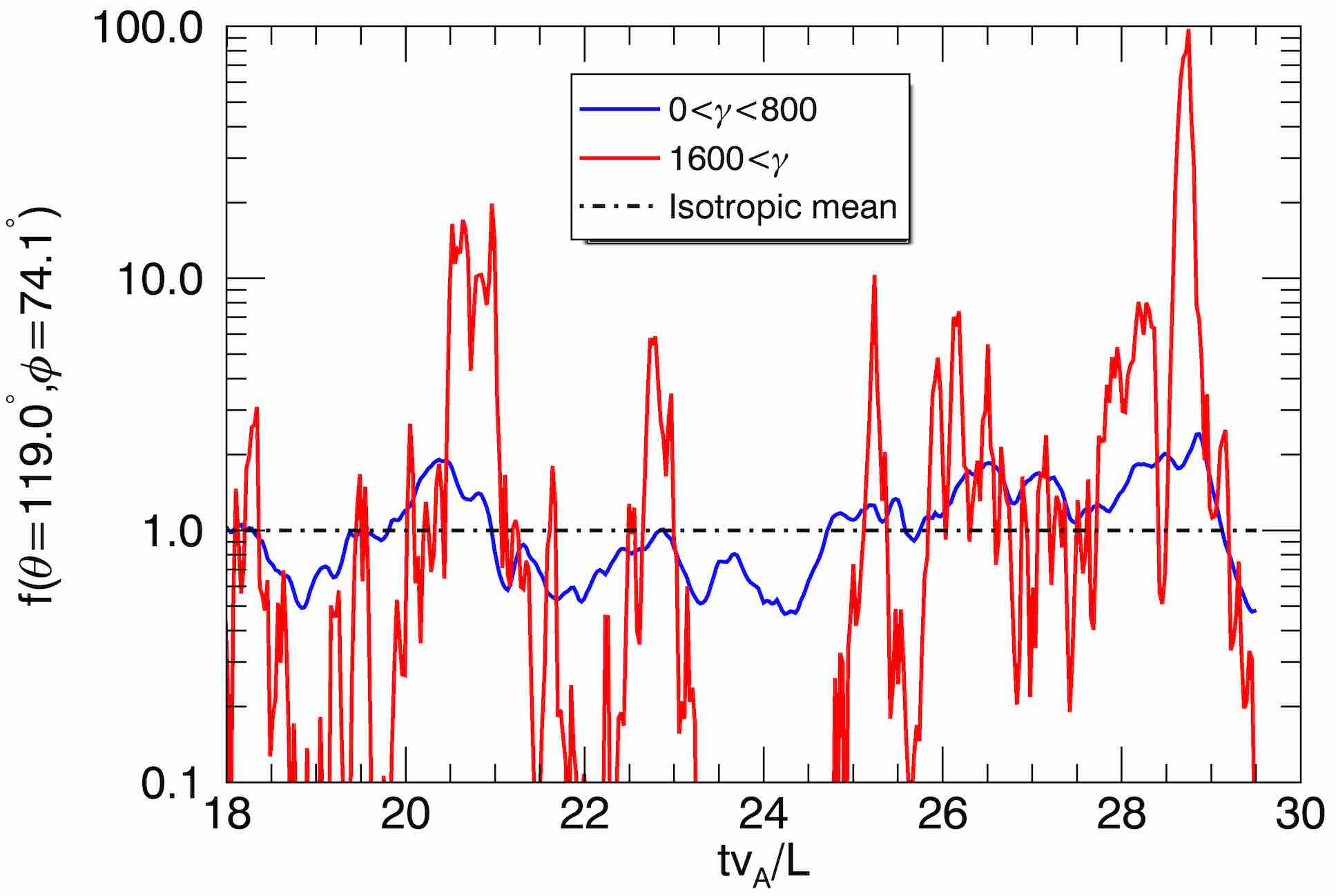}
  \centering
   \caption{\label{fig:dist_angular_loc} The evolution of the momentum anisotropy distribution $f(\theta,\phi|\gamma)$ versus time in two different energy bands ($\gamma < 800$, blue; $\gamma > 1600$, red), for the $512^3$, $\langle\sigma\rangle = 3.4$ case, taken in two different directions. Top panel: $\theta = 122^\circ$ and $\phi = 174^\circ$, coinciding with the beam from the fiducial event at $t v_A/L = 26.0$ (shown in Fig.~\ref{fig:dist_angular_event}). Bottom panel: $\theta = 119^\circ$ and $\phi = 74^\circ$, coinciding with the strongest beam in the simulation at $t v_A/L = 28.7$.}
 \end{figure}

In the top panel of Fig.~\ref{fig:dist_angular_loc}, we show the evolution of the momentum anisotropy distribution in the direction of the beam from the fiducial event, $f(\theta = 122^\circ, \phi = 174^\circ | \gamma)$. The two curves correspond to different energy bands: low energy ($\gamma < 800$) and high energy ($\gamma > 1600$). The low-energy population typically varies by a factor of $\sim 2$ on timescales comparable to $\sim L/v_A$, while the high-energy population exhibits a much more intermittent evolution, characterized by rapid beaming events. Nevertheless, enhancements in both high-energy particle and low-energy particle populations are strongly correlated, indicating that beaming is enhanced by bulk motion. We find that during the fiducial event, the population in the high-energy band increases by a factor of $\sim 40$ over the time- and direction-averaged value. For comparison, in the bottom panel of Fig.~\ref{fig:dist_angular_loc}, we show the evolution of the anisotropy distribution in the direction of the strongest beam in the simulation (occurring at $t v_A/L = 28.7$), $f(\theta = 119^\circ, \phi = 74^\circ | \gamma)$, which attains an even stronger peak, $\sim 100$ greater than the average. In terms of radiative signatures, these results imply that kinetic beams can appear as flares that are orders of magnitude more intense than the average emission.

\subsection{System size dependence} \label{sec:sys}

\begin{figure}
\includegraphics[width=\columnwidth]{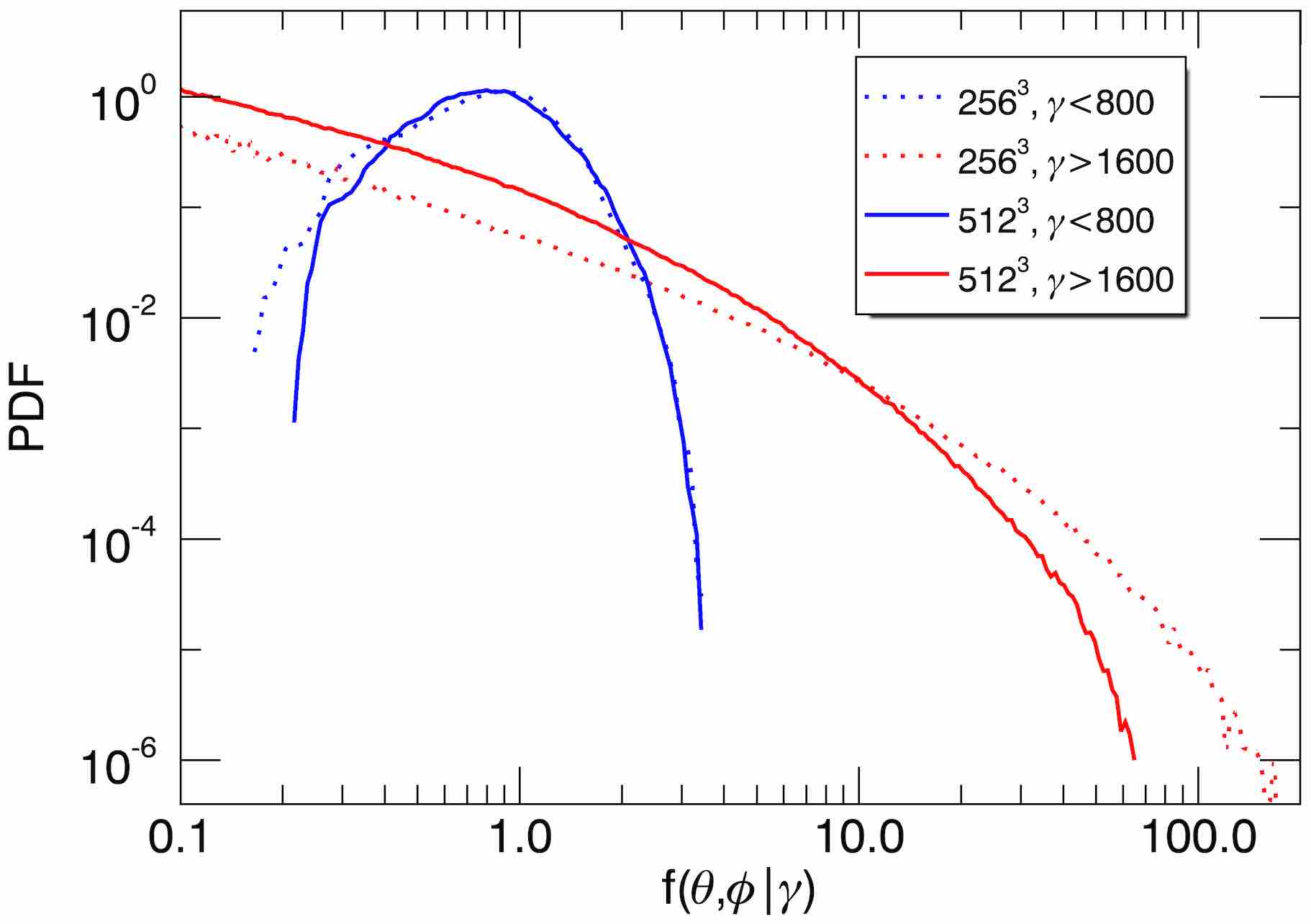}
  \centering
   \caption{\label{fig:angular_system_size} The probability distribution function for the momentum anisotropy distribution $f(\theta, \phi | \gamma)$ bin values, for low-energy particles ($\gamma < 800$, blue) and high-energy particles ($\gamma > 1600$, red), in $\langle\sigma\rangle = 3.4$ simulations with two different system sizes,  $L/2\pi\langle\rho_e\rangle = 18.0$ ($256^3$; dashed) and $L/2\pi\langle\rho_e\rangle = 39.1$ ($512^3$; solid).}
 \end{figure}
 
An important question is whether the properties of intermittent high-energy particle beams depend on system size. Indeed, if the structures responsible for the beaming have sizes comparable to the kinetic scales, one may anticipate that the momentum anisotropy distribution will become increasingly homogeneous for larger systems, due to the beaming events becoming weaker or due to the large number of superimposed beaming events being washed out in the global anisotropy. As a first step toward answering this question, we consider the PDF for the momentum anisotropy distribution bin values for simulations of varying system size (similar as we did for varying $\langle\sigma\rangle$ in Fig.~\ref{fig:dist_angular_sigma}). The measured PDF is shown in Fig.~\ref{fig:angular_system_size} for $\langle \sigma \rangle = 3.4$ cases with two different sizes: $L/2\pi\langle\rho_e\rangle = 18.0$ ($256^3$) and $L/2\pi\langle\rho_e\rangle = 39.1$ ($512^3$). We find that low-energy particles (with $\gamma < 800$) have similar PDFs in both cases, but the high-energy particles (with $\gamma > 1600$) have a narrower PDF in the larger simulation, with a high-energy cutoff that is roughly a factor of two smaller in energy. This suggests that the beams are less intense at larger system sizes, and the PDF in the limit of asymptotically large sizes is unclear. A more careful, complete investigation of the system size dependence of kinetic beaming is left to future work.

\subsection{High-energy particle density profile} \label{sec:nhe}

To better understand the physical origin of the kinetic beams, we next consider the spatial profile of the high-energy particle density. In the following, we define the high-energy particles as particles with $\gamma > 800$ (roughly $2.5 \overline{\gamma}$), and subsequently describe the number density field of these high-energy particles, which we denote by $n_{\rm he}(\boldsymbol{x})$.

\begin{figure}
\includegraphics[width=\columnwidth]{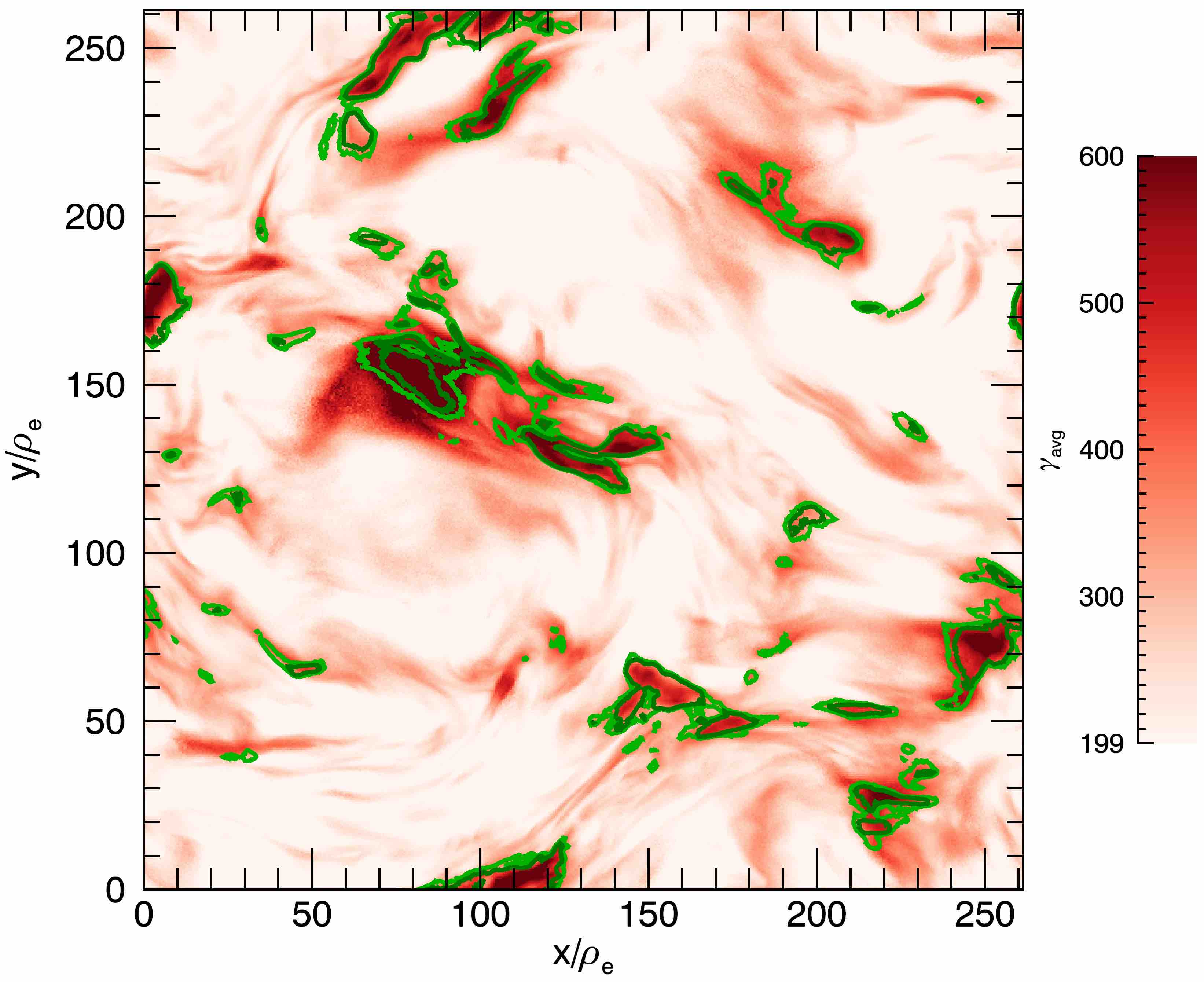}
\includegraphics[width=\columnwidth]{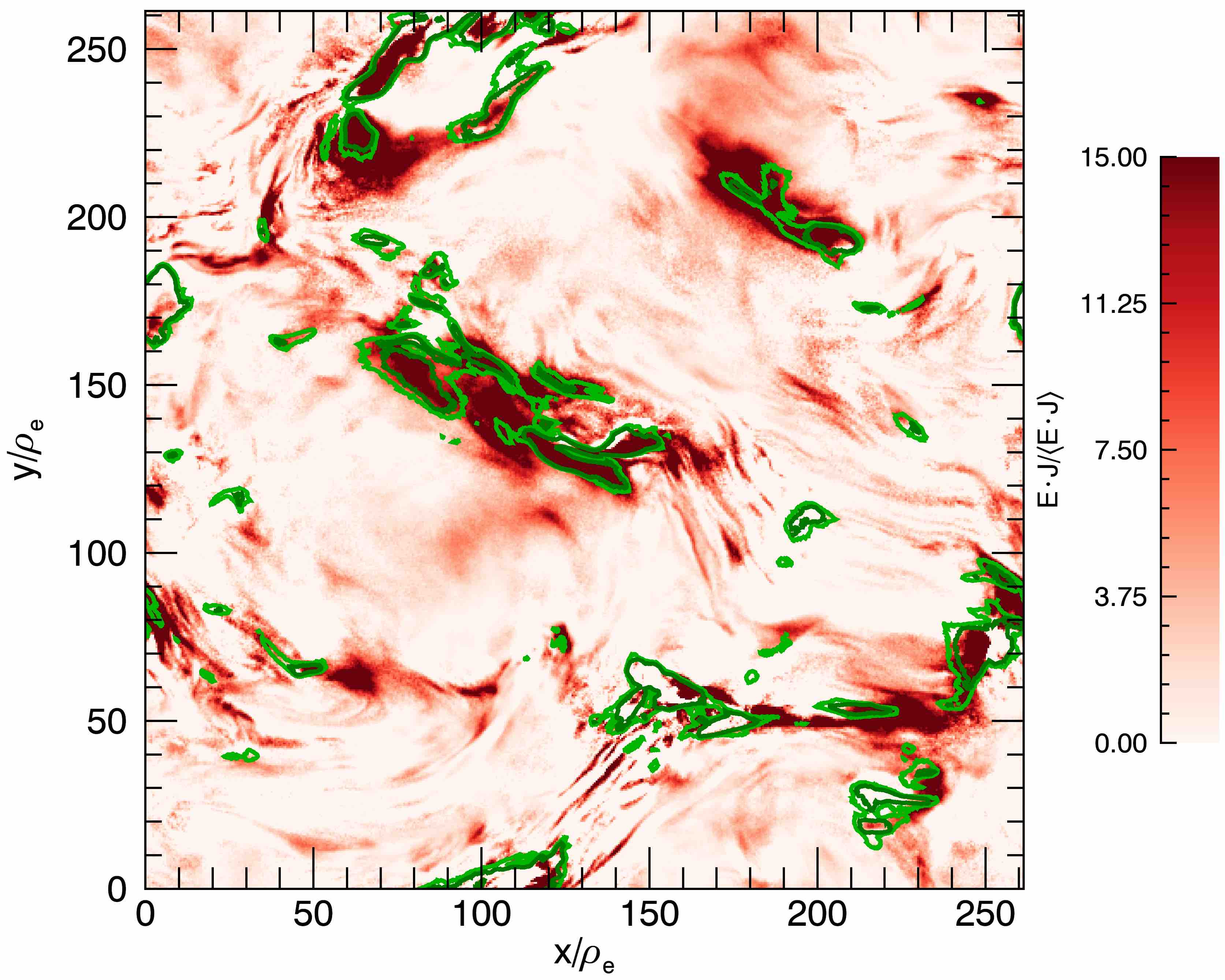}
\includegraphics[width=\columnwidth]{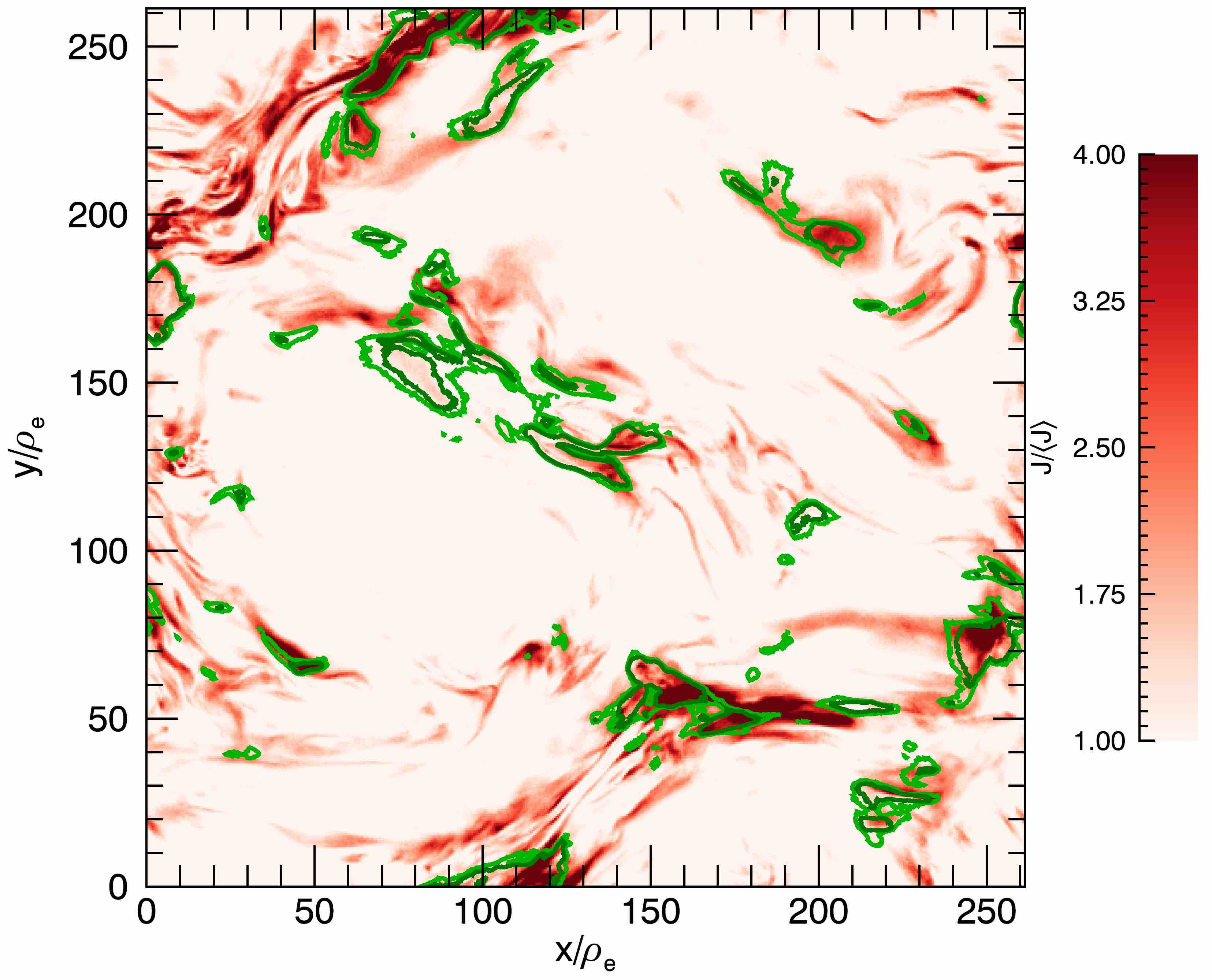}
  \centering
   \caption{\label{fig:nhe_visual} The cell-averaged particle energy $\gamma_{\rm avg}$ (top), heating rate proxy $\boldsymbol{E}\cdot\boldsymbol{J}$ (center), and current density $J$ (bottom), all with overlaid contours of the high-energy particle number density, $n_{\rm he}$ (with green contours at 4 and 8 times $\overline{n}_{\rm he}$). All quantities are averaged in the $z$ direction over $L/32$ of a snapshot from the $\langle\sigma\rangle = 3.4$, $512^3$ case.}
 \end{figure}
 
In the three panels of Fig.~\ref{fig:nhe_visual}, we show contours of the high-energy particle density $n_{\rm he}$ in a snapshot of the $512^3$, $\langle\sigma\rangle = 3.4$ case, averaged in the $z$ direction across $L/32$ of the domain (to reduce noise from the small population size). These contours are overlaid on top of images of the cell-averaged particle Lorentz factor $\gamma_{\rm avg}$ (top panel), the heating rate proxy $\boldsymbol{E} \cdot \boldsymbol{J}$ (center panel), and the magnitude of the current density $J$ (bottom panel). We find that $n_{\rm he}$ is strongly localized in space and strongly correlated with all three quantities ($\gamma_{\rm avg}$, $\boldsymbol{E} \cdot \boldsymbol{J}$, and $J$). The strong correlation of $n_{\rm he}$ with $\gamma_{\rm avg}$ suggests that high-energy particles are energized locally at the same sites as where much of the overall bulk plasma heating occurs. The correlation of $n_{\rm he}$ with $\boldsymbol{E} \cdot \boldsymbol{J}$ indicates that this energization is rapid, with high-energy particles confined to locations near the sites of energization. The correlation of $n_{\rm he}$ with intense, intermittent current sheets (and filaments) hints that magnetic reconnection may play a role in energizing and beaming the particles. Note, however, that the correlation is imperfect: there are some current sheets which lack high-energy particles \citep[perhaps indicating that reconnection is inefficient in those structures, e.g.,][]{zhdankin_etal_2013}, and some clusters of high-energy particles that are far away from current sheets (possibly a consequence of alternative channels of energy dissipation). High-energy particles are less well localized in the low magnetization simulations, consistent with weaker kinetic beaming in that regime (not shown).

\subsection{Tracked particle properties} \label{sec:track}

\begin{figure}
  \includegraphics[width=\columnwidth]{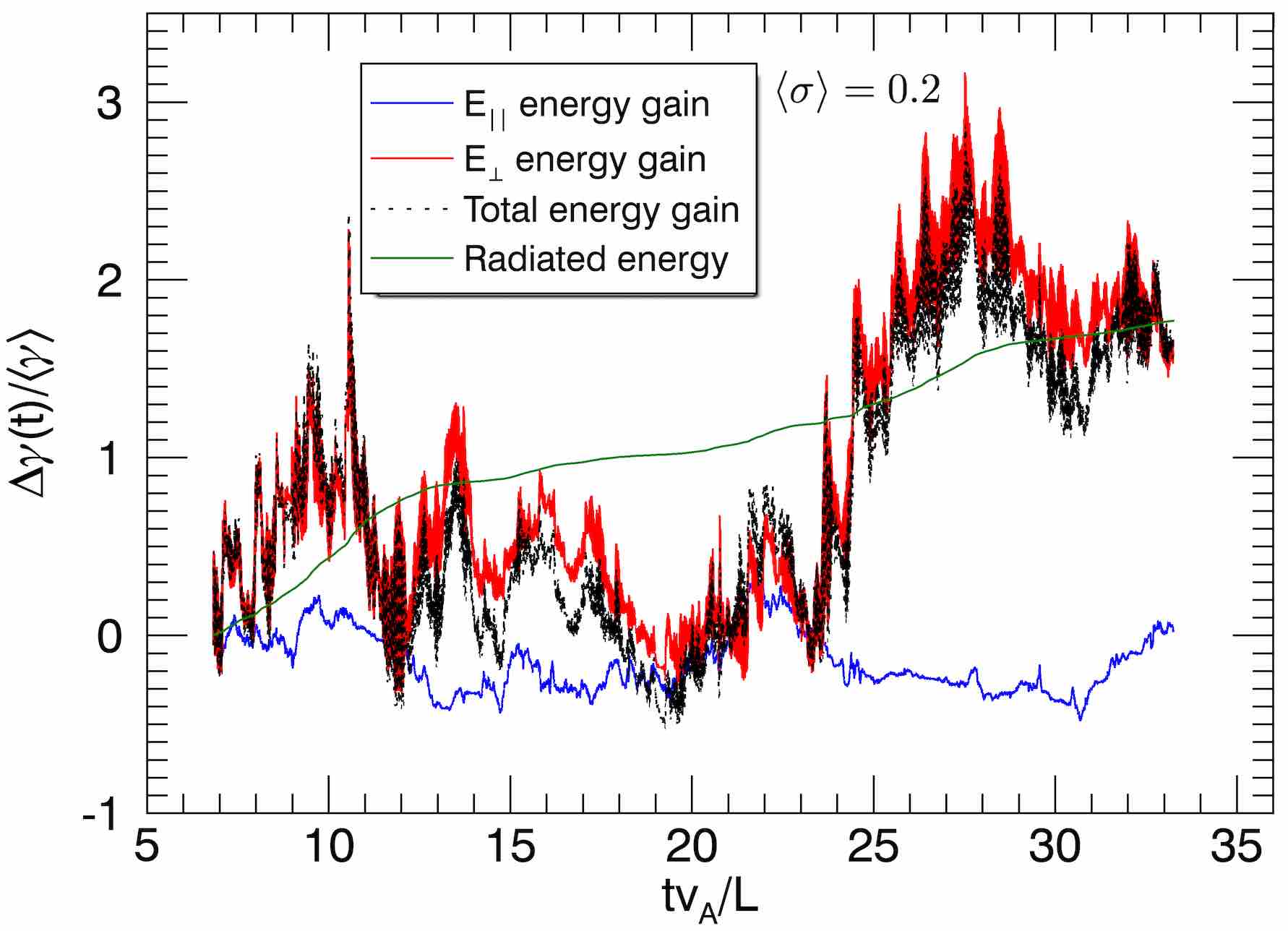}
    \includegraphics[width=\columnwidth]{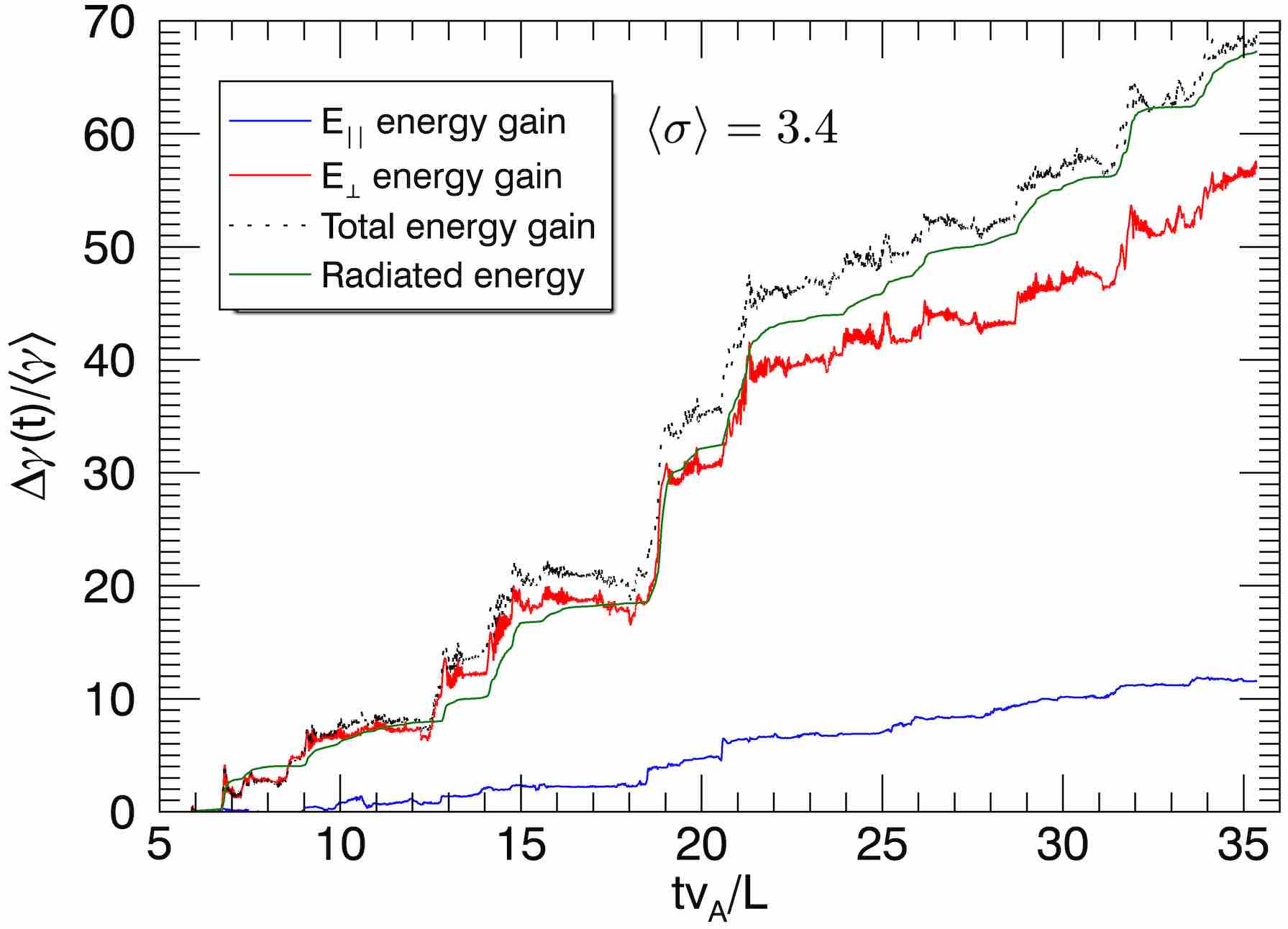}
  \centering
   \caption{\label{fig:track} The different contributions to the energy gain $\Delta\gamma$ for representative tracked particles in the $512^3$ simulations with $\langle\sigma\rangle =0.2$ (top) and $\langle\sigma\rangle = 3.4$ (bottom). The lines correspond to contributions from the parallel electric field (blue), perpendicular electric field (red), and overall electric field (black, dashed). We also show the energy radiated by the particle (green).}
 \end{figure}
 
Finally, to gain insight into the underlying charged particle dynamics, we consider some properties of tracked particles in the simulations. We randomly track $20,000$ of the simulated particles in each case, recording their positions, momenta, and field values starting at a time~$t_0$ that is during statistical steady state. With this information, we measure the amount of energy that each particle gains from the electric field component parallel to the magnetic field ($\boldsymbol{E}_{||} = \boldsymbol{E} \cdot \hat{\boldsymbol{B}} \hat{\boldsymbol{B}}$, absent in ideal MHD) and from the component perpendicular to it ($\boldsymbol{E}_{\perp} = \boldsymbol{E} - \boldsymbol{E}_{||}$, representative of the ideal MHD component), along with the energy lost to radiation. These energy contributions can be expressed, respectively, for the $i$th tracked particle, as
\begin{align}
\Delta\gamma_{||,i}(t) &= \frac{q_i}{m_e c^2} \int_{t_0}^t dt' \boldsymbol{E}_{||}(\boldsymbol{x}_i(t'),t') \cdot \boldsymbol{v}_i(t')\, , \nonumber \\
\Delta\gamma_{\perp,i}(t) &= \frac{q_i}{m_e c^2} \int_{t_0}^t dt' \boldsymbol{E}_{\perp}(\boldsymbol{x}_i(t'),t') \cdot \boldsymbol{v}_i(t') \, , \nonumber \\
\Delta\gamma_{{\rm rad},i}(t) &= - \frac{1}{m_e c^2} \int_{t_0}^t dt' \boldsymbol{F}_{{\rm IC}}(\boldsymbol{v}_i(t')) \cdot \boldsymbol{v}_i(t') \, ,
\end{align}
obtained by integrating Eq.~\ref{eq:eom} from the reference time~$t_0$ to the given time~$t$.

We first show, in Fig.~\ref{fig:track}, these contributions to the particle energy change $\Delta\gamma$ for two representative electrons: one from the $512^3$, $\langle\sigma\rangle = 0.2$ case and the other from the $512^3$, $\langle\sigma\rangle = 3.4$ case. In the low $\sigma$ case, changes in the particle energy are very slow and irregular, and are almost entirely due to $E_\perp$. In the high $\sigma$ case, the energization is also dominated by $E_\perp$, and occurs from a mixture of gradual heating and rapid, isolated energization events. It is tempting to associate the latter events with energization in current sheets, and thus with the kinetic beams. Indeed, these acceleration episodes are often preceded by a small energy kick from $E_{||}$, reminiscent of results from 2D relativistic kinetic turbulence that suggest a two-stage acceleration process in which $E_{||}$ (within intermittent current sheets) injects particles to energies where they can resonantly interact with the turbulent fluctuations \citep{comisso_sironi_2018}. In contrast to the rough evolution of energization from the electric fields, the amount of radiated energy shows a relatively smooth increase in time. On average, the amount of radiated energy balances out the energy gain from electric fields.

\begin{figure}
  \includegraphics[width=\columnwidth]{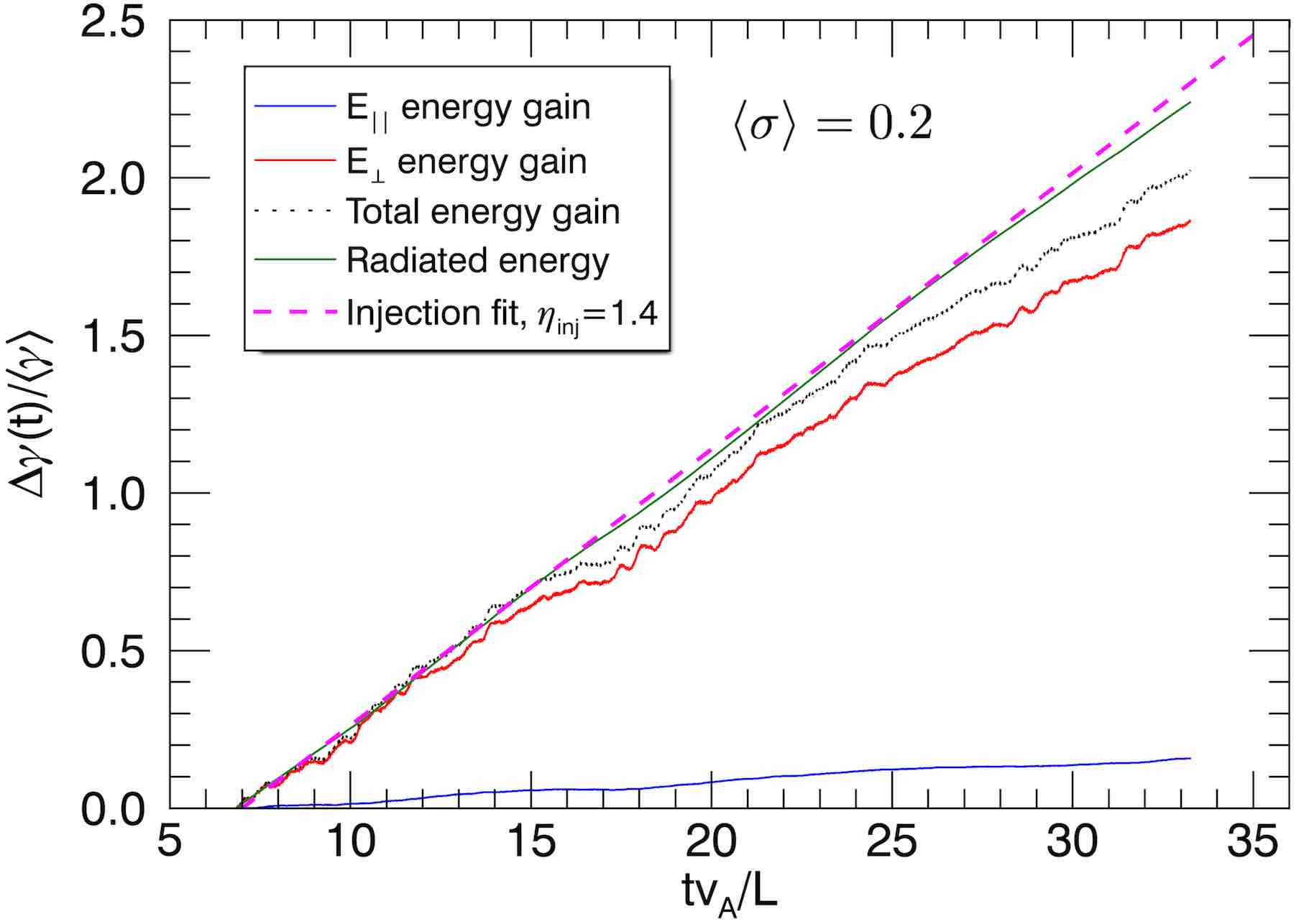}
    \includegraphics[width=\columnwidth]{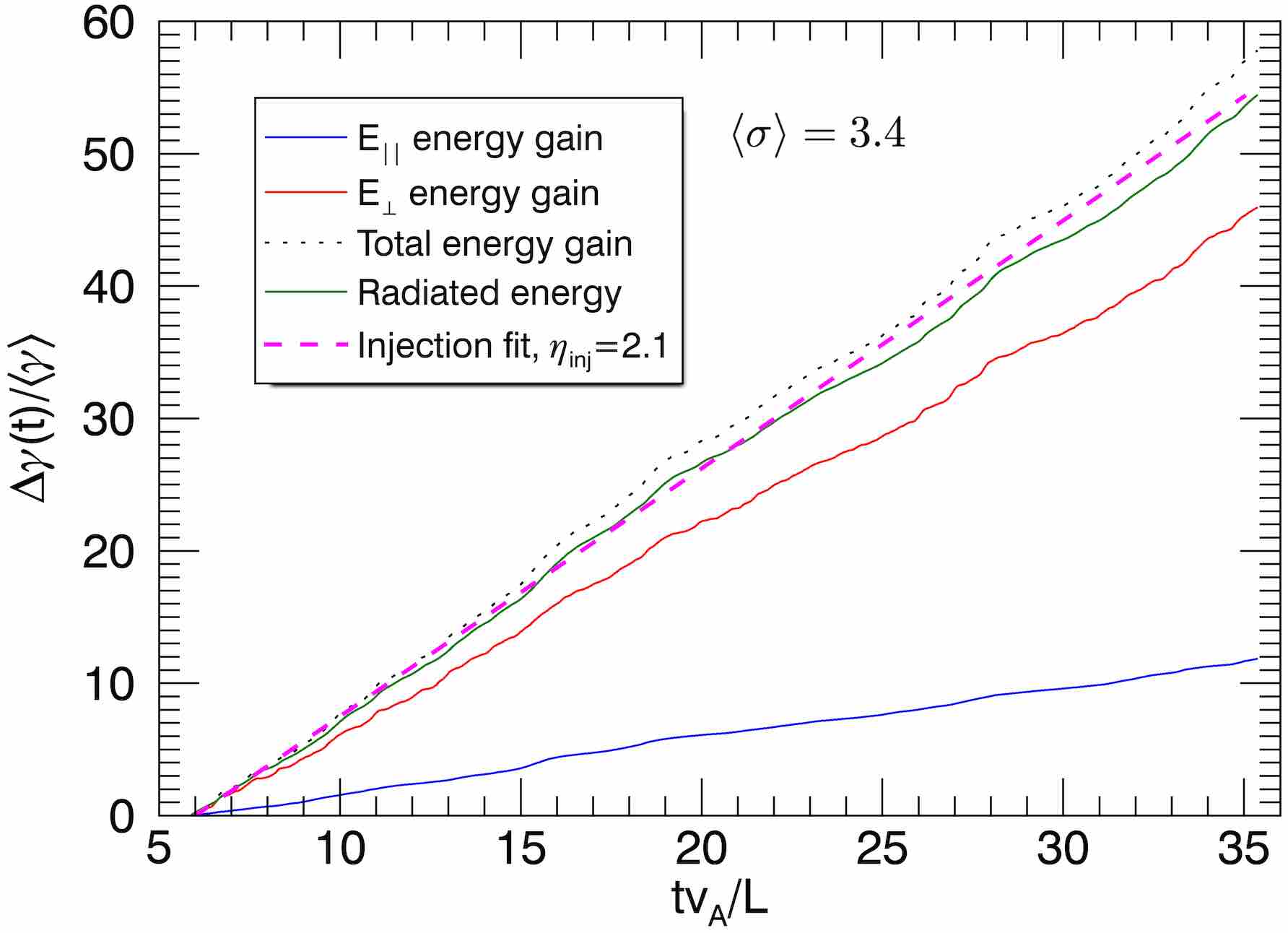}
  \centering
   \caption{\label{fig:track2} The different contributions to the particle energy gain $\Delta\gamma$, similar to Fig.~\ref{fig:track} but averaged for all tracked particles, in the $512^3$ simulations with $\langle\sigma\rangle =0.2$ (top) and $\langle\sigma\rangle = 3.4$ (bottom). The expected energy increase due to external driving (Eq.~\ref{eq:inj}) is shown for comparison, with $\eta_{\rm inj} = 1.4$ and $\eta_{\rm inj} = 2.1$, respectively (dashed magenta line).}
 \end{figure}
 
We next show the different contributions to $\Delta\gamma$ averaged for all tracked particles, again for the $512^3$, $\langle\sigma\rangle = 0.2$ and $\langle\sigma\rangle = 3.4$ cases, in Fig.~\ref{fig:track2}. We find that these contributions increase linearly in time, consistent with the statistical steady state and constant rate of energy injection. We compare the radiated energy with Eq.~\ref{eq:inj} to extract an injection efficiency of $\eta_{\rm inj} = 1.4$ and $\eta_{\rm inj} = 2.1$ for the two cases, respectively. The first value is consistent with $\eta_{\rm inj}$ inferred from the thermal equilibrium condition in Section~\ref{subsec:equil}, while the second is somewhat larger. The differences between the two approaches at high $\sigma$ may be attributed to significant deviations from a uniform Maxwell-J\"{u}ttner distribution (an assumption required to estimate the equilibrium temperature). 
 
 \begin{figure}
  \includegraphics[width=\columnwidth]{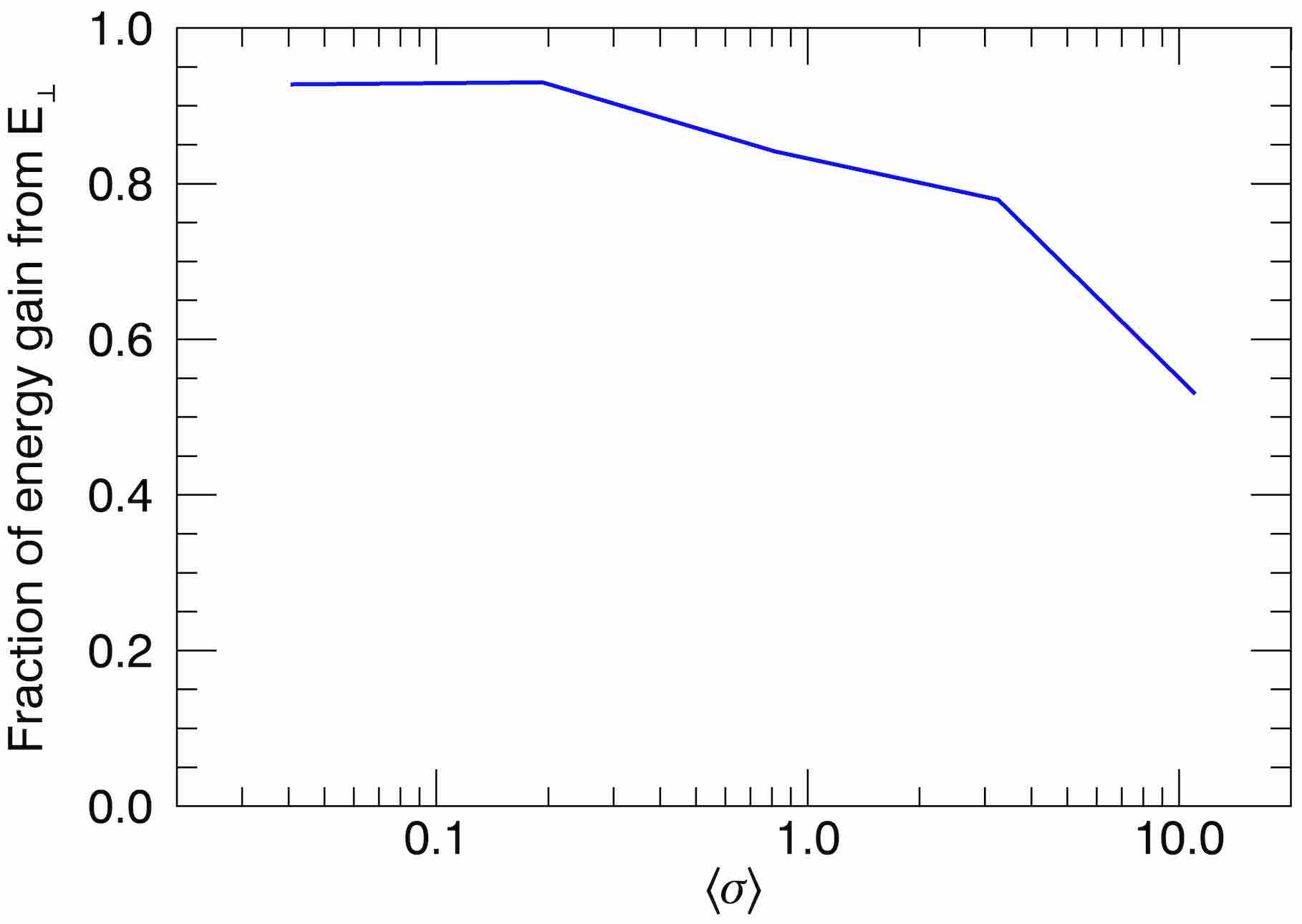}
  \centering
   \caption{\label{fig:eperp} The fraction of overall particle energy gain from perpendicular (rather than parallel) electric fields versus $\langle\sigma\rangle$, for the $384^3$ simulation series.}
 \end{figure}
 
To get a sense of the relative importance of the different electric field components at varying magnetizations, in Fig.~\ref{fig:eperp} we show the relative energy gain from perpendicular electric fields, $\Delta\gamma_{\perp,i}/(\Delta\gamma_{\perp,i} + \Delta\gamma_{||,i})$, versus $\langle\sigma\rangle$, averaged for all tracked particles in the $384^3$ simulation series. We find that at low $\langle\sigma\rangle$, more than $90\%$ of the energization is due to $E_\perp$. At high $\langle\sigma\rangle$, however, this percentage decreases, with only $\sim 50\%$ of the energization from $E_\perp$, and the rest from $E_{||}$, at $\langle\sigma\rangle \sim 10$. The preferential perpendicular energization is qualitatively similar to recent results from hybrid kinetic simulations of non-relativistic turbulence \citep{arzamasskiy_etal_2019} and from PIC simulations of the kink instability \citep{alves_etal_2018, alves_etal_2019}.

The energization of particles by $E_\perp$ is generally consistent with diffusive acceleration mechanisms (e.g., second-order Fermi or gyroresonant scattering), further supporting results in Section~\ref{sec:fp-fits}. The energization by $E_{||}$ in the high $\sigma$ regime, on the other hand, may be attributed to particles that are directly accelerated by the electric field within the current layers during magnetic reconnection \citep[also previously suggested in the low $\beta$ non-relativistic regime, e.g.,][]{makwana_etal_2017}. 
 
\section{Discussion} \label{sec5}

\subsection{Generic blazar emission} \label{sec:jet}

In this section, we consider the implications of our numerical results for high-energy astrophysical systems, choosing to focus on blazar jets. The broadband radiation spectra of blazar jets (and other high-energy astrophysical systems such as pulsar wind nebulae and black-hole accretion flows) have been interpreted to arise from a nonthermal population of electrons (and positrons) with a power-law energy distribution. Some studies have suggested that the underlying distributions may have modest deviations from a pure power law, as fit by, e.g., the log-parabola distribution \citep{sambruna_etal_1996, massaro_etal_2006, tramacere_etal_2011, fraschetti_pohl_2017}.

The generally thermal character of the particle distributions obtained from radiatively balanced driven turbulence suggests that a single-zone model of this kind does not account for the bulk of the relatively stable power-law emission spectrum from blazar jets. If the radiative cooling is strong, these spectra likely reflect the cumulative particle acceleration from numerous transient events (such as discrete reconnection events or decaying turbulence), distributed within the source, each contributing a burst of nonthermal particles which cool radiatively at later times (i.e., after the end of the energization event).

However, an intriguing alternative is that a power-law spectrum could result from the superposition of thermal peaks from an array of radiatively balanced turbulent regions with a power-law distribution of relativistic temperatures along the blazar jet \citep[see, e.g.,][]{henri_sauge_2006, boutelier_etal_2008}. For example, consider a Poynting flux-dominated jet, comprising pair plasma, carrying a luminosity  
\begin{align}
{\mathcal L}_j = 10^{46}{\mathcal L}_{46} \ {\rm erg \ s}^{-1} = \Omega_j r^2 \Gamma_j^2 \frac{B_0^2}{4\pi} c \, , \label{eq:lum}
\end{align}
where $\Omega_j(r)$ is the jet solid angle, $\Gamma_j(r) = 10 \Gamma_1(r)$ is the bulk Lorentz factor, $r = 10^{18} r_{18}$ cm is the distance from the black hole, and $B_0(r)$ is the typical magnetic field in the jet comoving frame.  We consider photons upscattered from an external radiation field $U_{\rm ph}(r)$ (e.g., supplied by the broad emission line region or dusty torus, and expressed in the jet comoving frame),
\begin{align}
U_{\rm ph} = \delta_{\rm ph} \Gamma_j^2 \frac{{\mathcal L}_j}{4\pi c r^2} \, , \label{eq:gam}
\end{align}
where $\delta_{\rm ph} \sim 0.01 - 0.1 $ is the fraction of jet power in the seed photon field.
Following the analysis in Section~\ref{sec:theory}, assuming a pair plasma composition, we then obtain the average particle Lorentz factor,
\begin{align}
\overline{\gamma}  = \frac{3\pi}{4} \sigma \frac{\eta_{\rm inj}}{\delta_{\rm ph}}  \left(\frac{r}{\Gamma_j L_{\rm turb}}\right)  \frac{m_e c^3 r}{\sigma_T {\mathcal L}_j \Gamma_j} \sim \sigma \frac{\eta_{\rm inj}}{\delta_{\rm ph}}  \left(\frac{r}{\Gamma_j L_{\rm turb}}\right) \frac{r_{18}}{\mathcal{L}_{46} \Gamma_1} \, , \label{eq:gam1}
\end{align}
where $L_{\rm turb}(r)$ represents the size of a turbulent region inside the jet, presumably driven by internal instabilities such as the kink mode \citep{begelman_1998, das_begelman_2019}. For the sake of argument, if we assume $L_{\rm turb} \sim 0.1 r/\Gamma_j$ and $\eta_{\rm inj} \sim 0.1$, then from Eq.~\ref{eq:gam1} we would estimate
\begin{equation}
\overline{\gamma} \sim \frac{\sigma}{\delta_{\rm ph}} \frac{ r_{18}}{{\mathcal L}_{46} \Gamma_1} \, . \label{eq:gamdim}
\end{equation}

If we next assume that~$\Omega_j$ and~$\Gamma_j$ are independent of~$r$, then conservation of both Poynting flux and particle flux implies that $\overline{\gamma} \propto r^{1/2}$ (solving Eqs.~\ref{eq:lum} and \ref{eq:gamdim}) and the energy of produced Compton photons is $\epsilon \propto r$ (given a fixed seed photon energy). The volume emissivity in the comoving frame is then $n_0 \dot{\mathcal E}/4\pi \propto r^{-3} \propto dI/dr$, the derivative of the frequency-integrated intensity entering the line of sight from radius $r$.  If we assume that the solid angle contributing to this intensity is proportional to $r^2$ (consistent with $\Omega_j$ independent of $r$), then the frequency integrated flux satisfies $dF/dr \propto r^{-1}$.  We can then deduce the observed spectrum by taking $dF/d\nu = dF/dr / (d\nu/dr ) \propto r^{-1} \propto \nu^{-1}$, which gives a spectral index comparable to observed values in the $\gamma$-ray band \citep{madejski_sikora_2016}. Elaboration of this model to more realistic turbulent jet propagation will have to await further work.

\subsection{Blazar flares}

The strong anisotropies and flaring behavior obtained from radiatively cooled kinetic turbulence at $\sigma \gg 1$ provide an appealing model for high-energy flares from blazars.  In standard models, where the flare variability timescale is assumed to be associated with the bulk dynamics of the jet or a localized energy release process within it, there is a tension between the short flare timescales and constraints that place the emitting region rather far from the black hole.  This tension is particularly acute when considering very-high-energy gamma-ray (TeV) flares from flat-spectrum radio quasars (FSRQs), where the flares must be produced on parsec scales in order to avoid being absorbed in pair production against the ambient diffuse radiation background \citep{begelman_etal_2008, nalewajko_etal_2014b, madejski_sikora_2016}. Such flares have been observed to be as short as a few minutes \citep{aharonian_etal_2007, albert_etal_2007}. If, however, the flares represent extremely relativistic beams of particles swinging past the observer's line of sight, the timescale and energetics are largely decoupled from the size and location of the emitting region.

Returning to the model in Section~\ref{sec:jet}, to create a flare at $\epsilon_{\rm TeV}$ TeV energies by upscattering seed photons of energy $\epsilon_{\rm eV}$ eV, typical of the broad line region, we need $\overline{\gamma} \approx 10^5 \Gamma_1^{-1} (\epsilon_{\rm TeV}/ \epsilon_{\rm eV})^{1/2}$ (assuming scattering by particles from a narrow quasi-thermal distribution), implying $\sigma \sim 10^3 -10^4$ for fiducial values of ${\mathcal L}_j$ and $r$. Thus, a self-consistent model invoking radiatively cooled turbulence in an FSRQ would have to have an enormous $\sigma$ and could therefore, extrapolating our numerical results, presumably produce extremely short and intense flares even at large distances from the black hole.  

Since we know very little about how mass is loaded onto the base of a jet, it is quite possible that jets start out with enormous values of~$\sigma$, especially if they are powered by black hole spin via the Blandford-Znajek mechanism (in which case they are also likely to be pair-dominated) \citep{blandford_znajek_1977}. Whether such high values of $\sigma$ can be sustained out to parsec-scale distances is a separate question.  In ideal MHD models for relativistic jet production assuming cold plasma, with a ratio of jet power to its rest mass energy flux $\mu \gg 1$, the flow typically accelerates to Lorentz factor $\Gamma \sim \mu^{1/3}$ near the light-cylinder radius \citep[near the fast magnetosonic point: e.g.,][]{begelman_li_1994} and can accelerate (more slowly) subsequently to $\Gamma \sim O(\mu/2)$, corresponding to approximate equipartition between Poynting flux and relativistic enthalpy flux, $\sigma \sim O(1)$.  However, these results apply primarily to cold, non-dissipative flows, whereas the entire premise of our analysis here is that internal instabilities lead to dissipation while drag in an external radiation field hampers acceleration.  If these effects are operating at relatively small radii within the flow, it would seem possible for a large value of $\sigma$ to get ``frozen'' into the jet flow, since once the magnetic field twists into an overwhelmingly toroidal configuration, as measured in the comoving jet frame, it is very difficult for it to share additional energy with the particle flow \citep{begelman_li_1994}.

The above results apply to a pair-dominated jet; the effect of an ion component on radiatively cooled turbulence is outside the scope of this paper.  The ions would effectively not cool at all, but would probably receive the majority of dissipated energy \citep{zhdankin_etal_2019}, suggesting that the true value of $\sigma$ could be much lower due to a hot ion component. A self-consistency condition that needs to be met in order for external IC to dominate over synchrotron cooling, $B_0^2 /8\pi < U_{\rm ph}$, could be violated if $\delta_{\rm ph}$ or $\Gamma_j$ is too small.

A cooled turbulence model is also attractive for explaining flares from BL Lac-type blazars, which lack an intense ambient radiation field.  Here it is plausible that the jet emission is self-contained and likely due to the SSC process, and even larger values of $\sigma$ (at least the $\sigma$ associated with the electron or pair content) would be required.

\section{Conclusion} \label{sec6}

In this work, we conducted and studied the results from a series of PIC simulations of driven turbulence in relativistic collisionless pair plasma with external IC radiation. We focused on understanding the statistical properties of turbulent fluctuations and particle distributions during the statistical steady state, where injected energy from external driving is balanced by radiative energy losses. We performed a parameter scan over the steady-state magnetization $\langle\sigma\rangle$ and system size $L/2\pi\langle\rho_e\rangle$, to the extent that computational resources allowed. We now summarize our main results:
\begin{enumerate}
\item Regardless of initial conditions, turbulence eventually acquires a rigorous statistical steady state (by heating or cooling) with an equilibrium temperature that is close to analytic predictions (Eq.~\ref{eq:ss}) and can be controlled by adjusting the energy density of the external photon bath, $U_{\rm ph}$. Radiative turbulence thus provides an ideal scenario for studying the fundamental properties of kinetic turbulence.
\item The properties of turbulence change from low $\langle\sigma\rangle$ to high $\langle\sigma\rangle$. The former looks like standard non-relativistic MHD turbulence, apart from the tendency to have rotational magnetic fluctuations. The latter has fluctuations that are highly relativistic, compressible, and electromagnetically dominated.
\item The steady-state particle energy distributions are always close to the thermal (Maxwell-J\"{u}ttner) distribution, with only a modest nonthermal population (broadening the tail by a factor of $\sim 2$ in energy). The nonthermal population is more temporally variable and energetically significant at high $\langle \sigma \rangle$.
\item These steady-state particle energy distributions are well fit by analytic solutions to the Fokker-Planck equation describing stochastic particle acceleration, for simple forms of the momentum advection and diffusion coefficients (linear and quadratic in energy, respectively). The diffusion coefficient scales with $\langle\sigma\rangle$ in a way that is consistent with second-order Fermi acceleration or gyroresonant interactions with Alfv\'{e}n modes. Tracked particles are mainly energized by the perpendicular electric field (for the accessible $\langle\sigma\rangle$), lending further evidence for this mechanism.
\item In the high magnetization regime, the nonthermal particle population is intermittently beamed in random directions. This produces a fine-scale anisotropy in the global momentum distribution, which can be manifest as highly variable radiative signatures. The beaming becomes weaker as the system size is increased, however, raising questions about beaming statistics in the limit of large system size.
\item The beamed high-energy particles are spatially correlated with intermittent current sheets, suggesting that magnetic reconnection may locally energize the nonthermal particles.
\end{enumerate}
We considered the applicability of our numerical results on radiative turbulence to blazar jets. The quasi-thermal distribution obtained in our model is at odds with the broadband emission spectra observed from blazars (as well as other systems), which require extended nonthermal distributions of electrons (and positrons). One way to reconcile these results is to posit that blazars may be in a weak cooling regime, where the distribution of electrons will lie somewhere in between the radiative steady-state solution (found in this work) and the extended power-law distributions seen in non-radiative kinetic turbulence simulations \citep{zhdankin_etal_2017, comisso_sironi_2018, nattila_2019}. Alternatively, we suggested that a multi-zone model incorporating a superposition of quasi-thermal distributions may be a viable way to obtain broadband emission spectra. Further observational constraints on the statistical properties of the blazar jet flows will be valuable for testing such theoretical models \citep[e.g.,][]{guo_mao_wang_2017}.

Our results may also be relevant for GRBs. The Fokker-Planck formalism with radiative cooling was previously employed to model GRB spectra \citep[e.g.,][]{asano_terasawa_2009, xu_zhang_2017}. If a power-law distribution of particles is injected, it can be substantially modified by strong radiative cooling (as well as adiabatic expansion), which has been suggested to cause a steeper power law distribution \citep[e.g.,][]{kardashev_1962, sari_piran_narayan}. PIC simulations of radiative turbulence will validate and inform such models.

This work takes the first step in numerically investigating radiative turbulence in collisionless plasmas, but it is far from complete or comprehensive. This work has established the existence of intermittent beaming in turbulence, but understanding the beam statistics and corresponding radiative signatures requires a deeper investigation. It is particularly important to understand these properties for larger systems and higher magnetization. Some astrophysical systems (e.g., blazar jets and pulsar winds) may have magnetizations higher than the maximum value ($\langle\sigma\rangle=11$) considered in our simulations. At these high magnetizations, magnetic reconnection may play the dominant role in particle energization \citep{comisso_sironi_2019}. Particle acceleration in an isolated magnetic reconnection site may manifest as a first-order mechanism in the Fokker-Planck equation \citep{lazarian_etal_2012, kowal_etal_2012, guo_etal_2014, guo_etal_2015, delValle_etal_2016}; however, it has also been suggested that the diffusion of particles through an ensemble of reconnection sites may appear as a second-order mechanism \citep[e.g.,][]{brunetti_lazarian_2016}. It is also conceivable that reconnection lies outside of the Fokker-Planck formalism, in which case a separate model may need to be developed to describe the steady state \citep[e.g.,][]{isliker_etal_2017}. Whether an extended power law distribution can form in this reconnection-dominated regime despite strong radiative cooling is an interesting question that awaits future simulations \citep[see, e.g.,][for a discussion of this possibility]{sobacchi_lyubarsky_2020}. In the high magnetization regime, we anticipate that a majority of the dissipated energy may be beamed. 

There are several other future directions in which to proceed. One direction is studying the effect of synchrotron cooling, which is self-consistently determined from the magnetic energy density rather than an imposed external photon field. We expect that the overall steady state for synchrotron cooling would closely resemble that for external IC radiation, but the additional spatial dependence due to the turbulent magnetic field may modify some features of the kinetic beams. A second future direction is studying the weak cooling regime, where the plasma temperature is significantly below the equilibrium value. In this case, particles will be accelerated to a power-law energy distribution, but cooling may impose a high-energy cutoff to the energy distribution. A third future direction is studying radiative cooling in electron-ion plasmas, where a radiative steady state does not necessarily exist due to ions being inefficient radiators. In this case, a steady state can occur only if there is an efficient collisionless thermal coupling mechanism between ions and electrons; we will investigate this problem in a follow-up paper.


\section*{Acknowledgements}

VZ thanks Matthew Kunz, Jonathan Squire, and Kai Wong for helpful discussions. The authors acknowledge support from NSF grants AST-1411879, AST-1806084, and AST-1903335; and NASA ATP grants NNX16AB28G and NNX17AK57G. VZ acknowledges support for this work from NASA through the NASA Hubble Fellowship grant \#HST-HF2-51426.001-A awarded by the Space Telescope Science Institute, which is operated by the Association of Universities for Research in Astronomy, Inc., for NASA, under contract NAS5-26555. An award of computer time was provided by the Innovative and Novel Computational Impact on Theory and Experiment (INCITE) program. This research used resources of the Argonne Leadership Computing Facility, which is a DOE Office of Science User Facility supported under Contract DE-AC02-06CH11357. This work also used the Extreme Science and Engineering Discovery Environment (XSEDE), which is supported by National Science Foundation grant number ACI-1548562. This work used the XSEDE supercomputer Stampede2 at the Texas Advanced Computer Center (TACC) through allocation TG-PHY160032 \citep{xsede}.




\bibliographystyle{mnras}
\bibliography{refs_all}

\begin{thebibliography}{}
\makeatletter
\relax
\def\mn@urlcharsother{\let\do\@makeother \do\$\do\&\do\#\do\^\do\_\do\%\do\~}
\def\mn@doi{\begingroup\mn@urlcharsother \@ifnextchar [ {\mn@doi@}
  {\mn@doi@[]}}
\def\mn@doi@[#1]#2{\def\@tempa{#1}\ifx\@tempa\@empty \href
  {http://dx.doi.org/#2} {doi:#2}\else \href {http://dx.doi.org/#2} {#1}\fi
  \endgroup}
\def\mn@eprint#1#2{\mn@eprint@#1:#2::\@nil}
\def\mn@eprint@arXiv#1{\href {http://arxiv.org/abs/#1} {{\tt arXiv:#1}}}
\def\mn@eprint@dblp#1{\href {http://dblp.uni-trier.de/rec/bibtex/#1.xml}
  {dblp:#1}}
\def\mn@eprint@#1:#2:#3:#4\@nil{\def\@tempa {#1}\def\@tempb {#2}\def\@tempc
  {#3}\ifx \@tempc \@empty \let \@tempc \@tempb \let \@tempb \@tempa \fi \ifx
  \@tempb \@empty \def\@tempb {arXiv}\fi \@ifundefined
  {mn@eprint@\@tempb}{\@tempb:\@tempc}{\expandafter \expandafter \csname
  mn@eprint@\@tempb\endcsname \expandafter{\@tempc}}}

\bibitem[\protect\citeauthoryear{Aharonian et~al.,}{Aharonian
  et~al.}{2007}]{aharonian_etal_2007}
Aharonian F.,  et~al., 2007, The Astrophysical Journal Letters, 664, L71

\bibitem[\protect\citeauthoryear{Albert et~al.,}{Albert
  et~al.}{2007}]{albert_etal_2007}
Albert J.,  et~al., 2007, The Astrophysical Journal, 669, 862

\bibitem[\protect\citeauthoryear{Alves, Zrake  \& Fiuza}{Alves
  et~al.}{2018}]{alves_etal_2018}
Alves E.~P.,  Zrake J.,   Fiuza F.,  2018, Physical review letters, 121, 245101

\bibitem[\protect\citeauthoryear{Alves, Zrake  \& Fiuza}{Alves
  et~al.}{2019}]{alves_etal_2019}
Alves E.~P.,  Zrake J.,   Fiuza F.,  2019, Physics of Plasmas, 26, 072105

\bibitem[\protect\citeauthoryear{{Arzamasskiy}, {Kunz}, {Chandran}  \&
  {Quataert}}{{Arzamasskiy} et~al.}{2019}]{arzamasskiy_etal_2019}
{Arzamasskiy} L.,  {Kunz} M.~W.,  {Chandran} B.~D.~G.,   {Quataert} E.,  2019,
  The Astrophysical Journal, 879, 53

\bibitem[\protect\citeauthoryear{Asano \& Terasawa}{Asano \&
  Terasawa}{2009}]{asano_terasawa_2009}
Asano K.,  Terasawa T.,  2009, The Astrophysical Journal, 705, 1714

\bibitem[\protect\citeauthoryear{Bale, Kasper, Howes, Quataert, Salem  \&
  Sundkvist}{Bale et~al.}{2009}]{bale_etal_2009}
Bale S.,  Kasper J.,  Howes G.,  Quataert E.,  Salem C.,   Sundkvist D.,  2009,
  Physical review letters, 103, 211101

\bibitem[\protect\citeauthoryear{Begelman}{Begelman}{1998}]{begelman_1998}
Begelman M.~C.,  1998, The Astrophysical Journal, 493, 291

\bibitem[\protect\citeauthoryear{Begelman \& Li}{Begelman \&
  Li}{1994}]{begelman_li_1994}
Begelman M.~C.,  Li Z.-Y.,  1994, The Astrophysical Journal, 426, 269

\bibitem[\protect\citeauthoryear{Begelman, Fabian  \& Rees}{Begelman
  et~al.}{2008}]{begelman_etal_2008}
Begelman M.~C.,  Fabian A.~C.,   Rees M.~J.,  2008, Monthly Notices of the
  Royal Astronomical Society: Letters, 384, L19

\bibitem[\protect\citeauthoryear{Beresnyak}{Beresnyak}{2015}]{beresnyak_2015}
Beresnyak A.,  2015, The Astrophysical Journal Letters, 801, L9

\bibitem[\protect\citeauthoryear{Beresnyak, Lazarian  \& Cho}{Beresnyak
  et~al.}{2005}]{beresnyak_etal_2005}
Beresnyak A.,  Lazarian A.,   Cho J.,  2005, The Astrophysical Journal Letters,
  624, L93

\bibitem[\protect\citeauthoryear{Blandford \& Eichler}{Blandford \&
  Eichler}{1987}]{blandford_eichler_1987}
Blandford R.,  Eichler D.,  1987, Physics Reports, 154, 1

\bibitem[\protect\citeauthoryear{Blandford \& Znajek}{Blandford \&
  Znajek}{1977}]{blandford_znajek_1977}
Blandford R.~D.,  Znajek R.~L.,  1977, Monthly Notices of the Royal
  Astronomical Society, 179, 433

\bibitem[\protect\citeauthoryear{Blumenthal \& Gould}{Blumenthal \&
  Gould}{1970}]{blumenthal_gould_1970}
Blumenthal G.~R.,  Gould R.~J.,  1970, Reviews of Modern Physics, 42, 237

\bibitem[\protect\citeauthoryear{Boldyrev, Chen, Xia  \& Zhdankin}{Boldyrev
  et~al.}{2015}]{boldyrev_etal_2015}
Boldyrev S.,  Chen C.~H.,  Xia Q.,   Zhdankin V.,  2015, The Astrophysical
  Journal, 806, 238

\bibitem[\protect\citeauthoryear{Boutelier, Henri  \& Petrucci}{Boutelier
  et~al.}{2008}]{boutelier_etal_2008}
Boutelier T.,  Henri G.,   Petrucci P.-O.,  2008, Monthly Notices of the Royal
  Astronomical Society: Letters, 390, L73

\bibitem[\protect\citeauthoryear{Brunetti \& Lazarian}{Brunetti \&
  Lazarian}{2016}]{brunetti_lazarian_2016}
Brunetti G.,  Lazarian A.,  2016, Monthly Notices of the Royal Astronomical
  Society, 458, 2584

\bibitem[\protect\citeauthoryear{Cerutti \& Philippov}{Cerutti \&
  Philippov}{2017}]{cerutti_etal_2017}
Cerutti B.,  Philippov A.~A.,  2017, Astronomy \& Astrophysics, 607, A134

\bibitem[\protect\citeauthoryear{Cerutti, Uzdensky  \& Begelman}{Cerutti
  et~al.}{2012}]{cerutti_etal_2012}
Cerutti B.,  Uzdensky D.~A.,   Begelman M.~C.,  2012, The Astrophysical
  Journal, 746, 148

\bibitem[\protect\citeauthoryear{Cerutti, Werner, Uzdensky  \&
  Begelman}{Cerutti et~al.}{2013}]{cerutti_etal_2013}
Cerutti B.,  Werner G.~R.,  Uzdensky D.~A.,   Begelman M.~C.,  2013, The
  Astrophysical Journal, 770, 147

\bibitem[\protect\citeauthoryear{Cerutti, Werner, Uzdensky  \&
  Begelman}{Cerutti et~al.}{2014a}]{cerutti_etal_2014b}
Cerutti B.,  Werner G.~R.,  Uzdensky D.~A.,   Begelman M.~C.,  2014a, Physics
  of Plasmas, 21, 056501

\bibitem[\protect\citeauthoryear{Cerutti, Werner, Uzdensky  \&
  Begelman}{Cerutti et~al.}{2014b}]{cerutti_etal_2014}
Cerutti B.,  Werner G.~R.,  Uzdensky D.~A.,   Begelman M.~C.,  2014b, The
  Astrophysical Journal, 782, 104

\bibitem[\protect\citeauthoryear{{Cerutti}, {Philippov}  \&
  {Spitkovsky}}{{Cerutti} et~al.}{2016}]{cerutti_etal_2016}
{Cerutti} B.,  {Philippov} A.~A.,   {Spitkovsky} A.,  2016, Monthly Notices of
  the Royal Astronomical Society, 457, 2401

\bibitem[\protect\citeauthoryear{Chandran}{Chandran}{2000}]{chandran_2000}
Chandran B.~D.,  2000, Physical Review Letters, 85, 4656

\bibitem[\protect\citeauthoryear{Chen, Leung, Boldyrev, Maruca  \& Bale}{Chen
  et~al.}{2014}]{chen_etal_2014b}
Chen C.,  Leung L.,  Boldyrev S.,  Maruca B.,   Bale S.,  2014, Geophysical
  research letters, 41, 8081

\bibitem[\protect\citeauthoryear{Cho}{Cho}{2005}]{cho_2005}
Cho J.,  2005, The Astrophysical Journal, 621, 324

\bibitem[\protect\citeauthoryear{Cho \& Lazarian}{Cho \&
  Lazarian}{2013}]{cho_lazarian_2013}
Cho J.,  Lazarian A.,  2013, The Astrophysical Journal, 780, 30

\bibitem[\protect\citeauthoryear{Cho \& Vishniac}{Cho \&
  Vishniac}{2000}]{cho_vishniac_2000}
Cho J.,  Vishniac E.~T.,  2000, The Astrophysical Journal, 539, 273

\bibitem[\protect\citeauthoryear{Chou \& Hau}{Chou \&
  Hau}{2004}]{chou_hau_2004}
Chou M.,  Hau L.-N.,  2004, The Astrophysical Journal, 611, 1200

\bibitem[\protect\citeauthoryear{Comisso \& Sironi}{Comisso \&
  Sironi}{2018}]{comisso_sironi_2018}
Comisso L.,  Sironi L.,  2018, Physical Review Letters, 121, 255101

\bibitem[\protect\citeauthoryear{Comisso \& Sironi}{Comisso \&
  Sironi}{2019}]{comisso_sironi_2019}
Comisso L.,  Sironi L.,  2019, The Astrophysical Journal, 886, 122

\bibitem[\protect\citeauthoryear{Das \& Begelman}{Das \&
  Begelman}{2019}]{das_begelman_2019}
Das U.,  Begelman M.~C.,  2019, Monthly Notices of the Royal Astronomical
  Society, 482, 2107

\bibitem[\protect\citeauthoryear{Demidem, Lemoine  \& Casse}{Demidem
  et~al.}{2019}]{demidem_etal_2019}
Demidem C.,  Lemoine M.,   Casse F.,  2019, arXiv preprint arXiv:1909.12885

\bibitem[\protect\citeauthoryear{Dong, Wang, Huang, Comisso  \&
  Bhattacharjee}{Dong et~al.}{2018}]{dong_etal_2018}
Dong C.,  Wang L.,  Huang Y.-M.,  Comisso L.,   Bhattacharjee A.,  2018,
  Physical review letters, 121, 165101

\bibitem[\protect\citeauthoryear{Fermi}{Fermi}{1949}]{fermi_1949}
Fermi E.,  1949, Physical Review, 75, 1169

\bibitem[\protect\citeauthoryear{Fermi}{Fermi}{1954}]{fermi_1954}
Fermi E.,  1954, The Astrophysical Journal, 119, 1

\bibitem[\protect\citeauthoryear{{Franci}, {Landi}, {Matteini}, {Verdini}  \&
  {Hellinger}}{{Franci} et~al.}{2016}]{franci_etal_2016}
{Franci} L.,  {Landi} S.,  {Matteini} L.,  {Verdini} A.,   {Hellinger} P.,
  2016, The Astrophysical Journal, 833, 91

\bibitem[\protect\citeauthoryear{Fraschetti \& Pohl}{Fraschetti \&
  Pohl}{2017}]{fraschetti_pohl_2017}
Fraschetti F.,  Pohl M.,  2017, Monthly Notices of the Royal Astronomical
  Society, 471, 4856

\bibitem[\protect\citeauthoryear{Giannios, Uzdensky  \& Begelman}{Giannios
  et~al.}{2009}]{giannios_etal_2009}
Giannios D.,  Uzdensky D.~A.,   Begelman M.~C.,  2009, Monthly Notices of the
  Royal Astronomical Society: Letters, 395, L29

\bibitem[\protect\citeauthoryear{Giannios, Uzdensky  \& Begelman}{Giannios
  et~al.}{2010}]{giannios_etal_2010}
Giannios D.,  Uzdensky D.~A.,   Begelman M.~C.,  2010, Monthly Notices of the
  Royal Astronomical Society, 402, 1649

\bibitem[\protect\citeauthoryear{Goldreich \& Sridhar}{Goldreich \&
  Sridhar}{1995}]{goldreich_sridhar_1995}
Goldreich P.,  Sridhar S.,  1995, The Astrophysical Journal, 438, 763

\bibitem[\protect\citeauthoryear{Guo, Li, Daughton  \& Liu}{Guo
  et~al.}{2014}]{guo_etal_2014}
Guo F.,  Li H.,  Daughton W.,   Liu Y.-H.,  2014, Physical Review Letters, 113,
  155005

\bibitem[\protect\citeauthoryear{Guo, Liu, Daughton  \& Li}{Guo
  et~al.}{2015}]{guo_etal_2015}
Guo F.,  Liu Y.-H.,  Daughton W.,   Li H.,  2015, The Astrophysical Journal,
  806, 167

\bibitem[\protect\citeauthoryear{Guo, Mao  \& Wang}{Guo
  et~al.}{2017}]{guo_mao_wang_2017}
Guo X.,  Mao J.,   Wang J.,  2017, The Astrophysical Journal, 843, 23

\bibitem[\protect\citeauthoryear{Hakobyan, Philippov  \& Spitkovsky}{Hakobyan
  et~al.}{2019}]{hakobyan_etal_2019}
Hakobyan H.,  Philippov A.,   Spitkovsky A.,  2019, The Astrophysical Journal,
  877, 53

\bibitem[\protect\citeauthoryear{Henri \& Saug{\'e}}{Henri \&
  Saug{\'e}}{2006}]{henri_sauge_2006}
Henri G.,  Saug{\'e} L.,  2006, The Astrophysical Journal, 640, 185

\bibitem[\protect\citeauthoryear{Isliker, Vlahos  \& Constantinescu}{Isliker
  et~al.}{2017}]{isliker_etal_2017}
Isliker H.,  Vlahos L.,   Constantinescu D.,  2017, Physical Review Letters,
  119, 045101

\bibitem[\protect\citeauthoryear{Jaroschek \& Hoshino}{Jaroschek \&
  Hoshino}{2009}]{jaroschek_hoshino_2009}
Jaroschek C.~H.,  Hoshino M.,  2009, Physical review letters, 103, 075002

\bibitem[\protect\citeauthoryear{Kagan, Nakar  \& Piran}{Kagan
  et~al.}{2016a}]{kagan_etal_2016a}
Kagan D.,  Nakar E.,   Piran T.,  2016a, The Astrophysical Journal, 826, 221

\bibitem[\protect\citeauthoryear{Kagan, Nakar  \& Piran}{Kagan
  et~al.}{2016b}]{kagan_etal_2016b}
Kagan D.,  Nakar E.,   Piran T.,  2016b, The Astrophysical Journal, 833, 155

\bibitem[\protect\citeauthoryear{Kardashev}{Kardashev}{1962}]{kardashev_1962}
Kardashev N.,  1962, Soviet Astronomy, 6, 317

\bibitem[\protect\citeauthoryear{Kirk \& Reville}{Kirk \&
  Reville}{2010}]{kirk_reville_2010}
Kirk J.~G.,  Reville B.,  2010, The Astrophysical Journal Letters, 710, L16

\bibitem[\protect\citeauthoryear{Kowal, Lazarian  \& Beresnyak}{Kowal
  et~al.}{2007}]{kowal_etal_2007}
Kowal G.,  Lazarian A.,   Beresnyak A.,  2007, The Astrophysical Journal, 658,
  423

\bibitem[\protect\citeauthoryear{Kowal, Dal~Pino  \& Lazarian}{Kowal
  et~al.}{2012}]{kowal_etal_2012}
Kowal G.,  Dal~Pino E. M. d.~G.,   Lazarian A.,  2012, Physical Review Letters,
  108, 241102

\bibitem[\protect\citeauthoryear{Kumar \& Narayan}{Kumar \&
  Narayan}{2009}]{kumar_narayan_2009}
Kumar P.,  Narayan R.,  2009, Monthly Notices of the Royal Astronomical
  Society, 395, 472

\bibitem[\protect\citeauthoryear{Kunz, Schekochihin  \& Stone}{Kunz
  et~al.}{2014}]{kunz_etal_2014}
Kunz M.~W.,  Schekochihin A.~A.,   Stone J.~M.,  2014, Physical Review Letters,
  112, 205003

\bibitem[\protect\citeauthoryear{Landau \& Lifshitz}{Landau \&
  Lifshitz}{1975}]{landau_lifshitz_1975}
Landau L.~D.,  Lifshitz E.~M.,  1975, Course of Theoretical Physics, 2

\bibitem[\protect\citeauthoryear{Lazarian \& Vishniac}{Lazarian \&
  Vishniac}{1999}]{lazarian_vishniac_1999}
Lazarian A.,  Vishniac E.,  1999, The Astrophysical Journal, 517, 700

\bibitem[\protect\citeauthoryear{Lazarian, Vlahos, Kowal, Yan, Beresnyak  \&
  Dal~Pino}{Lazarian et~al.}{2012}]{lazarian_etal_2012}
Lazarian A.,  Vlahos L.,  Kowal G.,  Yan H.,  Beresnyak A.,   Dal~Pino E.
  d.~G.,  2012, Space science reviews, 173, 557

\bibitem[\protect\citeauthoryear{Longair}{Longair}{2011}]{longair_2011}
Longair M.~S.,  2011, High energy astrophysics.
cambridge university Press

\bibitem[\protect\citeauthoryear{Lyutikov}{Lyutikov}{2006}]{lyutikov_2006}
Lyutikov M.,  2006, New Journal of Physics, 8, 119

\bibitem[\protect\citeauthoryear{Madejski \& Sikora}{Madejski \&
  Sikora}{2016}]{madejski_sikora_2016}
Madejski G.,  Sikora M.,  2016, Annual Review of Astronomy and Astrophysics,
  54, 725

\bibitem[\protect\citeauthoryear{{Makwana}, {Li}, {Guo}  \& {Li}}{{Makwana}
  et~al.}{2017}]{makwana_etal_2017}
{Makwana} K.,  {Li} H.,  {Guo} F.,   {Li} X.,  2017, in Journal of Physics
  Conference Series. p. 012004

\bibitem[\protect\citeauthoryear{Massaro, Tramacere, Perri, Giommi  \&
  Tosti}{Massaro et~al.}{2006}]{massaro_etal_2006}
Massaro E.,  Tramacere A.,  Perri M.,  Giommi P.,   Tosti G.,  2006, Astronomy
  \& Astrophysics, 448, 861

\bibitem[\protect\citeauthoryear{Matthaeus, Wan, Servidio, Greco, Osman,
  Oughton  \& Dmitruk}{Matthaeus et~al.}{2015}]{matthaeus_etal_2015}
Matthaeus W.,  Wan M.,  Servidio S.,  Greco A.,  Osman K.,  Oughton S.,
  Dmitruk P.,  2015, Philosophical Transactions of the Royal Society of London
  A: Mathematical, Physical and Engineering Sciences, 373, 20140154

\bibitem[\protect\citeauthoryear{Medvedev \& Spitkovsky}{Medvedev \&
  Spitkovsky}{2009}]{medvedev_spitkovsky_2009}
Medvedev M.~V.,  Spitkovsky A.,  2009, The Astrophysical Journal, 700, 956

\bibitem[\protect\citeauthoryear{Mehlhaff, Werner, Uzdensky  \&
  Begelman}{Mehlhaff et~al.}{2019}]{mehlhaff_etal_inprep}
Mehlhaff J.,  Werner G.,  Uzdensky D.,   Begelman M.,  2019, In preparation

\bibitem[\protect\citeauthoryear{Miller, Guessoum  \& Ramaty}{Miller
  et~al.}{1990}]{miller_etal_1990}
Miller J.~A.,  Guessoum N.,   Ramaty R.,  1990, The Astrophysical Journal, 361,
  701

\bibitem[\protect\citeauthoryear{Nalewajko, Begelman, Cerutti, Uzdensky  \&
  Sikora}{Nalewajko et~al.}{2012}]{nalewajko_etal_2012}
Nalewajko K.,  Begelman M.~C.,  Cerutti B.,  Uzdensky D.~A.,   Sikora M.,
  2012, Monthly Notices of the Royal Astronomical Society, 425, 2519

\bibitem[\protect\citeauthoryear{Nalewajko, Begelman  \& Sikora}{Nalewajko
  et~al.}{2014}]{nalewajko_etal_2014b}
Nalewajko K.,  Begelman M.~C.,   Sikora M.,  2014, The Astrophysical Journal,
  789, 161

\bibitem[\protect\citeauthoryear{Nalewajko, Yuan  \&
  Chru{\'s}li{\'n}ska}{Nalewajko et~al.}{2018}]{nalewajko_etal_2018}
Nalewajko K.,  Yuan Y.,   Chru{\'s}li{\'n}ska M.,  2018, Journal of Plasma
  Physics, 84

\bibitem[\protect\citeauthoryear{Narayan \& Kumar}{Narayan \&
  Kumar}{2009}]{narayan_kumar_2009}
Narayan R.,  Kumar P.,  2009, Monthly Notices of the Royal Astronomical
  Society: Letters, 394, L117

\bibitem[\protect\citeauthoryear{N{\"a}ttil{\"a}}{N{\"a}ttil{\"a}}{2019}]{nattila_2019}
N{\"a}ttil{\"a} J.,  2019, arXiv preprint arXiv:1906.06306

\bibitem[\protect\citeauthoryear{Nishikawa et~al.,}{Nishikawa
  et~al.}{2011}]{nishikawa_etal_2011}
Nishikawa K.-I.,  et~al., 2011, Advances in Space Research, 47, 1434

\bibitem[\protect\citeauthoryear{Philippov \& Spitkovsky}{Philippov \&
  Spitkovsky}{2018}]{philippov_spitkovsky_2018}
Philippov A.~A.,  Spitkovsky A.,  2018, The Astrophysical Journal, 855, 94

\bibitem[\protect\citeauthoryear{Roytershteyn, Karimabadi  \&
  Roberts}{Roytershteyn et~al.}{2015}]{roytershteyn_etal_2015}
Roytershteyn V.,  Karimabadi H.,   Roberts A.,  2015, Philosophical
  Transactions of the Royal Society of London A: Mathematical, Physical and
  Engineering Sciences, 373, 20140151

\bibitem[\protect\citeauthoryear{Rybicki \& Lightman}{Rybicki \&
  Lightman}{2008}]{rybicki_lightman_2008}
Rybicki G.~B.,  Lightman A.~P.,  2008, Radiative processes in astrophysics.
John Wiley \& Sons

\bibitem[\protect\citeauthoryear{Sambruna, Maraschi  \& Urry}{Sambruna
  et~al.}{1996}]{sambruna_etal_1996}
Sambruna R.~M.,  Maraschi L.,   Urry C.~M.,  1996, The Astrophysical Journal,
  463, 444

\bibitem[\protect\citeauthoryear{{Sari}, {Piran}  \& {Narayan}}{{Sari}
  et~al.}{1998}]{sari_piran_narayan}
{Sari} R.,  {Piran} T.,   {Narayan} R.,  1998, The Astrophysical Journal
  Letters, 497, L17

\bibitem[\protect\citeauthoryear{Schlickeiser}{Schlickeiser}{1984}]{schlickeiser_1984}
Schlickeiser R.,  1984, Astronomy and Astrophysics, 136, 227

\bibitem[\protect\citeauthoryear{Schlickeiser}{Schlickeiser}{1985}]{schlickeiser_1985}
Schlickeiser R.,  1985, Astronomy and Astrophysics, 143, 431

\bibitem[\protect\citeauthoryear{Schlickeiser}{Schlickeiser}{1989}]{schlickeiser_1989}
Schlickeiser R.,  1989, The Astrophysical Journal, 336, 243

\bibitem[\protect\citeauthoryear{Schoeffler, Grismayer, Uzdensky, Fonseca  \&
  Silva}{Schoeffler et~al.}{2019}]{schoeffler_etal_2019}
Schoeffler K.,  Grismayer T.,  Uzdensky D.,  Fonseca R.,   Silva L.,  2019, The
  Astrophysical Journal, 870, 49

\bibitem[\protect\citeauthoryear{Servidio, Osman, Valentini, Perrone, Califano,
  Chapman, Matthaeus  \& Veltri}{Servidio et~al.}{2014}]{servidio_etal_2014b}
Servidio S.,  Osman K.,  Valentini F.,  Perrone D.,  Califano F.,  Chapman S.,
  Matthaeus W.,   Veltri P.,  2014, The Astrophysical Journal Letters, 781, L27

\bibitem[\protect\citeauthoryear{Sironi \& Spitkovsky}{Sironi \&
  Spitkovsky}{2009}]{sironi_spitkovsky_2009}
Sironi L.,  Spitkovsky A.,  2009, The Astrophysical Journal Letters, 707, L92

\bibitem[\protect\citeauthoryear{Sobacchi \& Lyubarsky}{Sobacchi \&
  Lyubarsky}{2019}]{sobacchi_lyubarsky_2019}
Sobacchi E.,  Lyubarsky Y.,  2019, Monthly Notices of the Royal Astronomical
  Society, 484, 1192

\bibitem[\protect\citeauthoryear{Sobacchi \& Lyubarsky}{Sobacchi \&
  Lyubarsky}{2020}]{sobacchi_lyubarsky_2020}
Sobacchi E.,  Lyubarsky Y.~E.,  2020, Monthly Notices of the Royal Astronomical
  Society, 491, 3900

\bibitem[\protect\citeauthoryear{Squire, Schekochihin, Quataert  \&
  Kunz}{Squire et~al.}{2019}]{squire_etal_2019}
Squire J.,  Schekochihin A.~A.,  Quataert E.,   Kunz M.~W.,  2019, Journal of
  Plasma Physics, 85

\bibitem[\protect\citeauthoryear{Stawarz \& Petrosian}{Stawarz \&
  Petrosian}{2008}]{stawarz_petrosian_2008}
Stawarz {\L}.,  Petrosian V.,  2008, The Astrophysical Journal, 681, 1725

\bibitem[\protect\citeauthoryear{Takamoto \& Lazarian}{Takamoto \&
  Lazarian}{2016}]{takamoto_lazarian_2016}
Takamoto M.,  Lazarian A.,  2016, The Astrophysical Journal Letters, 831, L11

\bibitem[\protect\citeauthoryear{TenBarge, Howes, Dorland  \& Hammett}{TenBarge
  et~al.}{2014}]{tenbarge_etal_2014}
TenBarge J.,  Howes G.~G.,  Dorland W.,   Hammett G.~W.,  2014, Computer
  Physics Communications, 185, 578

\bibitem[\protect\citeauthoryear{Tenerani \& Velli}{Tenerani \&
  Velli}{2018}]{tenerani_velli_2018}
Tenerani A.,  Velli M.,  2018, The Astrophysical Journal Letters, 867, L26

\bibitem[\protect\citeauthoryear{Teraki \& Takahara}{Teraki \&
  Takahara}{2011}]{teraki_takahara_2011}
Teraki Y.,  Takahara F.,  2011, The Astrophysical Journal Letters, 735, L44

\bibitem[\protect\citeauthoryear{Thompson \& Blaes}{Thompson \&
  Blaes}{1998}]{thompson_blaes_1998}
Thompson C.,  Blaes O.,  1998, Physical Review D, 57, 3219

\bibitem[\protect\citeauthoryear{Towns et~al.,}{Towns et~al.}{2014}]{xsede}
Towns J.,  et~al., 2014, Computing in Science \& Engineering, 16, 62

\bibitem[\protect\citeauthoryear{Tramacere, Massaro  \& Taylor}{Tramacere
  et~al.}{2011}]{tramacere_etal_2011}
Tramacere A.,  Massaro E.,   Taylor A.,  2011, The Astrophysical Journal, 739,
  66

\bibitem[\protect\citeauthoryear{Uzdensky}{Uzdensky}{2018}]{uzdensky_2018}
Uzdensky D.~A.,  2018, Monthly Notices of the Royal Astronomical Society, 477,
  2849

\bibitem[\protect\citeauthoryear{Walker, Boldyrev  \& Loureiro}{Walker
  et~al.}{2018}]{walker_etal_2018}
Walker J.,  Boldyrev S.,   Loureiro N.~F.,  2018, Physical Review E, 98, 033209

\bibitem[\protect\citeauthoryear{{Weinberg}}{{Weinberg}}{1972}]{weinberg_1972}
{Weinberg} S.,  1972, {Gravitation and Cosmology: Principles and Applications
  of the General Theory of Relativity}

\bibitem[\protect\citeauthoryear{{Werner}, {Philippov}  \& {Uzdensky}}{{Werner}
  et~al.}{2019}]{werner_etal_2019}
{Werner} G.~R.,  {Philippov} A.~A.,   {Uzdensky} D.~A.,  2019, Monthly Notices
  of the Royal Astronomical Society: Letters, 482, L60

\bibitem[\protect\citeauthoryear{Wong, Zhdankin, Uzdensky, Werner  \&
  Begelman}{Wong et~al.}{2019}]{wong_etal_2019}
Wong K.,  Zhdankin V.,  Uzdensky D.~A.,  Werner G.~R.,   Begelman M.~C.,  2019,
  arXiv preprint arXiv:1901.03439

\bibitem[\protect\citeauthoryear{Xu \& Zhang}{Xu \&
  Zhang}{2017}]{xu_zhang_2017}
Xu S.,  Zhang B.,  2017, The Astrophysical Journal Letters, 846, L28

\bibitem[\protect\citeauthoryear{{Yuan}, {Nalewajko}, {Zrake}, {East}  \&
  {Blandford}}{{Yuan} et~al.}{2016}]{yuan_etal_2016}
{Yuan} Y.,  {Nalewajko} K.,  {Zrake} J.,  {East} W.~E.,   {Blandford} R.~D.,
  2016, The Astrophysical Journal, 828, 92

\bibitem[\protect\citeauthoryear{Zhdankin, Uzdensky, Perez  \&
  Boldyrev}{Zhdankin et~al.}{2013}]{zhdankin_etal_2013}
Zhdankin V.,  Uzdensky D.~A.,  Perez J.~C.,   Boldyrev S.,  2013, The
  Astrophysical Journal, 771, 124

\bibitem[\protect\citeauthoryear{Zhdankin, Boldyrev  \& Uzdensky}{Zhdankin
  et~al.}{2016}]{zhdankin_etal_2016b}
Zhdankin V.,  Boldyrev S.,   Uzdensky D.~A.,  2016, Physics of Plasmas, 23,
  055705

\bibitem[\protect\citeauthoryear{Zhdankin, Werner, Uzdensky  \&
  Begelman}{Zhdankin et~al.}{2017}]{zhdankin_etal_2017}
Zhdankin V.,  Werner G.~R.,  Uzdensky D.~A.,   Begelman M.~C.,  2017, Physical
  Review Letters, 118, 055103

\bibitem[\protect\citeauthoryear{{Zhdankin}, {Uzdensky}, {Werner}  \&
  {Begelman}}{{Zhdankin} et~al.}{2018a}]{zhdankin_etal_2018a}
{Zhdankin} V.,  {Uzdensky} D.~A.,  {Werner} G.~R.,   {Begelman} M.~C.,  2018a,
  Monthly Notices of the Royal Astronomical Society, 474, 2514

\bibitem[\protect\citeauthoryear{{Zhdankin}, {Uzdensky}, {Werner}  \&
  {Begelman}}{{Zhdankin} et~al.}{2018b}]{zhdankin_etal_2018b}
{Zhdankin} V.,  {Uzdensky} D.~A.,  {Werner} G.~R.,   {Begelman} M.~C.,  2018b,
  The Astrophysical Journal Letters, 867, L18

\bibitem[\protect\citeauthoryear{{Zhdankin}, {Uzdensky}, {Werner}  \&
  {Begelman}}{{Zhdankin} et~al.}{2019}]{zhdankin_etal_2019}
{Zhdankin} V.,  {Uzdensky} D.~A.,  {Werner} G.~R.,   {Begelman} M.~C.,  2019,
  Physical Review Letters, 122, 055101

\bibitem[\protect\citeauthoryear{Zrake}{Zrake}{2014}]{zrake_2014}
Zrake J.,  2014, The Astrophysical Journal Letters, 794, L26

\bibitem[\protect\citeauthoryear{Zrake \& East}{Zrake \&
  East}{2016}]{zrake_east_2016}
Zrake J.,  East W.~E.,  2016, The Astrophysical Journal, 817, 89

\bibitem[\protect\citeauthoryear{Zrake \& MacFadyen}{Zrake \&
  MacFadyen}{2011}]{zrake_macfadyen_2011}
Zrake J.,  MacFadyen A.~I.,  2011, The Astrophysical Journal, 744, 32

\bibitem[\protect\citeauthoryear{Zrake, Beloborodov  \& Lundman}{Zrake
  et~al.}{2019}]{zrake_etal_2019}
Zrake J.,  Beloborodov A.~M.,   Lundman C.,  2019, The Astrophysical Journal,
  885, 30

\bibitem[\protect\citeauthoryear{del Valle, de Gouveia Dal~Pino  \& Kowal}{del
  Valle et~al.}{2016}]{delValle_etal_2016}
del Valle M.~V.,  de Gouveia Dal~Pino E.,   Kowal G.,  2016, Monthly Notices of
  the Royal Astronomical Society, 463, 4331

\makeatother
\end{thebibliography}



\appendix

\section{Steady-state Fokker-Planck model} \label{app1}

Given a stochastic scattering process, nonthermal particle acceleration can generally be modeled with a Fokker-Planck diffusion-advection equation for the global momentum distribution $F(\boldsymbol{p}, t)$ \citep[e.g.,][]{blandford_eichler_1987}. The Fokker-Planck equation, along with a radiative cooling term, can be written as
\begin{align}
\partial_t F = \frac{\partial}{\partial \boldsymbol{p}} \cdot \left( {\bf D}_{pp} \cdot \frac{\partial F}{\partial \boldsymbol{p}} \right) - \frac{\partial}{\partial \boldsymbol{p}} \cdot \left( \boldsymbol{A}_p F + \boldsymbol{F}_{\rm IC} F \right) \, ,
\end{align}
where $\boldsymbol{F}_{\rm IC} = - (4/3) \sigma_T U_{\rm ph} (p/m_ec)^2 \hat{\boldsymbol{p}}$ is the IC radiation backreaction force. Assuming isotropy for simplicity, the diffusion coefficient becomes~${\bf D}_{pp}(\boldsymbol{p}) = D_{pp}(p) {\bf I}$ and the advection coefficient becomes $\boldsymbol{A}_p(\boldsymbol{p}) = A_p(p) \hat{\boldsymbol{p}}$. The energy distribution (assuming ultra-relativistic particles, $\gamma = p/m_e c$) can be obtained by~$f(\gamma) = 4 \pi p^2 m_e c F(\boldsymbol{p})$, allowing us to write the Fokker-Planck equation for~$f(\gamma)$,
\begin{align}
\partial_t f &= \partial_\gamma \left( \gamma^2 D_{pp} \partial_\gamma \frac{f}{\gamma^2} \right) - \partial_\gamma \left(A_{p} f - \frac{\gamma^2}{\gamma_0 \tau_c} f \right) \, , \label{eq:fp}
\end{align}
where $\gamma_0$ is a characteristic (reference) energy and $\tau_c = 3 m_e c^2/(4 \sigma_T U_{\rm ph} \gamma_0)$ describes the radiative cooling timescale. We now consider the steady state, $\partial_t f = 0$. Eq.~\ref{eq:fp} can be integrated in~$\gamma$ using the boundary condition that~$D_{pp}(\gamma)$ and~$A_p(\gamma)$ approach a constant value (or zero) as~$\gamma \to 0$. We can then write
\begin{align}
\partial_\gamma \log{f} &= \frac{2}{\gamma} + \frac{A_p - \gamma^2/\gamma_0\tau_c}{D_{pp}} \, .
\end{align}
In this work, we consider the following simple form for $D_{pp}(\gamma)$ and $A_p(\gamma)$:
\begin{align}
A_p &= \frac{\gamma_0}{\tau_h} + \frac{\gamma}{\tau_a} \, , \nonumber \\
D_{pp} &= \frac{\gamma_0^2}{\tau_0} + \frac{\gamma_0 \gamma}{\tau_1} + \frac{\gamma^2}{\tau_2} \, .
\end{align}
Letting~$x = \gamma/\gamma_0$, and defining the rates compared to the cooling timescale~$\Gamma_s = \tau_c/\tau_s$, we have
\begin{align}
\partial_x \log{f} &= \frac{2}{x} + \frac{\Gamma_h + \Gamma_a x - x^2}{\Gamma_0 + \Gamma_1 x + \Gamma_2 x^2} \, .
\end{align}
This can be analytically integrated to obtain the general, five-parameter solution for $f(x)$:
\begin{align}
f(x) \propto x^2 \left( \Gamma_0 + \Gamma_1 x + \Gamma_2 x^2 \right)^{(\Gamma_a \Gamma_2 + \Gamma_1)/2 \Gamma_2^2} \exp{\left(- \frac{x}{\Gamma_2}\right)} \times \nonumber \\ 
 \exp{\left[ \frac{2\Gamma_0 \Gamma_2 - \Gamma_1^2 + 2 \Gamma_h \Gamma_2^2 - \Gamma_a \Gamma_1 \Gamma_2}{\Gamma_2^2 \sqrt{4 \Gamma_0 \Gamma_2 - \Gamma_1^2}} \tan^{-1}{\left( \frac{\Gamma_1 + 2 \Gamma_2 x}{\sqrt{4 \Gamma_0 \Gamma_2 - \Gamma_1^2}} \right)} \right]} \, . \label{eq:fpsol}
\end{align}
As~$x \to 0$, as long as~$\Gamma_0 > 0$, $f(x) \sim x^2$ as required by continuity. At intermediate values of $x$, for nonzero $\Gamma_1 > 0$, if the ordering $\Gamma_0 \ll \Gamma_1 x \ll \Gamma_2 x^2$ is satisfied, there may be a power law with index $2 + (\Gamma_a \Gamma_2 + \Gamma_1)/2 \Gamma_2^2$ (which is thus increasing with energy unless $\Gamma_a$ is sufficiently negative). Finally, at large $x \gtrsim \Gamma_2$, there is an exponential cutoff. If all~$\Gamma_s$ are zero except for~$\Gamma_2$, then a Maxwell-J\"{u}ttner distribution is recovered, with temperature $\Theta = \Gamma_2 \gamma_0$. In general, adding the other $\Gamma_s$ produces modest nonthermal corrections to the Maxwell-J\"{u}ttner distribution. \blue{Thus, within this steady-state Fokker-Planck framework, we do not expect to see an extended nonthermal population, even if there is an efficient first-order acceleration mechanism.}

In this work, we restrict ourselves to $\Gamma_1 = 0$ and $\Gamma_0 = \Gamma_2$ (anticipating that the diffusion process becomes independent of energy for particles with energies less than the mean energy). Thus, the advection and diffusion coefficients are
\begin{align}
A_p &= \frac{\gamma_0}{\tau_c} \left( \Gamma_h + \Gamma_a x \right) \, , \nonumber \\
D_{pp} &= \frac{\gamma_0^2}{\tau_c} \Gamma_2 \left( 1 + x^2 \right) \, .
\end{align}
Eq.~\ref{eq:fpsol} then becomes the three-parameter solution
 \begin{align}
f(x) \propto x^2 \left( 1 + x^2 \right)^{\Gamma_a/2 \Gamma_2}\exp{\left(- \frac{x}{\Gamma_2} + \frac{\Gamma_h + 1}{ \Gamma_2} \tan^{-1}{x} \right)} \, . \label{eq:fpsolred}
\end{align}
Note that since fitting to a physical distribution requires choosing $\gamma_0$, the actual number of fitting parameters is four ($\gamma_0$, $\Gamma_h$, $\Gamma_a$, and $\Gamma_2$) .

We choose the remaining parameters as follows. We set $\gamma_0 = 300$, approximately the mean energy measured in the simulations. We suppose that the diffusive acceleration timescale is given by (isotropic) second-order Fermi acceleration \citep{longair_2011} or gyroresonant interactions with Alfv\'{e}nic fluctuations \citep[e.g.,][]{schlickeiser_1989}, which can be expressed by
\begin{align}
\tau_2 \sim \frac{3 \lambda_{\rm mfp} c}{u_A^2} \sim \frac{3 L}{\eta_2 \sigma c} \, ,
\end{align}
where $u_A = v_A (1-v_A^2/c^2)^{-1/2}$ is the Alfv\'{e}n four-velocity, $\lambda_{\rm mfp}$ is the scattering mean free path, and $\eta_2$ is the efficiency of scattering. Using the radiative steady-state condition (Eq.~\ref{eq:ss:gamma}), assuming $\Theta_{ss} = \gamma_0/3$, we also have
\begin{align}
\tau_c &= \frac{4}{\eta_{\rm inj}} \frac{L}{\sigma v_A} \, .
\end{align}
Combining the two, we obtain
\begin{align}
\Gamma_2 = \frac{\tau_c}{\tau_2} \sim \frac{4 \eta_2}{3 \eta_{\rm inj}} \frac{c}{v_A} \propto \frac{c}{v_A} \, . \label{eq:gamma2}
\end{align}
We thus take $\Gamma_2 \propto c/v_A$ when comparing different simulations with varying $\langle\sigma\rangle$. We then have two unconstrained parameters, $\Gamma_h$ and $\Gamma_a$, which we choose by hand. These two terms control the effect of heating processes and first-order acceleration processes, respectively, which have not (to our best knowledge) been rigorously modeled by past analytic works.


\bsp	
\label{lastpage}
\end{document}